\documentclass[12pt]{spieman}  
\usepackage{amsmath,amsfonts,amssymb}
\usepackage{graphicx}
\usepackage{setspace}
\usepackage{tocloft}
\usepackage{lineno}
\usepackage{multirow}
\usepackage{hyperref}
\hypersetup{breaklinks=true}

\usepackage{url}

\newcommand{\minus}{\scalebox{0.75}[1.0]{$-$}}

\usepackage{supertabular}
\usepackage{caption}

\usepackage[dvipsnames]{xcolor}

\title{Leveraging Photometry for Deconfusion of Directly Imaged Multi-Planet Systems}

\author[a,*]{Samantha N.~Hasler}
\author[b]{Leonid Pogorelyuk}
\author[c]{Riley Fitzgerald}
\author[a,d]{Kerri L.~Cahoy}
\author[e]{Rhonda Morgan}
\affil[a]{Massachusetts Institute of Technology, Department of Earth, Atmospheric and Planetary Sciences, 77 Massachusetts Avenue, Cambridge, MA 02139, USA}
\affil[b]{Rensselaer Polytechnic Institute, Department of Mechanical, Aerospace and Nuclear Engineering, 110 8th Street, Troy, NY 12180,  USA}
\affil[c]{Virginia Tech, Department of Aerospace and Ocean Engineering, 
            1600 Innovation Drive, Blacksburg, VA 24061}
\affil[d]{Massachusetts Institute of Technology, Department of Aeronautics and Astronautics, 77 Massachusetts Avenue, Cambridge, MA 02139}
\affil[e]{Jet Propulsion Laboratory, 4800 Oak Grove Drive, Pasadena, CA, 91109}

\cftpagenumbersoff{figure}
\cftpagenumbersoff{table} 
\begin{document} 
\maketitle

\begin{abstract}
Future missions, including the Habitable Worlds Observatory, will aim to image Earth-like exoplanets around Sun-like stars in reflected light. 
Determining whether an exoplanet is in the habitable zone of its star may be difficult in multi-planet systems when the observer does not know in advance which detection corresponds to which planet. 
This ``confusion'' problem  will be a concern for future missions due to the high occurrence rate of multi-planet systems, and will be exacerbated by lack of prior knowledge about planets' orbital parameters or characteristics.
We address the exoplanet confusion problem by applying a photometry model to update an orbit ranking scheme for a ``deconfuser'' tool . This helps to account for phase variation of planets throughout their orbits.
We demonstrate the updated ranking scheme as a proof-of-concept on a subset of known to be confused simulated multi-planet systems among three inclination groupings  (low, medium, and high).
We find that incorporating photometry improves correctly interpreting previously confused orbits in more than half of these particularly challenging cases.
These results emphasize that photometry is useful for orbit discrimination and deconfusion of directly imaged multi-planet systems, providing a framework for including photometry alongside astrometry when fitting orbits to detections. 
\end{abstract}


{\noindent \footnotesize\textbf{*}Corresponding author,  \linkable{shasler@mit.edu} }

\begin{spacing}{2}   

\section{Introduction}
\label{sec:intro}  

    The Astro2020 decadal survey recommended a large IR/O/UV strategic mission to yield direct detections of exoplanets in reflected light at close-in orbital separations \cite{astro2020}. 
    Such a mission, now called the ``Habitable Worlds Observatory'' (HWO), will directly image exoplanets in reflected light and characterize potentially habitable worlds around Sun-like stars. 
    The goal for such a mission is to reach planet-star contrast levels of at least $1\mathrm{e}$\minus$10$ \cite{astro2020, Feinberg2024SPIE13092E..1NF}. While the instrument specifications are not yet defined, the inner working angle (IWA) will likely be similar to that of the HabEx Mission Concept \cite{habex2020arXiv} (2.4 $\lambda/D$).
    A coronagraph technology demonstration that will enable HWO's capabilities will be flown on the Nancy Grace Roman Space Telescope (Roman) \cite{Mennesson2020arXiv200805624M}. 
    The Roman Coronagraph is expected to achieve contrast ratios \cite{Bailey2023SPIE12680E..0TB} of at least $1\mathrm{e}$\minus$8$ at separations of 3 to 9 $\lambda/D$ .

    When observing with future direct imaging missions, we may encounter challenges in identifying and characterizing new exoplanets in multi-planet systems. 
    A ``confusion'' problem can arise when multiple detections are made of a multi-planet system and it is not clear which detection belongs to which planet \cite{keithly2021solar, pogorelyuk2022deconfusing}. 
    Previous work modeling the Solar System demonstrates that our own system's planets may encounter many incidences of confusion in brightness and separation between planets, especially at higher viewing inclinations \cite{keithly2021solar}.
    In this work, we refer to ``confusion'' as the uncertainty in distinguishing between planet detections across multiple epochs of a set of direct imaging observations. 
    The confusion problem can make it difficult to differentiate planets in a system and will exacerbate the orbit-fitting and characterization problem. 
    
    To address the confusion problem, Pogorelyuk et al.~2022~\cite{pogorelyuk2022deconfusing} developed an algorithm called the ``deconfuser'' (\href{https://github.com/MIT-STARLab/deconfuser}{https://github.com/MIT-STARLab/deconfuser}) which efficiently fits orbits to observations of planets given no prior information on the system. 
    The deconfuser was developed to provide a faster orbit-fitting method for flight programs and to address the planet-detection confusion problem.
    The tool accepts unlabeled astrometric detections of planets and generates all possible combinations of orbit matches for a system.
    It specifically provides discrete solutions to match detections to planets in a set of images, within some tolerance set by the user. 
    This approach is different than other orbit-fitting tools, such as those that employ Monte Carlo (MC) methods (e.g., {\ttfamily Octofitter} \cite{thompson2023octofitter}, {\ttfamily orbitize!} \cite{blunt2020orbitize, Blunt2024JOSS....9.6756B}, {\ttfamily PyAstrOFit} \cite{wertz2017pyastrofit}, {\ttfamily orvara} \cite{Brandt2021AJ....162..186B}) to fit distributions of orbital parameters to planet detections.
    MC orbit-fitting tools also can often be relatively slow due to their iterative processes. 
    For example, testing one set of orbit combinations with {\ttfamily orbitize!}~took the authors $\sim$6.4 hours on a standard MacBook Pro with 20 temperatures, 500 walkers, and 10,000 steps per walker.
    For the same system (i.e., three planets, three detections), the deconfuser can process all orbit matches in less than 0.2\,seconds \cite{pogorelyuk2022deconfusing} (a direct comparison between deconfuser and {\ttfamily orbitize!}~results can be found in Appendix~\ref{appendixC_orbitize}).
    While traditional MCMC orbit-fitting methods provide posterior distributions and error bars on the fits, the deconfuser provides discrete fits without error ranges.
    
    Additionally, existing orbit-fitting tools require the user to know which detection belongs to which planet prior to fitting and do not address the detection-confusion problem.
    They also do not consider photometric variation due to orbital phase or atmospheric characteristics when returning orbit fits for a given set of detections. 
    Existing orbit-fitting tools have largely been applied to self-luminous giant planets (e.g., Mesa et al.~2023\cite{Mesa2023A&A...672A..93M}, Blunt et al.~2023\cite{Blunt2023AJ....166..257B}, Do Ó et al.~2025\cite{DoO2025ApJ...995..190D}), whose brightnesses do not vary in the same manner as reflected light planets as a function of orbital position.
    Roman Coronagraph and HWO will push us into a new regime with reflected light observations of mature exoplanets, highlighting the need for tools that account for exoplanet photometry.

    For multi-planet systems, there may be many ways to assign detections to planets, which can lead to numerous combinations of orbits per system that each describe the data equally well. 
    In the deconfuser, orbit matches are initially ranked based on the combinations of orbits that best describe the detections.
    While relative astrometry is good enough to deconfuse many cases with three to four epochs of observation \cite{pogorelyuk2022deconfusing}, there are still problematic corner cases where confusion remains and relative astrometric information does not provide enough insight for orbit differentiation. 
    This is true even in the case of perfect knowledge of the detections' positions (i.e., zero uncertainty). 
    Confusion is a fundamental issue, which can leave many possible combinations of orbits for a set of detections and would benefit from additional ranking steps.

    In this work, we build upon the work presented in Pogorelyuk et al.~2022\cite{pogorelyuk2022deconfusing} (hereafter, the ``astrometric deconfuser'') by developing an additional orbit-ranking step for the deconfuser with the goal of reducing confusion for difficult systems by leveraging the phase variation of orbiting planets, using intensity-only variation. 
    We apply the updated orbit ranking scheme (the ``combined deconfuser'') on a set of highly confused simulated systems across low ($i = 0-45^\circ$), medium ($i = 45-70^\circ$), and high inclination ($i = 70-90^\circ$) ranges to assess how we can use photometric information and planetary orbital phase to deconfuse observations of multi-planet systems across inclination ranges.
    We explore here whether photometry is sufficient for deconfusion in highly confused systems, or if other techniques may still be required for orbit differentiation.

    Incorporating photometric variation into simulated observations and planet-planet system identification is an involved process; many parameters influence our observations ranging from factors including phase angle variation, individual planet's atmospheric parameters, instrument systematics, and joint dependence of variables (e.g., astrometry and photometry). 
    Other physical parameters such as rings and moons can also influence exoplanetary spectra and photometry \cite{Arnold2004A&A...420.1153A, Dyudina2005ApJ...618..973D, Coulter2022ApJS..263...15C, Limbach2024AJ....168...57L}.
    The scope of this paper is to assume that we have perfect knowledge of planets' relative astrometric locations in an image and do not yet include the joint dependence of photometry and relative astrometry. 
    That is, the brightness of a planet detection does not influence our knowledge of the planet's location in terms of astrometric error.
    We do not yet include rings or moons in our simulations.
    In this work, we assess if and when single-band photometric variation due to orbital phase can assist in ranking exoplanet orbit fits using an example subset of highly confused simulated systems.
    The method presented in this work uses all information that would be immediately available to observers from a set of photometric observations.
    
    Accurate orbit determination may also support improved exoplanet characterization. 
    Previous work showed that prior knowledge of a planet's orbit and phase angle can improve the constraints on the range of possible radii \cite{Salvador2024ApJ...969L..22S}. 
    Carrión-González et al.~2020\cite{carrion-gonzalez2020A&A...640A.136C} concluded that unknown planetary radius will prevent the detection of clouds and blur other atmospheric information. 
    Improving orbit determination may therefore trickle down to support atmospheric characterization by reducing radius uncertainties.

\subsection{Photometry from Direct Imaging}\label{subsec:bkg_photometry}

\begin{figure*}[h!]
      \centering
      \includegraphics[width=0.9\textwidth]{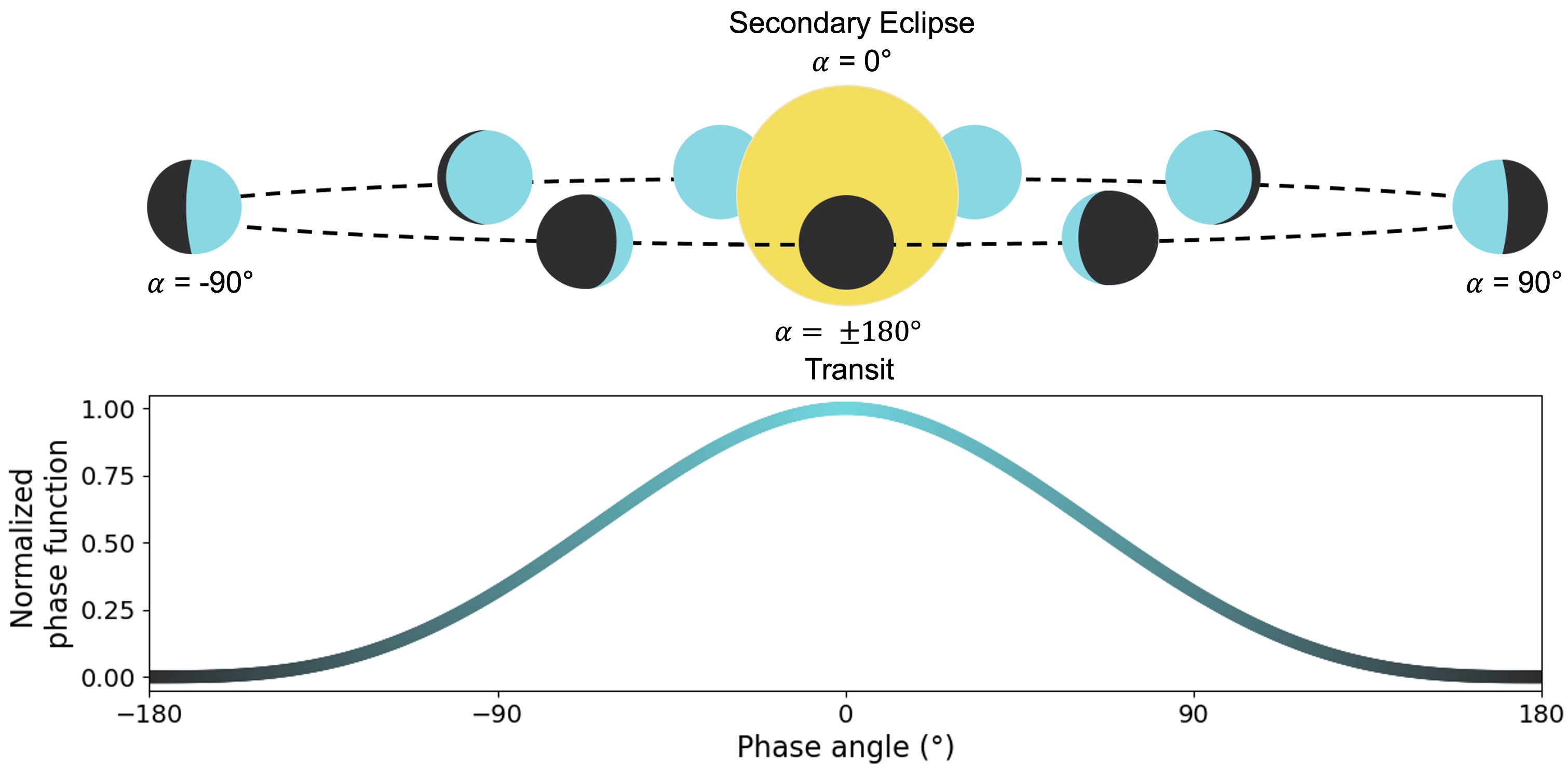}
      \caption{Planetary orbital phase illustration for an edge-on system ($i=90^\circ$) (top), with the phase angle convention used in this work and a simulated Lambertian phase curve (bottom) (see Sec.~\ref{subsubsec:phase_function} and Eq.~(\ref{eqn:lambertian}) \cite{brown2005single}). Full phase, or $\alpha=0^\circ$, occurs behind the star, relative to the observer (secondary eclipse); $\alpha=180^\circ$ occurs during the planet's primary eclipse of the star. $\pm 180^\circ$ are the same point on the orbit. We note that the planetary systems explored in this work will not necessarily transit their host star and this figure is for illustrative purposes.} 
      \label{fig:phase_illustration}
    \end{figure*}

    This section presents a brief overview of exoplanet photometry in the context of our work.
    Direct imaging provides detections of the light reflected from a planet's surface and atmosphere, or the self-luminosity of a planet due to internal heat sources, depending on the wavelength of observation. 
    In this work, we assume visible band observations of reflected light. 
    Combining reflected light observations with the orbital motion of a planet results in variations in brightness and color due to changing planetary phase angle with respect to the observer (see Fig.~\ref{fig:phase_illustration}; e.g., Traub et al.~2010\cite{traub2010direct}, Cahoy et al.~2010\cite{cahoy2010exoplanet}, Madhusudhan et al.~2012\cite{madhusudhan2012analytic}). 
    Previous studies have shown that a planet's brightness, color, and reflected light spectra change based on the type of planet being observed and the planet's location in its orbit (e.g., Traub 2003\cite{traub2003extrasolar}, Mallama 2007\cite{Mallama2007Icar..192..404M}, Mallama 2009\cite{Mallama2009Icar..204...11M}, Cahoy et al.~2010\cite{cahoy2010exoplanet}, Madhusudhan et al.~2012\cite{madhusudhan2012analytic}, Batalha et al.~2018\cite{batalha2018color}, Mallama et al.~2018\cite{Mallama2018A&C....25...10M}, Smith et al.~2020\cite{smith2020utilizing}). 
    These effects are observed because the properties depend on the geometric albedo and the planet's phase angle, which are illustrated in Fig.~\ref{fig:phase_illustration} and through the following equation for the planet-to-star flux ratio \cite{sobolev1975light},
    
    \begin{equation}\label{eqn:fluxratio}
        \frac{F_p(\lambda, \alpha)}{F_s(\lambda)} = A_g(\lambda)(\frac{R_p}{d})^2\Phi(\lambda, \alpha),
    \end{equation}

    \noindent where $\lambda$ is the wavelength of observation. $\alpha$ is the phase angle, where $\alpha = 180^\circ$ is the angle when the planet is transiting with respect to the observer, and $\alpha = 0^\circ$ is the angle at secondary eclipse (see Fig.~\ref{fig:phase_illustration}). $A_g(\lambda)$ is the wavelength-dependent geometric albedo of the planet at $\alpha=0^\circ$ (full-phase), $R_p$ is the planet's radius, $d$ is the planet-star separation, and $\Phi(\lambda, \alpha)$ is the planet's phase function. 
    The geometric albedo describes the reflectivity of an object at full phase ($\alpha = 0^\circ$) relative to an identically sized isotropically reflecting surface under the same incident flux. 
    As a planet orbits its host star, the planetary phase change appears to the observer as a change in brightness. 
    In Fig.~\ref{fig:phase_illustration}, the model phase function is a Lambertian, which assumes diffuse reflection in all directions (see Sec.~\ref{subsubsec:phase_function} and Lambert 1760\cite{lambert1760photometria}). 
    We can leverage this photometric variation in the combined deconfuser to gather information about a planet as it progresses along its orbit, and to assist in breaking ties between orbit options that would otherwise be close to equally ranked.

    Including atmospheric characteristics or phase variation in the deconfuser could support differentiation of orbit assignments.
    Previous studies have highlighted the dependence of planetary brightness on orbital separation, location, and atmospheric composition (e.g., Traub 2003\cite{traub2003extrasolar}, Mallama 2007\cite{Mallama2007Icar..192..404M}, Mallama 2009\cite{Mallama2009Icar..204...11M}, Cahoy et al.~2010\cite{cahoy2010exoplanet}, Mallama \& Schmude 2012\cite{Mallama2012Icar..220..211M}, Gao et al.~2017\cite{Gao2017AJ....153..139G}, Batalha et al.~2018\cite{batalha2018color}, Madden \& Kaltenegger 2018\cite{Madden2018AsBio..18.1559M}, Mallama \& Hilton 2018\cite{Mallama2018A&C....25...10M}, Smith et al.~2020\cite{smith2020utilizing} 
    ). 
    Planetary albedo can be dependent on separation due to the amount of flux received and the instellation influencing various condensable species and photochemistry in atmospheres. 
    Previous works have expressed these results through model spectra (e.g., Gao et al.~2017\cite{Gao2017AJ....153..139G}, Hu et al.~2019\cite{Hu2019ApJ...887..166H}) and Solar System observations (e.g., Mallama et al.~2006\cite{Mallama2006Icar..182...10M}, Mallama \& Hilton 2018\cite{Mallama2018A&C....25...10M}). 
    We develop an expanded version of the deconfuser with the goal of supporting the rejection of geometrically valid orbits that are inconsistent with the expected variation in intensity with orbital phase.

    In this work we model brightness variations throughout an orbit with a single assumed planetary albedo and a Lambertian phase function (see Sec.~\ref{sec:phot_model}).
    However, Solar System observations have shown that planetary albedos may vary across epochs (e.g., Lockwood et al.~2006\cite{Lockwood2006Icar..180..442L}, Palle et al.~2016\cite{Palle2016GeoRL..43.4531P}, Irwin et al.~2025\cite{Irwin2025MNRAS.540.1719I}). 
    Solar System phase functions also deviate slightly from a Lambertian assumption (e.g., Mayorga et al.~2016\cite{Mayorga2016AJ....152..209M}, Mallama et al.~2018\cite{Mallama2018A&C....25...10M}, Mayorga et al.~2020\cite{Mayorga2020AJ....160..238M}). 
    Regardless, Lambertian reflectance generally approximates the shape of a reflected light planetary phase function as a planet orbits its star.
    We expand on these limitations in Sec.~\ref{subsec:lambertian_assumption} and describe avenues of future work to address these limitations in Sec.~\ref{sec:conclusion}. 

    The confused systems presented in this work are three-planet systems for which the astrometric deconfuser\cite{pogorelyuk2022deconfusing} is unable to find a single solution to match the detections given only their relative astrometry.
    We intentionally examine only a few known difficult scenarios to demonstrate a proof-of-concept with the photometry ranking scheme, since not all system cases will require additional help with deconfusion (see confusion rates presented in Pogorelyuk et al.~2022\cite{pogorelyuk2022deconfusing}).     
    We demonstrate in Sec.~\ref{sec:results} that single-band photometry shows promise for differentiating orbits in more than half of these representative highly confused cases.

 \subsection{Organization}
    
    We outline our approach to including additional ranking metrics in the deconfuser and describe how we test whether photometry is sufficient for differentiating orbits in Sec.~\ref{sec:approach}. 
    In Sec.~\ref{sec:results}, we present the results of the combined deconfuser on the confused systems. 
    We describe the impact of photometry on deconfusion for these challenging systems and the performance of the combined deconfuser in estimating the correct orbital parameters in Sec.~\ref{sec:discussion}. 
    In Sec.~\ref{sec:conclusion}, we summarize the results of this work and present directions for future investigations and improvements with the deconfuser.

\section{Approach}\label{sec:approach}

    In this work, we investigate the influence of photometry on deconfusion by applying a photometry model and a new likelihood orbit-ranking scheme to the astrometric deconfuser.
    We demonstrate the utility of photometry for differentiating orbits using a representative set of highly confused corner cases across three inclination groups (low, medium, and high inclination), assuming all planets are visible at all times. 
    We adopt the definition of confusion from Pogorelyuk et al.~2022\cite{pogorelyuk2022deconfusing}, where a system is considered to be confused if the astrometric deconfuser provides two or more closely ranked orbit assignments per system. 
   When fit by the astrometric deconfuser, orbit options may be considered closely ranked if they fall within the ``acceptable'' RMS error, which is related to the user-defined tolerance parameter that is set to 0.05\,AU in this work.

\begin{figure*}[h!]
        \centering
        \includegraphics[width=\linewidth]{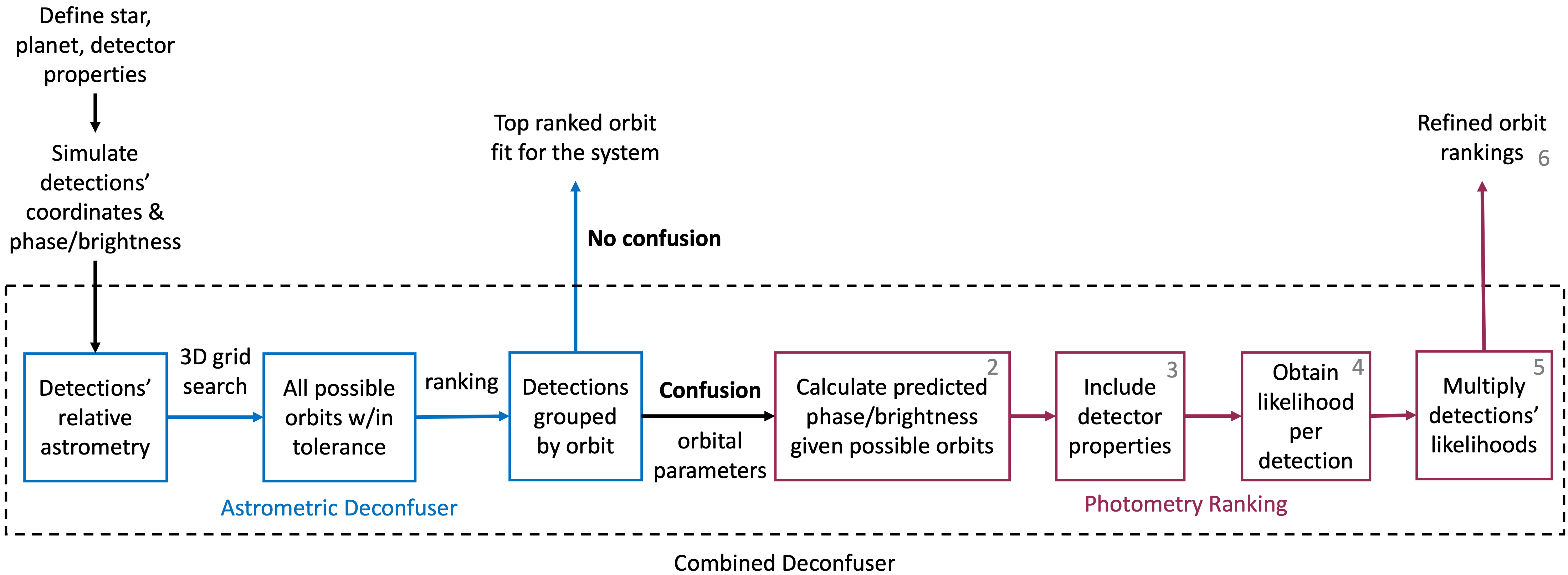}
        \caption{Flow of the photometry ranking algorithm as it fits with the deconfuser. The photometry model accepts the orbital parameters for each confused orbit option from the astrometric deconfuser and calculates the phase angle and expected planetary brightness for each detection. Detector noise is added to obtain a more realistic brightness for each detection. We compare the likelihood of each confused orbit option given the planet detections, which translates to a ranking value for each confused system option. }
        \label{fig:code_flow}
    \end{figure*}

    The error returned by the astrometric deconfuser describes the RMS of the fit in the X and Y directions of the plane of the sky\cite{pogorelyuk2022deconfusing}, while the tolerance defines how fine the grid search is during the orbit fitting step -- \textit{if} the observations can be perfectly fit with an orbit on the grid, the RMS is guaranteed to be smaller than that tolerance.
    However, cases that push the limits of the grid search during orbit fitting (e.g., eccentricity above the maximum search value) may return RMS errors greater than the tolerance threshold.
    Thus, it is possible to have confused orbit fits from the astrometric deconfuser that are within the tolerance but have relatively higher RMS error that still cannot be easily differentiated using relative astrometry alone. 
    After fitting orbits and ranking assignments based on the number of matched detections with the astrometric deconfuser, we use an additional ranking scheme to break ties between orbits if they have a more statistically likely brightness result.
    Figure~\ref{fig:code_flow} illustrates the combined flow of the algorithms presented in this work.

    In this work, the photometry ranking is applied as a secondary step after the astrometric orbit fitting step rather than being incorporated directly into the orbit search step. 
    The astrometric deconfuser algorithm is designed to operate on astrometric detections with a single tolerance parameter that corresponds to astrometric uncertainty \cite{pogorelyuk2022deconfusing}, which is currently constant across all epochs.
    Because photometry introduces additional epoch-dependent uncertainties and a joint dependence between brightness and relative astrometric precision, incorporating photometry directly into the orbit search requires a more complex statistical treatment of the two.
    Here we instead treat photometry as an additional ranking metric applied only when relative astrometry alone cannot identify one solution, providing a modular proof-of-concept for leveraging intensity variation due to phase angle in orbit discrimination.
    
\subsection{Photometry Model}\label{sec:phot_model}

    The photometry ranking scheme begins with a photometry model at one bandpass, which returns the phase angle and brightness of a planet at any location on its orbit and calculates the brightness measured at the instrument for each detection. 
    We have both the truth parameters of the planetary system, as well as the multiple possible parameters of the confused system options.
    To determine the expected contrast of a planet detection, we first calculate the planet's separation from its host star. 
    The deconfuser provides the three-dimensional position (x, y, z) of a simulated planet's location (see Fig.~\ref{fig:code_flow}, step 1). 
    The position is used to calculate the planet's separation from its host star. 
    Then, we use the planet position to determine the phase angle of the planet detection with respect to the observer (see Fig.~\ref{fig:code_flow}, step 2). 
    The phase angle is calculated using Eq.~(\ref{eqn:phase_angle}),
    
    \begin{equation}\label{eqn:phase_angle}
            \alpha = \cos^{-1}\left(\frac{\vec{r} \cdot \vec{s}}{|\vec{r} | |\vec{s} |}\right),
    \end{equation}

    \noindent where $\vec{r}$ is the vector describing the planet's separation from the host star and $\vec{s}$ is the vector describing the system's separation from the observer. 
    For the corner cases described here (details given in Table~\ref{tab:all_system_simulations}), the planetary system locations are set to a distance of ten parsecs, placing them well within the range of the most accessible targets presented in NASA Exoplanet Exploration Program's Mission Target Star List for HWO \cite{Mamajek2024arXiv240212414M}. 
    
    \subsubsection{Phase Function}\label{subsubsec:phase_function}
    
    The phase angle calculated in Eq.~(\ref{eqn:phase_angle}) is used in the phase function of the planet at each orbital phase. For simplicity, we assume a Lambertian phase function and adopt the following equation (as presented in Brown 2005\cite{brown2005single}):

    \begin{equation}\label{eqn:lambertian}
        \Phi(\alpha) = \frac{\sin(\alpha) + (\pi - \alpha)\cos(\alpha)}{\pi}.
    \end{equation}

    \noindent Combining Eqs.~(\ref{eqn:fluxratio}) and~(\ref{eqn:lambertian}) brings us to an ``expected'' brightness of an observation. 
    This is the planet-star flux ratio that we would expect to receive at the observing system.

    Assuming Lambertian reflectance is a simple approximation that is commonly chosen when modeling and comparing reflected light from planets (e.g., Zugger et al.~2010\cite{Zugger2010ApJ...723.1168Z}, Fujii \& Kawahara 2012\cite{Fujii2012ApJ...755..101F}, Cowan et al.~2013\cite{Cowan2013MNRAS.434.2465C}, Morgan et al.~2019\cite{Morgan2019sdet.rept.....M}, DeCock et al.~2022\cite{DeCock2022AJ....163....5D}), but the Lambertian assumption neglects the dependence of scattering on albedo, wavelength, and complex atmospheric properties (e.g., Greco \& Burrows 2015\cite{Greco2015ApJ...808..172G}, Mayorga et al.~2016\cite{Mayorga2016AJ....152..209M}).
    For an initial study of how planetary phase variation can inform orbit differentiation, a Lambertian approximates the overall shape of a phase curve well to model phase angle variation.
    Future work could incorporate phase functions that describe more complex processes, such as the one- or two-term Henyey-Greenstein phase functions, which introduce additional parameters to model forward- and back-scattering processes \cite{Henyey1941ApJ....93...70H, Irvine1965ApJ...142.1563I}. 

    \subsubsection{Noise Model}
    
    We also applied a noise model of detector parameters given the planet flux returned by the photometry model (see Fig.~\ref{fig:code_flow}, step 3).
    We adopt the instrument parameters defined for the Roman Coronagraph Instrument, as Roman's Coronagraph has been extensively modeled and tested in preparation for its expected launch in Fall 2026 \cite{Perkins2024SPIE13092E..0RP}. 
    The instrument parameters for the Roman Coronagraph Instrument are available online through IPAC at Caltech \cite{roman_ipac_website}, and the parameters that we use in our code (from Phase C, ``Beginning of Life'') are listed in Table~\ref{tab:parameters}. 
    
    \begin{table*}[ht!]
        \caption[Telescope and instrument parameter values for simulations]{Telescope and instrument parameter values for simulations (see Section~\ref{sec:phot_model}).}                
        \label{tab:parameters}    
        \centering                        
        \begin{tabular}{l l c}      
        \hline\hline               
        Parameter & Description & Value \\         
        \hline                      
        $\lambda$ & central wavelength for bandpass (nm) & 573.8 \\
        $\Delta \lambda$ & bandpass width (nm) & 56.5 \\
        $D$ & telescope diameter (m) & 2.36 \\
        $R_{e-}$ & read noise (e-) & 120 \\
        $D_{e-}$ & dark current (e- pixel$^{-1}$ s$^{-1}$) & 1.39$\mathrm{e}$\minus 5 \\ 
        $CIC$ & clock-induced charge (e- pixel$^{-1}$ frame$^{-1}$) & 0.016 \\
        $q$ & detector quantum efficiency & 0.837 \\
        $f_p$ & fraction of planetary light that falls within aperture & 0.039 \\
        $\mathcal{T}$ & telescope and instrument throughput & 0.38 \\
        $\Delta t$ & observation time (s) & 3600 \\
        $g$ & EMCCD gain (e- photoelectron$^{-1}$) & 1000 \\
        \hline                                  
        \end{tabular}
    \end{table*}
    
    \subsubsection{Detector Model}
    
    We convert the planet flux density to the planet photon count rate at the telescope following the method described in Robinson et al.~2016\cite{robinson2016characterizing}, which was developed with a telescope like Roman in mind and modeled on the work presented in Brown 2005~\cite{brown2005single}. 
    The detector noise function accepts the calculated planet photon rate for each observation and propagates it through a simplified Electron-Multiplying Charge-Coupled Device (EMCCD) detector model, which adds noise properties to the signal. 
    The model includes shot noise, dark current, read noise, and EMCCD gain. 
    The result is a planet signal with detector noise added for each simulated detection, which amounts to the output electron counts from the detector due to the incoming planet signal. 
    For each orbit option returned by the deconfuser, we calculate both the ``observed'' noisy photometry and the ``expected'' noisy photometry of the simulated planet detection. 
    The ``observed'' photometry simulates the instrument's observation of a planet.
    In other words, this is the measured photometry that a user would provide to the algorithm for a real observation. 
    The ``expected'' photometry is the value one expects to receive from a detected planet based on the assumed phase angle and planet parameters, including detector noise.

\subsection{Photometry Likelihood Ranking}\label{sec:ranking}

    After we have included the detector noise for each observation, we implement a ranking scheme to sort confused orbit options based on the likelihood of the calculated photon rates per system option, given the assigned orbital parameters.
    We obtain the likelihood of each detection from the probability distribution of the brightness. This value is returned by Fig.~\ref{fig:code_flow}, step 4. 
    
    To generate the probability distribution for each detection's brightness, the detector noise model first draws samples from a Poisson distribution describing the expected number of photons reaching the detector. 
    The incident photons are multiplied by the detector quantum efficiency before the dark current is added to the signal. 
    We account for the EMCCD gain ($g$) by drawing samples from a Gamma distribution, following the model described in Hirsch et al.~2013\cite{Hirsch2013-wg} and applied in \textit{LOWFSSim} (the Roman Coronagraph Instrument optical model of low-order wavefront sensing and control \cite{lowfssim_bdube}; \href{https://github.com/nasa-jpl/lowfssim}{https://github.com/nasa-jpl/lowfssim}). 
    The shape parameter of the gamma distribution is the number of input photoelectrons ($n_{ie}$), and the scale parameter ($\theta$) is

    \begin{equation}
        \theta = g - 1 + (1/n_{ie}).
    \end{equation}
    
    \noindent The final addition is the detector's read noise, which is sampled from a normal distribution. 
    All parameter values adopted in our model are listed in Table~\ref{tab:parameters}. 
    
    The contributions to the noise model are a combination of three probability distributions: a Poisson distribution, $P(n_p)$, describing the incident photons (Eq.~\ref{eqn:poisson}), a Gamma distribution, $\Gamma(n_{ie}, \theta)$, describing the EM register of the EMCCD (Eq.~\ref{eqn:EM_gain}), and a normal distribution, $\mathcal{N}(R_{e^\mathrm{-}})$, describing the read noise (Eq.~\ref{eqn:read_noise}) to convert the signal to image values. 
    The following equations describe these distributions.

    \begin{equation}\label{eqn:poisson}
        p(n_{ie}) = P(n_p) \cdot q + (D_{e\mathrm{-}} \Delta t + CIC )
    \end{equation}

    \begin{equation}\label{eqn:EM_gain}
        p(n_{oe}) = \Gamma(n_{ie}, \theta) + n_{ie} - 1
    \end{equation}

    \begin{equation}\label{eqn:read_noise}
        p(n_r) = \mathcal{N}(0, R_{e^\mathrm{-}})
    \end{equation}

    \noindent where $q$ is the detector quantum efficiency, $D_{e\mathrm{-}}$ is the dark current, $\Delta t$ is the observation duration, and $CIC$ is the detector clock-induced charge (see also Table~\ref{tab:parameters}).  
    $p(n_{ie})$ is the probability to get $n_{ie}$ input electrons, $p(n_{oe})$ is the probability to get $n_{oe}$ output electrons, and $p(n_r)$ is the probability to get $n_r$ readout electrons.
    Combining Eqs.~(\ref{eqn:poisson}),~(\ref{eqn:EM_gain}), and~(\ref{eqn:read_noise}) gives the probability of measuring a certain number of counts for a detection, $p(n_c)$:

    \begin{equation}\label{eqn:count_likelihood}
        p(n_c) = p(n_{ie}) + p(n_{oe}) + p(n_r).
    \end{equation}

    The output of the noise model is the probability density function ($f(y|n_p)$ in Eq.~\ref{A_eqn:prob_function}) describing the number of photoelectrons read out by the detector, from which we can determine the likelihood. 
    The likelihood of a detection, given the orbital parameters of the orbit fit, is obtained by comparing where the observed brightness falls along this probability density function.
    We assume each planet detection is an independent event, so the likelihood of measuring all planet detections given the parameters of the matched orbits is the product of each individual detection. 
    The resulting likelihood, $L_\mathrm{system}$, is used as a new ranking score for each set of confused orbit options per system.
    A more detailed mathematical description of this process is given in Appendix~\ref{appendix_A}.
    We note that in some cases, it may be beneficial to examine likelihoods on a per-planet basis to avoid a biased likelihood due to one poorly fit planet in the system. 
    We expand on this in more detail in Section~\ref{sec:discussion}.

\subsection{Simulated Systems}\label{sec:sims}

    This section describes our method for identifying representative highly confused systems to demonstrate the photometry ranking scheme on.
    We simulate ten test case systems across three inclination groupings, low ($i < 45^\circ$), medium ($45^\circ < i < 70^\circ$), and high ($i > 70^\circ$), to investigate how exoplanet photometry can support deconfusion of multi-planet systems. 
    The ten simulated systems are each different and only inclination is changing across the three inclination groupings, creating a total of thirty  cases for study.
    Each of the simulated systems contains three planets.
    We chose the first ten highly confused cases from a much larger sample of simulated systems.
    The methods for simulating and identifying these confused systems are described in Sec.~\ref{subsec:identifying}.
    
    We limit this analysis to three-planet systems to demonstrate the photometry ranking scheme based on Pogorelyuk et al.~2022's\cite{pogorelyuk2022deconfusing} findings regarding confusion in systems with three or more planets.
    The cases in this work were intentionally chosen because they are clearly challenging cases. 
    That is, each of the simulated systems examined here results in confusion every time they are passed through the astrometric deconfuser at each of the low, medium, and high inclinations.
    The purpose of this work is not to present large number statistics, but rather to explore if photometry is sufficient for deconfusing these representative known highly confused scenarios.
    We also demonstrate the photometry ranking scheme on a few Solar System planet analogs in Appendix~\ref{appendix_B}.

    The systems presented here are not explicitly evaluated for dynamical stability. 
    We simulate systems with orbital separations greater than 0.3\,AU, with mutual inclinations and low eccentricities. 
    For planets with masses similar to Earth, separations of 0.3\,AU would place planets at separations well outside several times the Hill radii of the other planets in the system (see e.g., Gladman et al.~1993\cite{Gladman1993Icar..106..247G} or Dulz et al.~2020\cite{Dulz2020ApJ...893..122D} for discussion of using mutual Hill radii as a basic stability criterion).  
    Dynamical simulations indicate that Earth-mass planets in three-planet systems are stable for much longer than one billion years if their initial orbital separation is at least 7.5 times their mutual Hill radii\cite{Smith2009Icar..201..381S}. 
    Simulations also show that stable multi-planet systems generally have lower median eccentricities and mutual inclinations \cite{He2020AJ....160..276H}.

\subsubsection{Identifying Confused Systems}\label{subsec:identifying}

    The simulated systems were chosen by first identifying confused systems with inclinations of less than 45$^\circ$. 
    Due to the low probability of confusion for systems with $i < 45^\circ$, it is more challenging to identify confused low inclination cases. 
    For this reason, we identify test cases for this work starting with low inclination simulations and working our way up.
    Systems with inclinations less than $\sim$20$^\circ$ are also rarely confused \cite{pogorelyuk2022deconfusing}. 
    Thus, there are no systems in our sample set which have an inclination less than 20$^\circ$.
    
    Low inclination systems are simulated with three co-planar planets with random orbital parameters drawn from uniform distributions with semimajor axes between 0.5 and 6.0\,AU, eccentricities less than 0.1, inclinations less than 45$^\circ$, and minimum orbital separations between planets of 0.3\,AU. 
    These systems were simulated using the system generation function (\texttt{sample\allowbreak\_planets.random\allowbreak\_planet\allowbreak\_elements}) contained within the deconfuser framework.
    
    Once a low inclination system is found to be confused with the first version of the deconfuser, that system was passed through the deconfuser again with only the inclination changed. 
    Updated inclination values were randomly sampled from uniform arrays ranging between 45$^\circ$ to $<$70$^\circ$ until a confused system was identified in the medium inclination range through empirical trial and error testing. 
    The same method was used for the high inclination grouping, but inclinations were sampled between 70$^\circ$ and 90$^\circ$.
    If a system returned confused options with inclinations in the low, medium, and high inclination groupings, it was included in the sample of test systems for this work, until we reached ten systems total in each inclination grouping. 
    By varying only inclination across otherwise identical systems, we are also able to assess whether the inclusion of brightness information more strongly supports deconfusion for low, medium, or high inclination systems.
    
    The systems identified and described in this work are systems for which the astrometric deconfuser is unable to identify a single answer for the corresponding detections within an RMS error of 0.05\,AU or within the tolerance limits of the orbit parameter grid search, resulting in confusion. 
    We reiterate that the astrometric deconfuser may return confused orbit options with RMS errors greater than the tolerance threshold if the proposed orbit fits push the limits of the grid search (see Sec.~\ref{sec:approach}). 
    Examples of this for the systems in this work can be seen where confused orbit options had eccentricities near 0.1, which is where the maximum eccentricity threshold was set for the orbit search (e.g., Low inclination Option B in Table~\ref{tab:systen_confused_rankings}).

    The orbital parameters describing the shape (semimajor axis, eccentricity) and inclination of the planets' orbits in each of the ten systems are listed in Table~\ref{tab:all_system_simulations}. 
    We focus on the shape parameters and inclination here as these are the primary orbital parameters that will be used to initially characterize an exoplanet's orbit. 
    The semimajor axis is essential for constraining whether or not an observed exoplanet lies within the host star's projected habitable zone, which is a defining requirement for deciding whether an observatory like HWO should allocate additional time and resources for follow-up observations \cite{luvoir2019arXiv}.

\subsubsection{Simulating Detections}\label{subsubsec:sim_detections}

\begin{figure*}[h!]
   \centering
    \includegraphics[width=\textwidth]{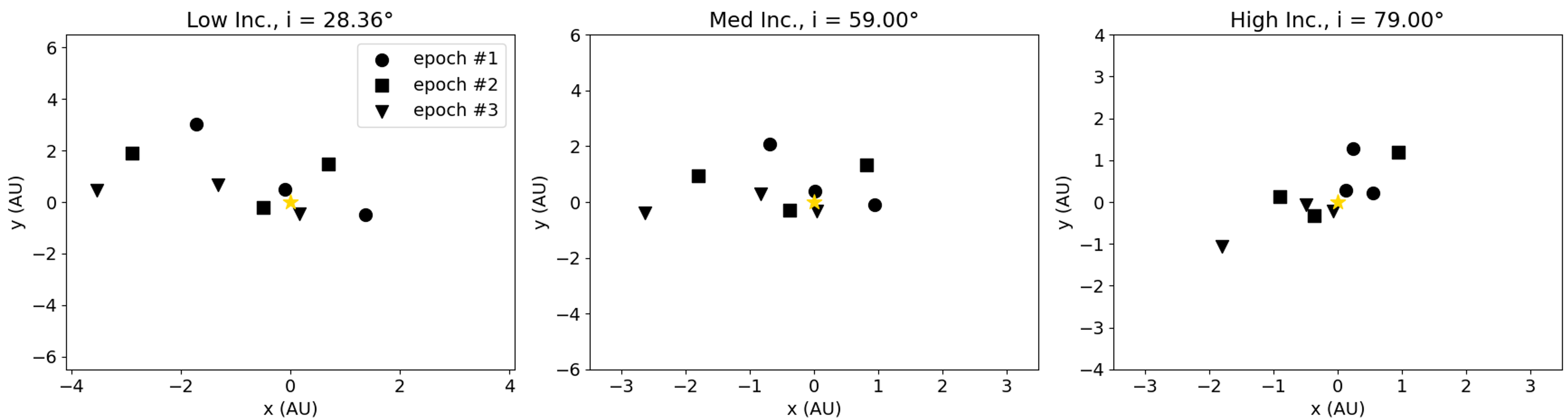}
      \caption{Simulated detections for test case system ten across the low (left), medium (center), and high (right) inclination ranges (groups described in Sec.~\ref{subsec:identifying}). Observations were spaced equally over one Earth year and each epoch of observation is represented by a different marker (circle, square, and triangle). Orbital parameters for the test case systems are given in Table~\ref{tab:all_system_simulations}.}
     \label{fig:simulated_detections}
   \end{figure*}

    Examples of simulated detections are shown in Fig.~\ref{fig:simulated_detections} for system ten. 
    We assume observations are equally spaced to occur three times over a period of one Earth year. 
    We choose to simulate three observations to find highly confused systems, and because previous work has demonstrated that the minimum number of observations required to sufficiently constrain a planet's orbit to below 10\% is three visits \cite{Guimond2019AJ....157..188G, horning2019minimum, Romero-Wolf2021JATIS...7b1219R}.
    We assume each system is located at a distance of 10\,parsecs from the observer with a Sun-like host star. 
    The best potential target stars identified by Mamajek \& Stapelfeldt 2024\cite{Mamajek2024arXiv240212414M} (``Tier A'') for observation with HWO have distances that fall within a range of 3 to 19 parsecs.
    In our simulations, we also assume all planets are visible at all times for this initial study. 
    This paints an overly optimistic picture of confusion rates and the ability to leverage photometry for high inclination systems; an investigation of null-detections is left for future work.
    Each simulated planet is prescribed a radius of 1\,$R_{\oplus}$ for the photometry calculations and a geometric albedo of 0.3. 
    We prescribe these parameters to emulate observations of rocky Earth-like planets \cite{Cox2000PhT....53j..77C, Mallama2009Icar..204...11M, Mallama2017Icar..282...19M}, which will be primary targets for HWO. 
    
\subsubsection{Confused System Orbit Options}

\begin{figure*}[h!]
   \centering
    \includegraphics[width=0.85\textwidth]{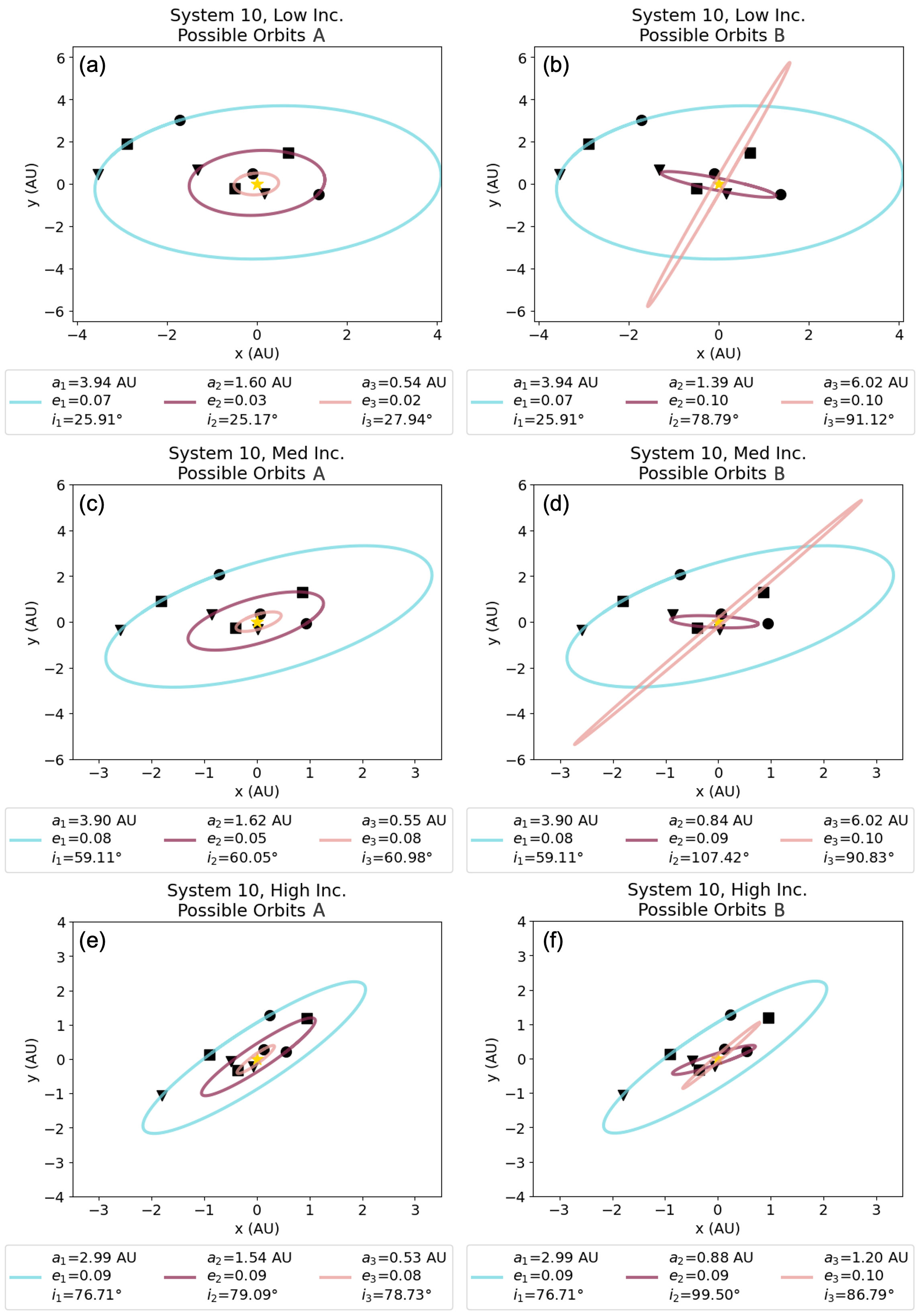}
      \caption{Confused orbit options for the low (a and b), medium (c and d), and high inclination (e and f) detections for system ten (detections shown in Fig.~\ref{fig:simulated_detections}).
      The rankings for each of the confused options across the inclination groupings are given in Table~\ref{tab:systen_confused_rankings}. 
      The orbit fits from each possible set of options are compared to the true simulated system's orbits in Fig.~\ref{fig:compare_top_ranked} for each inclination group.}
     \label{fig:all_confused_options}
   \end{figure*}

    The number of confused orbits per simulated system are listed in the three right columns of Table~\ref{tab:all_system_simulations}. 
    The system that we will use as an example to illustrate the proof-of-concept, system ten, has two confused orbit options for the low inclination case, two confused orbit options for the medium inclination case, and two confused orbit options for the high inclination case.
    The orbit options for each inclination grouping for system ten are shown in Fig.~\ref{fig:all_confused_options}. 
    
\begin{figure*}[h]
        \centering
        \includegraphics[width=\textwidth]{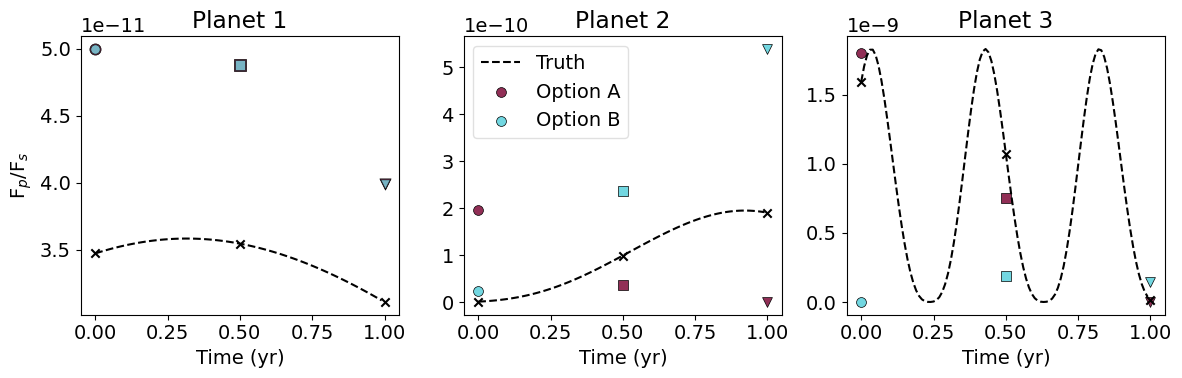}
        \caption{Examples of planet-star flux ratios at the epochs of detection for the confused orbit options (maroon and cyan) and true orbits (x markers and dashed black line) in high inclination system ten. The markers for options A and B for planet 1 are plotted on top of one another since the orbit fit is identical in both cases. Different orbital parameters can introduce significant variation in the expected brightness of a detection due to varying phase angles and separations.}\label{fig:phase_curves}
    \end{figure*}

    For each set of confused orbit options (i.e., for each system), we apply the new photometry likelihood ranking scheme by simulating the expected brightness of each detection given the orbital parameters fit by the astrometric deconfuser. 
    The planet-star flux ratios for high inclination system ten are shown compared to the true options in Fig.~\ref{fig:phase_curves}.
    The likelihoods for each confused system are compared to rank the matched orbits in order of most to least likely for each system.
    Rankings from the combined deconfuser for the low, medium, and high inclination cases of system ten are given in Table~\ref{tab:systen_confused_rankings}. Table~\ref{tab:systen_confused_rankings} also provides the semimajor axis, eccentricity, inclination, and RMS error of each planet's orbit option as fit by the astrometric deconfuser.

    We expect orbit discrimination with photometry to depend on viewing geometry. 
    Nearly face-on systems exhibit minimal phase variation and therefore provide limited information, while highly inclined systems show stronger phase variation.
    However, higher inclinations also introduce greater geometric degeneracies in their projected orbits. 
    In Section~\ref{sec:discussion}, we briefly discuss whether these representative systems show a ``sweet spot'' for deconfusion where phase-dependent brightness variation is detectable and astrometric orbit confusion is not at an extreme.

    \begin{table*}
        \caption[Orbital parameters of the simulated systems and the number of confused orbit options per system]{Orbital parameters of the simulated systems and the number of confused orbit options per system. The full set of orbital parameters for all simulated planets can be found in the accompanying supplementary file (see \textit{Code, Data, and Materials Availability} Section).}
        \label{tab:all_system_simulations} 
        \centering
        \footnotesize
        \begin{tabular}{cccccccccc}
        \hline\hline             
        System & Planet & a & e &``Low'' i & ``Med'' i & ``High'' i & No.~confused
        & No.~confused & No.~confused \\ 
        No. & No. & (AU) &  & (°) & (°) & (°) & (Low i) & (Med i) & (High i) \\
        \hline
        \multirow{3}{*}{1} & 1 & 4.616 & 0.050 & 33.6 & 60.0 & 80.0 & \multirow{3}{*}{4} & \multirow{3}{*}{4} & \multirow{3}{*}{4} \\
                           & 2 & 2.462 & 0.073 & 33.6 & 60.0 & 80.0 & & & \\
                           & 3 & 2.863 & 0.089 & 33.6 & 60.0 & 80.0 & & & \\
         \hline
        \multirow{3}{*}{2} & 1 & 1.325 & 0.053 & 36.9 & 65.0 & 85.0 & \multirow{3}{*}{2} & \multirow{3}{*}{2} & \multirow{3}{*}{4} \\
                           & 2 & 0.941 & 0.030 & 36.9 & 65.0 & 85.0 & &&  \\
                           & 3 & 2.632 & 0.041 & 36.9 & 65.0 & 85.0 & & &  \\
         \hline
        \multirow{3}{*}{3} & 1 & 2.811 & 0.051 & 35.5 & 68.0 & 76.0 & \multirow{3}{*}{2} & \multirow{3}{*}{3} & \multirow{3}{*}{3} \\
                           & 2 & 2.206 & 0.047 & 35.5 & 68.0 & 76.0 & & &  \\
                           & 3 & 2.507 & 0.077 & 35.5 & 68.0 & 76.0 & & &  \\
        \hline
        \multirow{3}{*}{4} & 1 & 4.330 & 0.072 & 37.6 & 63.0 & 89.0 & \multirow{3}{*}{2} & \multirow{3}{*}{2} & \multirow{3}{*}{3}  \\
                           & 2 & 1.769 & 0.008 & 37.6 & 63.0 & 89.0 & &&  \\
                           & 3 & 1.230 & 0.039 & 37.6 & 63.0 & 89.0 & & &  \\
        \hline
        \multirow{3}{*}{5} & 1 & 1.031 & 0.066 & 24.1 & 62.0 & 90.0 & \multirow{3}{*}{2} & \multirow{3}{*}{2} & \multirow{3}{*}{2} \\
                           & 2 & 2.475 & 0.075 & 24.1 & 62.0 & 90.0 & &  &  \\
                           & 3 & 2.915 & 0.038 & 24.1 & 62.0 & 90.0 & & &  \\
        \hline
        \multirow{3}{*}{6} & 1 & 1.284 & 0.027 & 40.7 & 55.0 & 89.0 & \multirow{3}{*}{2} & \multirow{3}{*}{2} & \multirow{3}{*}{4} \\
                           & 2 & 3.684 & 0.039 & 40.7 & 55.0 & 89.0 & & & \\
                           & 3 & 0.915 & 0.088 & 40.7 & 55.0 & 89.0 & & &  \\
        \hline
        \multirow{3}{*}{7} & 1 & 1.681 & 0.015 & 39.2 & 59.0 & 86.0 & \multirow{3}{*}{2} & \multirow{3}{*}{2}  & \multirow{3}{*}{2}  \\
                           & 2 & 2.090 & 0.005 & 39.2 & 59.0 & 86.0 & & &  \\
                           & 3 & 0.864 & 0.071 & 39.2 & 59.0 & 86.0 & & &  \\
        \hline
        \multirow{3}{*}{8} & 1 & 5.529 & 0.034 & 22.3 & 49.0 & 86.0 & \multirow{3}{*}{2} & \multirow{3}{*}{2} & \multirow{3}{*}{2} \\
                           & 2 & 0.804 & 0.049 & 22.3 & 49.0 & 86.0 & & &  \\
                           & 3 & 1.288 & 0.011 & 22.3 & 49.0 & 86.0 & & &   \\
        \hline
        \multirow{3}{*}{9} & 1 & 0.718 & 0.036 &  20.4 & 64.0 & 84.0 & \multirow{3}{*}{2} & \multirow{3}{*}{2} & \multirow{3}{*}{2} \\
                           & 2 & 1.647 & 0.035 & 20.4 & 64.0 & 84.0 & & &  \\
                           & 3 & 3.716 & 0.074 & 20.4 & 64.0 & 84.0 & & &  \\
        \hline
        \multirow{3}{*}{10} & 1 & 3.850 & 0.010 & 28.4 & 59.0 & 79.0 & \multirow{3}{*}{2} & \multirow{3}{*}{2} & \multirow{3}{*}{2}  \\
                            & 2 & 1.630 & 0.016 & 28.4 & 59.0 & 79.0 & & &  \\
                            & 3 & 0.538 & 0.013 & 28.4 & 59.0 & 79.0 & & &  \\
        \hline
        \end{tabular}
    \end{table*}

    \begin{table*}
        \caption{System ten photometry ranking results. Orbital parameters and updated rankings for the confused orbit options returned by the combined deconfuser for the low, medium, and high inclination system ten. The confused orbit options are shown in Fig.~\ref{fig:all_confused_options}. RMS errors for each individual orbit fit by the astrometric deconfuser are also listed. Results for all simulated systems are provided in Appendix~\ref{appendix_D_tables}.}
        \label{tab:systen_confused_rankings} 
        \centering
        \footnotesize
        \begin{tabular}{cccccccccc}
        \hline\hline             
         &  & & \multicolumn3c{``Confused'' orbit parameters} & Deconfuser &   &   & \\
        Incl. & System & Planet & $a$ & $e$ & $i$ &  RMS fit & Orbit & System & System \\
        group & option &  no. & (AU) & & (°) & error (AU) & likelihood & likelihood & ranking \\
        \hline
        &  & 1 & 3.938 & 0.065 & 25.91 & 0.004 & $5.684\mathrm{e}{\minus 2}$  & &  \\
        Low & A & 2 & 1.599 & 0.029 & 25.17 & 0.014 & $1.761\mathrm{e}{\minus 4}$  & $1.485\mathrm{e}{\minus 9}$ & 1 \\
        &  & 3 & 0.537 & 0.025 & 27.94 & 0.005 & $1.483\mathrm{e}{\minus 4}$  & &  \\
         \hline
         & & 1 & 3.938 & 0.065 & 25.91 & 0.004 & $6.708\mathrm{e}{\minus 2}$ & &   \\
        Low & B & 2 & 1.394 & 0.099 & 78.79 & 0.228 & $3.588\mathrm{e}{\minus 4}$ & $1.907\mathrm{e}{\minus 14}$ & 2 \\
         & & 3 & 6.020 & 0.097 & 91.12 & 0.285 & $7.924\mathrm{e}{\minus 10}$ & &  \\
         \hline
          & & 1 & 3.897 & 0.084 & 59.11 & 0.004 & $4.152\mathrm{e}{\minus 3}$ & &  \\
        Med & A & 2 & 1.622 & 0.052 & 60.05 & 0.008 & $2.717\mathrm{e}{\minus 3}$ & $4.621\mathrm{e}{\minus 8}$ & 1 \\
         & & 3 & 0.545 & 0.082 & 60.98 & 0.011 & $4.097\mathrm{e}{\minus 3}$ & & \\
        \hline
         & & 1 & 3.897 & 0.084 & 59.11 & 0.004 & $2.560\mathrm{e}{\minus 3}$ & &  \\
        Med & B & 2 & 0.838 & 0.090 & 107.42 & 0.002 & $2.330\mathrm{e}{\minus 3}$ & $3.018\mathrm{e}{\minus 16}$ & 2 \\
         & & 3 & 6.020 & 0.097 & 90.83 & 0.297 & $5.059\mathrm{e}{\minus 11}$ & & \\
        \hline
        &  & 1 & 2.994 & 0.095 & 76.71 & 0.011 & $9.733\mathrm{e}{\minus 3}$ & &  \\
        High & A & 2 & 1.555 & 0.092 & 79.09 & 0.006 & $8.314\mathrm{e}{\minus 6}$ & $4.468\mathrm{e}{\minus 10}$ & 1 \\
         & & 3 & 0.533 & 0.082 & 78.73 & 0.006 & $5.522\mathrm{e}{\minus 3}$ & &  \\
         \hline
         & & 1 & 2.994 & 0.095 & 76.71 & 0.011 & $1.577\mathrm{e}{\minus 2}$  & &   \\
        High & B & 2 & 0.881 & 0.095 & 99.50 & 0.038 & $1.029\mathrm{e}${\minus 3} & $2.297\mathrm{e}{\minus 13}$ & 2 \\
         & & 3 & 1.202 & 0.100 & 86.79 & 0.322 & $1.416\mathrm{e}{\minus8}$ & &  \\
        \hline
        \end{tabular}
    \end{table*}

\section{Results}\label{sec:results}

    Here we present findings regarding the performance of the updated ranking scheme (combined deconfuser) using planetary phase and brightness.
    The results are based on a representative set of corner case systems systems, assuming that all planets are visible at all times. 
    Simulated detections are passed through the combined deconfuser with the photometry model, without knowledge of which detection corresponds to which planet. 
    For each of the systems that we apply the photometry ranking to, the astrometric deconfuser finds orbit solutions that fit the detections, but is unable to identify a single solution for each system given only their position information (i.e., confusion). 
    The orbit parameters for the options between the astrometric deconfuser and combined deconfuser are identical.
    The resulting likelihoods from the photometry ranking describe which orbit option better matches the measured intensity information given the orbits fit by the astrometric deconfuser.
    We use test case system ten as an illustrative example in this section.
    Photometry ranking results for all systems and inclinations are provided in Appendix~\ref{appendix_D_tables}.

\subsection{Photometry Ranking for Deconfusion}\label{sec:deconf_w_phot}

    Here we describe the improvement in deconfusion for the representative systems using photometry (i.e., the combined deconfuser). 
    In the case of system ten, photometry aids in deconfusing the system by providing an updated ranking to each set of orbits for the system. 
    We compare the top-ranked orbit match to the true system by calculating the percent difference between individual orbit parameters for each planet.
    The orbit fits which more closely resemble the true simulated system parameters for system ten were ranked highest across all inclination groupings.
    ``Closely'' here refers to the minimum percent difference between orbit parameters across all confused orbit matches.
    
\begin{figure*}[h!]
   \centering
    \includegraphics[width=\textwidth]{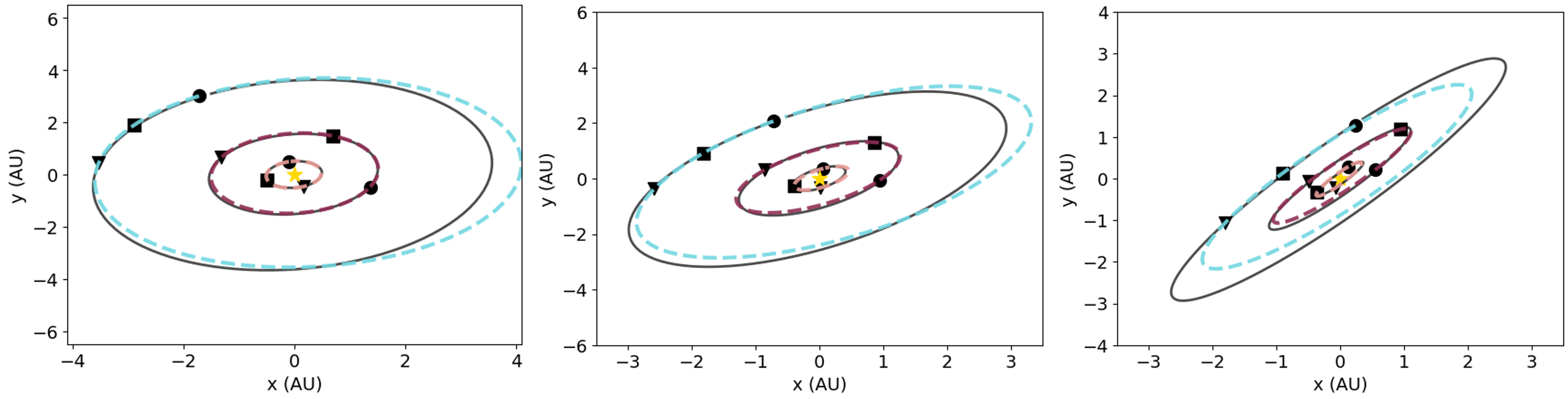}
      \caption{Top ranked orbit options (dashed lines) for system ten compared to the true simulated system's orbits (solid black lines) for each system: (left) low inclination, (center) medium inclination, and (right) high inclination.}
     \label{fig:compare_top_ranked}
   \end{figure*}

    For the low inclination case (detections in Fig.~\ref{fig:simulated_detections}, panel A), the astrometric deconfuser originally returned two equally likely fits for the system. 
    The first option (option A) depicted a relatively co-planar system with low eccentricities for all planets and inclinations of less than 28$^\circ$ (panel a in Fig.~\ref{fig:all_confused_options}). 
    Option B included a similar fit for planet one in the system, but planets two and three were fit with high inclination orbits and eccentricities of 0.1 (panel b in Fig.~\ref{fig:all_confused_options}). 
    The photometry ranking scheme in the combined deconfuser determined that option B was statistically less likely based on the detected planetary brightnesses.
    The same is true for the medium (Fig.~\ref{fig:all_confused_options}, Panels c and d) and high inclination (Fig.~\ref{fig:all_confused_options}, Panels e and f) cases of system ten. 
    The top ranked system options are shown compared to the true simulated system orbits in Fig.~\ref{fig:compare_top_ranked}.

\begin{figure*}[h!]
        \centering
        \includegraphics[width=0.98\textwidth]{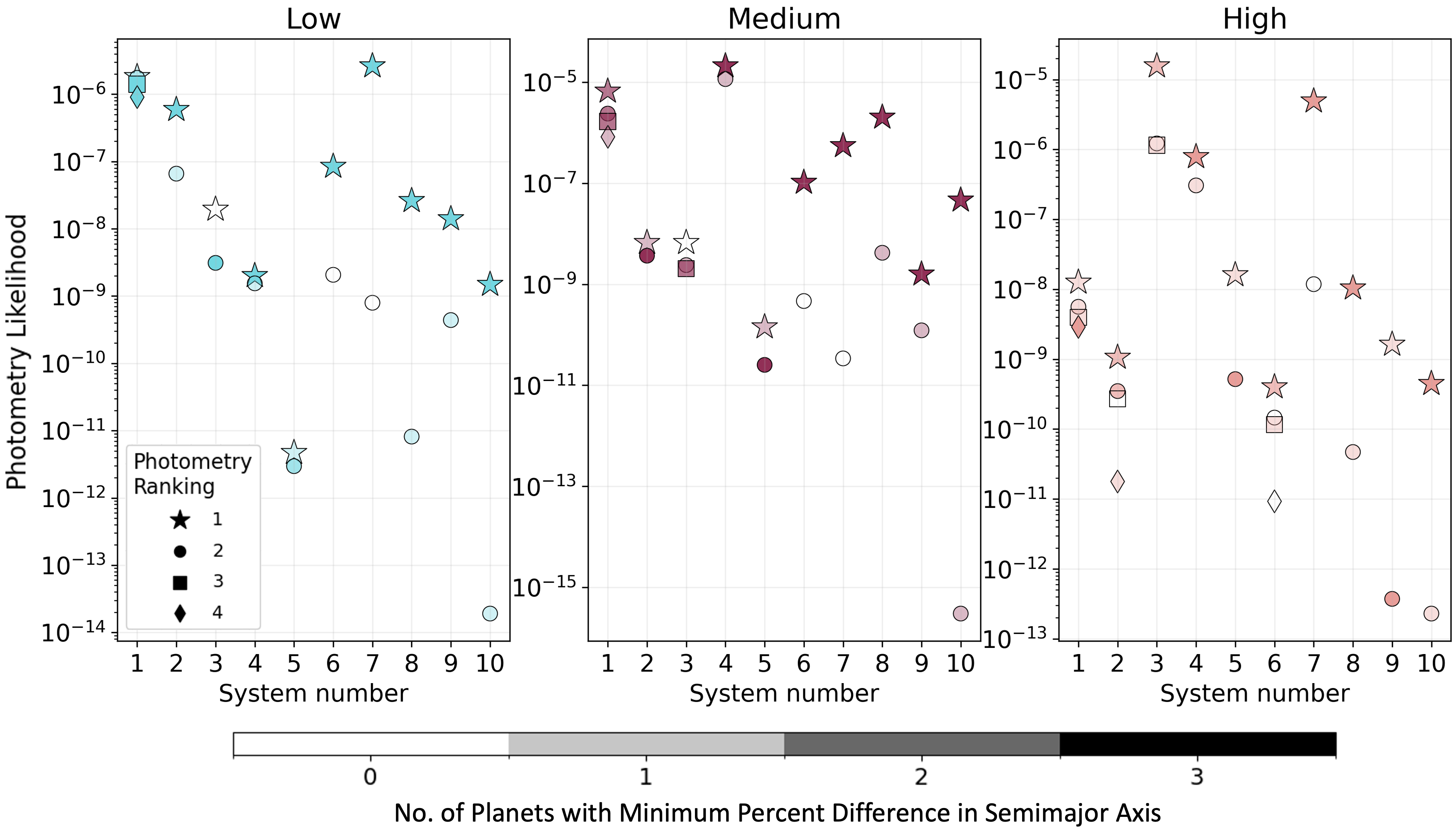}
        \caption{Comparison of combined deconfuser likelihood rankings for all systems across each inclination grouping: (left) low, (center) medium, and (right) high. The marker shapes indicate the photometry ranking order. Marker opacity corresponds to the number of individual planets per system which were also the ``best'' choice based on the semimajor axis fit compared to the true system (i.e., the semimajor axis which has the minimum percent difference compared to the true value).
        Prior to ranking with photometry, all systems were nearly equally ranked by the astrometric deconfuser.
        Improvements with photometry are evident where the star markers are also opaque (e.g., systems 7, 10).
        Photometry shows promise for orbit differentiation over geometric orbit-fitting alone. High inclination systems exhibit more phase variation, which typically allows for better distinction between orbit options (in likelihood space) for these example cases. }
        \label{fig:L_all_comparison}
    \end{figure*}

    For all test case systems in the low inclination grouping, the photometry ranking scheme ranked the ``best'' confused option higher in seven of the ten cases.
    The ``best'' option describes the orbit fit for which the semimajor axis has the minimum percent difference compared to the true value.
    The ``worst'' option refers to a match that does not have the minimum percent difference compared to the true value.
    For the medium and high inclination groupings, the photometry ranking scheme also ranked the best system option higher in six of the ten cases. 
    Fig.~\ref{fig:L_all_comparison} presents the likelihood results for all test cases examined in this work, with different marker shapes corresponding to the photometry ranking assigned to each orbit fit in the system.
    We emphasize that prior to ranking with photometry, all of the systems tested here were confused and these are improvements for more than half of the representative highly confused cases. 
    With only relative astrometry, the astrometric deconfuser was unable to choose just one set of orbits to fit the detections.
    We present a discussion of these results in Sec.~\ref{sec:discussion} and discuss further improving deconfusion with photometry in Sec.~\ref{sec:conclusion}. 

\section{Discussion}\label{sec:discussion}

    Here we present a more detailed discussion of the results, including how well the orbits are fit compared to the truth and the implications of the results for direct imaging missions.
    Our proof-of-concept suggests that including photometry benefits deconfusion for the example systems in this work, across all inclination ranges. 
    Across the low, medium, and high inclination cases simulated here, the photometry likelihood ranking scheme enabled the separation of confused orbit options for each of the systems. 
    Photometry aided in choosing the better orbit fit out of the confused options in more than half of the cases across inclination groups.

    \subsection{Photometry for Deconfusion}\label{subsec:disccussion_phot_deconf}
    
    There are several contributing factors which influence the utility of photometry for deconfusing orbit options in these example cases.
    The first is due to the statistical nature of photon detections when observing exoplanets and the limited number of photons which will be received when imaging at high contrast levels. 
    The likelihoods that are used as the ranking metric in this work are derived from a distribution of potential electron counts at the detector for each planet detection, which is a combination of a Poisson distribution, Gamma distribution, and Gaussian distribution (see detailed description in Sec.~\ref{sec:ranking}).  
    The combination of statistical distributions and low number of photons per observation may result in variations of estimated count rate per detection, which influences the likelihood value. 
    This explains the potential for variation in likelihoods on a per-orbit level.
    That is, if there are two system options for which one planet was fit with the same orbit in both, the options will return different likelihood values between the two confused systems. 
    This may ultimately affect the resulting system ranking for better or worse, even though all likelihood results fall within the distribution of possible expected values.

    An example of this can be seen with system one, which has four confused options across all inclinations. 
    For the outer planet, the detections sample a small fraction of the full orbit, and it is fit with the same orbit parameters across all confused matches (and inclinations).
    While the outer planet has the same fit parameters in each of the top two ranked cases, the likelihood for each of these options is different.
    In the low inclination case for system one, the difference between the likelihood for the outer planet (planet one) in the top-ranked match is four times higher than its likelihood in the second-ranked match, even though the second-ranked match would overall provide fits closer to the true system.
    
    Another commonality between systems for which the ``worst'' option is chosen appears when planet detections are close to each other in the plane of the sky. 
    In system one, the detections of the two inner planets appear to be less than 0.1\,AU from one another on the sky at each inclination. 
    Systems four and five also fall into this category. 
    In system four, the second detections of the inner two planets appear closer than 0.3\,AU (or 0.03$''$) of one another.
    For this reason, the difference in likelihoods between the low inclination cases are relatively small as the predicted phase angles are similar (in combination with the limited number of received photons) and spatially close together. 
    Figure~\ref{fig:L_all_comparison} shows the likelihoods for all systems and ranked options, and is shaded based on the number of planets per photometry ranking for which the semimajor axis was the ``best'' fit for the system (i.e., minimum percent difference between true and fit semimajor axes).
    The difficulty in separating options with detections with small separations can be seen in Fig.~\ref{fig:L_all_comparison} for systems one, four, and five, where the differences between system likelihoods are small.
    Section~\ref{sec:conclusion} expands on methods that could investigate reducing the confusion caused by closely separated detections.

    Additionally, there may be cases where the overall system likelihood is dominated by one planet being poorly fit out of all planets in the system.
    Because the likelihood of each system is a product of the likelihood of all planet detections in the system, the system likelihood can potentially be biased toward a worse overall system fit if one planet/detection dominates over the rest.
    For these reasons, it is advisable to examine the likelihood of the orbit for each planet in the system on an individual level. 

    Figure~\ref{fig:L_all_comparison} introduces some key points regarding photometry for orbit discrimination. 
    For the example systems presented here, it is clear that photometry shows promise for supporting orbit differentiation better than geometric orbit-fitting alone. 
    Previously, each of the system options presented were closely ranked using only relative astrometry.
    Likelihoods assigned to these systems by the photometry ranking scheme gave an updated ranking of orbit combinations in the combined deconfuser. 
    The photometry ranking scheme generally provides more distinction between orbits for the higher inclination systems in this work ($>$45$^\circ$), compared to lower inclination, because there is more phase variation in these observations. 
    This can be seen by the greater scatter in likelihood values in the medium and high inclination cases for the systems presented here (e.g., systems five, seven, eight, nine, and ten).

    \subsection{The Role of Geometric Albedo}
    
    We briefly comment on the role of geometric albedo for modeling photometry, as this is currently a constant parameter in our simulations.
    The choice in geometric albedo ($A_g$) for modeling the photometry will affect the overall likelihood values if the albedo is estimated incorrectly.
    A model albedo that is larger than the observed albedo would require a lower phase angle, decreased planet-star separation, or larger planet radius to produce the observed signal; a model albedo that is smaller than the observed albedo would require the opposite. 
    An incorrectly guessed albedo might preference a fit with larger/smaller planet-star separation if the model albedo is less/greater than the truth, if such an orbit-fit has been returned by the deconfuser. 
    However, if all other parameters are held constant, the general trend of a planet's phase variation remains the same for all values of $A_g$. 
    Because the photometry ranking scheme is choosing between and comparing orbits that have already been fit to detections, rather than influencing the orbit fitting process, we do not expect an incorrect guess for the geometric albedo to drastically change the performance of the photometry ranking. 

    \subsection{Phase Function and Albedo Assumptions}\label{subsec:lambertian_assumption}
    
    A limitation of this work lies in the choice of a Lambertian phase function and the constant albedo to model planetary reflectance.
    The assumption of Lambertian reflectance approximates the shape of a planetary phase function relatively well for the type of planets and temperature regimes that would be probed by a reflected light direct imaging mission. 
    Strong asymmetries in the reflected light curves of cool, mature planets are less likely because they are receiving relatively uniform irradiation from their host star (i.e., they are not tidally locked like known short-period transiting planets; e.g., Zellem et al.~2014\cite{Zellem2014ApJ...790...53Z}, Parmentier et al.~2021\cite{Parmentier2021MNRAS.501...78P}). 
    However, Solar System phase curves have been shown to deviate from Lambertian assumptions (by at least 25\%, in some cases)\cite{Mayorga2016AJ....152..209M, Mallama2018A&C....25...10M, Mayorga2020AJ....160..238M}. 

    Observations of Solar System planets have also demonstrated that real planetary albedos vary with wavelength and time, driven by factors such as clouds, aerosol scattering, surface features, and seasons.
    Mallama et al.~2018~\cite{Mallama2018A&C....25...10M} discusses some of these results in detail. 
    For example, analysis of models and observations show that the Earth may exhibit magnitude variation ranging from 0.12 to 0.76 between cloud-free and cloudy cases (e.g., Tinetti et al.~2006\cite{Tinetti2006AsBio...6..881T}, Stephens et al.~2015\cite{Stephens2015RvGeo..53..141S}, Mallama et al.~2018\cite{Mallama2018A&C....25...10M}). 
    The phase curve of Venus shows a reversal in brightness at high phase angles, which has been attributed to forward scattering of sulfuric acid droplets in the atmosphere \cite{Mallama2006Icar..182...10M}. 
    The giant planets in the Solar System exhibit albedo variation on a smaller scale at visible wavelengths, compared to the terrestrial planets \cite{Mallama2018A&C....25...10M}. 
    Jupiter observations near full phase show magnitude variations at the level of a few hundredths due to changes in the cloud bands \cite{Mallama2012Icar..220..211M}. 
    
    These complexities are not yet included in this work, but we describe future directions in Sec.~\ref{sec:conclusion} to incorporate more realistic reflectance behavior.
    In the scheme of the likelihood ranking, the true solution will always lie in the sample of possible solutions that are calculated, if the forward model assumptions are correct. 
    However, one might reject the best-fit orbit solution if the model lacks fidelity. 
    For example, in the case where the true planetary reflectance is non-Lambertian. 

    \subsection{Exo-moons and Exo-rings}
    
    Other factors that may influence exoplanetary photometry are moons and rings, which could have a negative impact on the photometry likelihood ranking scheme. 
    Rings alter the spectrum of a planet and would introduce asymmetries to the phase curve in reflected light \cite{Arnold2004A&A...420.1153A, Dyudina2005ApJ...618..973D, Coulter2022ApJS..263...15C}. 
    Previous work has suggested that an icy ring around an Earth-like planet could result in a brightness 100 times that of a ringless planet \cite{Arnold2004A&A...420.1153A}. 
    In such a case, additional observations at longer wavelengths would be important for breaking the albedo-radius degeneracy and inferring the presence of rings.
    
    For exomoons, previous work indicates that the influence on the spectrum of an Earth-like planet is minimal near visible wavelengths \cite{Limbach2024AJ....168...57L}. 
    Their study concluded that detecting a moon-Earth mutual event at 550\,nm with a bandwidth of 20\% with an 8-meter class telescope would require around two to twenty mutual events. 
    Such cases where observations measure anomalous photometry due to some influencing factors would need to be considered on a case-by-case basis. 
    Using a more complex phase function and incorporating additional filters or polarimetry would likely be more beneficial for ranking such systems where phase curve asymmetries are being induced by material in the circumplanetary environment. 

    \subsection{Effect of Planetary Motion on Astrometry and Photometry}
    
    The motion of a planet around its star and through the obscured regions of an observing instrument will affect the ability to measure a planet's position and photometry. 
    This depends to some extent on the orbit parameters of the planet and the observing system.
    If a planet goes inside the IWA or outside the OWA (outer-working angle) of the observing instrument, it will not be possible to get accurate astrometry or measure photometry.
    If the planet has an eccentric orbit, the cloud and surface properties may change substantially with orbital position.
    For example, varying instellation will cause different types of clouds to condense at different regions along the planet's orbit\cite{Marley2013cctp.book..367M, cahoy2010exoplanet}.
    It may also cause the surface to freeze over or thaw, therefore increasing or decreasing the surface albedo if the orbit takes a planet in/out of cooler and warmer conditions\cite{Venkatesan2025AsBio..25...42V}. 

    Missed detections (e.g., when a planet falls inside/outside the working angles of the instrument) are also likely to increase confusion, because this may increase the number of orbits that can be fit to the points that are visible to the observer \cite{pogorelyuk2022deconfusing}. 
    However, where a planet's photometry can be measured, the likelihood ranking scheme would operate in the same manner. 
    For the current model of photometric behavior and radiometric brightness, these keep-out zones and portions of the orbit where planets are not observable would not be input to the deconfuser.
    The user would only provide the observable points, and thus, the deconfuser would only fit orbits to the input points \cite{pogorelyuk2022deconfusing}.
    This would not change any of the outcomes of this work, as a user would have measured photometry for the input points that are provided to the deconfuser.
    Section~\ref{sec:conclusion} identifies future work to address cases where planets are unobservable in their orbit.

    \subsection{Comparing Photometry Ranking to Dynamical Assumptions}
    
    The combined deconfuser currently makes no assumptions about the dynamical stability of systems when fitting orbits to planet detections. 
    In the context of orbit deconfusion, one might consider discarding solutions with high mutual inclinations or eccentricities as being physically implausible.
    In many cases examined here, the photometrically favored solution corresponds to what could be considered the more ``dynamically probable'' solution (i.e., lower eccentricities and mutual inclinations), and rejecting the higher inclination/eccentricity options would lead to the same result. 
    This is true for all systems in the medium and low inclination groupings. 
    However, this is not true for three of the high inclination systems. 
    In systems like high inclination system seven, where both orbit solutions have comparable inclinations and eccentricities for all planets (within 3$^\circ$ in inclination and 0.04 in eccentricity), removing a system based on higher inclination/eccentricity would result in selecting the ``worst'' orbit configuration.

    A significant fraction of known exoplanet systems exhibit eccentricities greater than 0.1. 
    Of the systems with measured eccentricities, 46\% have $e>0.1$ and 27\% have $e > 0.2$ \cite{Christiansen2025PSJ.....6..186C}.
    Therefore, while dynamical stability is an important and necessary criterion for deconfusion, photometry provides a critical, independent discriminator for deconfusion decisions.

    \subsection{Observing Strategy}
    
    In this work, we have simulated three equally spaced observations of each system spread uniformly over one Earth year. 
    In practice, an exoplanet imaging mission may choose to schedule observations in a different manner due to many factors (e.g., pointing restrictions or slew time). 
    This could result in a different observing scheme for a system; for example, taking the first two images within two months of one another and following up for the third observation eight months later. 
    This would result in sampling a smaller region of the orbital path for longer-period planets in the first two observations, and thus a smaller range of phase angles. 
    Variation of observing epochs as in this example does not necessarily guarantee observing diverse phase angles. 
    However, we would expect the photometry ranking scheme presented here to perform equally as well with any observing scheme, given good-quality orbit-fits from the deconfuser, because it considers the phase angle and albedo of a planet.
    Of course, more complex atmospheric behavior or surface features could otherwise influence the photometric variation separately from the phase angle.

    As was presented in Pogorelyuk et al.~2022\cite{pogorelyuk2022deconfusing}, the level of confusion depends on the number of observations and the system architecture. 
    With three visits, the probability of confusion is still relatively high for high inclination systems ($>40\%$ \cite{pogorelyuk2022deconfusing}). 
    Increasing the number of visits to a system to four or five can significantly reduce ambiguity, but this comes at a cost to overall mission yield \cite{Stark2016JATIS...2d1204S, Morgan2022SPIE12180E..20M}. 
    Each observation of a system requires hours of integration time (dependent on the host star's brightness and the required contrast) that use up mission resources and time that could be spent characterizing other systems. 
    Due to these trade-offs, it is beneficial to overall mission yield to reduce the number of observations needed to characterize a system/planet and optimize the observing strategy for each system.
    Efficient deconfusion that leverages all available information from a series of observations is highly valuable in this regard.

    \subsection{The Potential Impact of Zodiacal and Exozodiacal Dust on Deconfusion}

    Confusion rates and the utility of photometry for deconfusion will also be impacted by astrophysical noise sources, including zodiacal (``zodi'') and exozodiacal (``exozodi'') dust.
    The Solar System's zodi will vary as a function of the observatory's pointing\cite{Levasseur-Regourd1980A&A....84..277L}. 
    Exozodi will contribute spatially varying background that may fluctuate as a function of position within its system, mimic a planet signal, mask subtle photometric variation as the planet orbits, or even conceal the planet entirely\cite{Roberge2012PASP..124..799R, Defrere2012SPIE.8442E..0MD, Currie2025arXiv250319932C}. 
    Both contribute additional sources of time-dependent noise and variations in the measured photometry that will introduce deviations from our assumed Lambertian model, potentially degrading the photometry ranking if the background contribution is not accurately modeled or subtracted before employing the deconfuser.

    \subsection{Accuracy of Deconfuser Orbit Fits}
    
    To perform a thorough assessment of the deconfuser's performance, we also calculated the accuracy of the semimajor axis fits returned by the deconfuser compared to the true semimajor axis for each planet, by comparing their percent difference.
    The accuracy of the fit for the semimajor axis is important for enabling follow-up observations and characterization of planets. 
    The tolerance for orbital parameter accuracy defined by the NASA Standards Definition and Evaluation Team for Yield Analysis \cite{Morgan2019sdet.rept.....M} is to better than ten percent accuracy. 
    Of the planets in the low inclination systems simulated here, 23 of the 30 planets were retrieved to within ten percent of the true semimajor axis. All planets except one were fit to an accuracy of better than 25 percent. 
    Of the medium inclination systems, 
    21 of the 30 planets were fit to within ten percent of the true value, and 28 planets to within 25 percent of the true value. 
    For the high inclination systems, the number of planets for which semimajor axes were retrieved to within ten percent of the true value dropped to 17 of the 30 planets, and 22 of the 30 were constrained within 25 percent of the true semimajor axis.
    We emphasize that these percentages result after selection of the ``best'' orbit using the new photometry ranking scheme in the combined deconfuser, where the best orbit is the one with the minimum percent difference between individual orbital parameters compared to the true system.
    In order to characterize potentially habitable planets with a future mission like HWO, observation schedulers will require accurate knowledge of whether a planet lies within the host star's potential habitable zone and where it will be in the future in order to schedule the next observation. 

    \subsection{Importance of Deconfusion for Observation Planning}
    
    Being able to deconfuse multiple orbit options using photometry will ultimately improve observation scheduling and enable characterization of exoplanet systems. 
    It will also inform decisions about how to allocate mission time to studying planets of interest. 
    For example, the confused orbit options for the inner two planets in the high inclination system ten give fits that change the number of planets with separations that fall within the boundaries of a Solar-type star's liquid-water habitable zone (defined here with an inner edge at 0.836\,AU and outer edge at 1.656\,AU as in Kane et al.~2012\cite{kane2012habitable}).
    In the first option for the high inclination system ten, only the second planet in the system would fall within the liquid-water habitable zone. 
    In the second option, both planet two and planet three would fall within the habitable zone. 
    The difference between these orbital separations and architectures may inform how we determine follow-up observations on such a system. 
    Observers may choose not to allocate valuable mission observing time and resources in a scenario where an orbit option places a planet outside of the habitable zone. 
    The combined deconfuser can support this decision making.

\section{Summary \& Future Work}\label{sec:conclusion}

    In this work, we introduce a method and updated tool for using orbital phase and planetary brightness variation to deconfuse orbit combinations.
    We present a proof-of-concept analysis of a limited set of cases to explore the utility of planetary phase for planet-detection deconfusion.
    We apply the updated orbit ranking scheme to thirty representative highly confused systems containing three planets, categorized into three groups of ten across low, medium, and high inclination groupings. 
    We adopt instrument parameters for the Roman Space Telescope Coronagraph Instrument to implement the photometric orbit-ranking scheme with realistic instrument values, but the results remain applicable to any direct imaging mission.
    
    The analysis presented here shows that combining photometry with astrometry shows promise for improving deconfusion of multi-planet systems and enabling orbit differentiation. 
    For the corner cases presented here, each system had multiple closely ranked orbit options using only relative astrometry, which results in confusion.
    The photometry ranking scheme helps to choose the system fit that most closely resembles the true simulated system in more than half of the representative cases across inclination groupings. 
    Future work will explore the potential for improving the error rates on orbit ranking with photometry by setting thresholds for comparing system likelihoods and using more realistic planetary reflectance treatments.
    
    On top of supporting observation scheduling and follow-up decisions by providing an estimate of where a planet resides in its system, orbit determination also supports placing better constraints on exoplanet properties. 
    Having \textit{a priori} knowledge of a planet's orbit supports accurate retrieval of the radius in reflected light observations\cite{Salvador2024ApJ...969L..22S}. 
    Orbit determination helps to break the degeneracy between planet radius and phase angle. 
    Knowing a planet's radius can help differentiate between clear and cloudy atmospheric scenarios, and supports placing constraints on atmospheric abundances of species like CH$_4$ \cite{carrion-gonzalez2020A&A...640A.136C}.
    An unknown planet radius reduces the accuracy of the atmospheric retrieval for reflected light spectroscopy. 
    Therefore, reducing the radius uncertainty on a planet through orbit determination is a useful prior for exoplanet characterization.
    
    Future work will explore additional ranking parameters that may also support deconfusion of multi-planet systems, including multi-band photometry (color). 
    Testing a wider parameter space of number of planets and planet types/radii is left for a follow-on paper, which will employ model albedo spectra in a study to assess the utility of multi-wavelength band observations and realistic planetary reflectance treatments for deconfusion.
    We will also update the detector model to include more robust assumptions, including modeling the detector QE with a binomial draw to capture extremely low photon rate conditions.
    To enable a more complete statistical assessment of confusion when photometry is accounted for, a broader sample size of systems with varying numbers of planets and planet types would be beneficial.
    Additional avenues of future work include an investigation of how incorrect assumptions about a planet's geometric albedo or spectrum would affect the photometry ranking scheme as biases in albedo might lead to misidentification of orbital solutions. 
    Additional sources like zodiacal and exozodiacal dust, and speckles\cite{pogorelyuk2022deconfusing}, will also contribute significant noise to future observations, and will be modeled in future iterations of this work for a robust understanding of confusion rates.
    
    Additional ranking and orbit-fitting considerations may support a reduction in confusion rates, such as considering only orbits which are mutually inclined or accounting for dynamical stability of systems. 
    Similar to the photometry ranking scheme, the dynamical stability of orbit fits returned by the deconfuser could be evaluated and assigned a likelihood value based on long-term stability.

    During direct imaging observations, planets may also be blocked by the inner or outer working angles of the observing instrument.
    This would result in missed detections when the planet falls within these obscured regions. 
    Detections might also be missed if they are at a phase angle that is too dim to observe. 
    In these cases, the probability of confusion for a system will be higher \cite{pogorelyuk2022deconfusing}. 
    However, with photometric information, we may be able to reject or assign a lower ranking to geometrically valid orbits which would require the planet to be at a phase angle which is inconsistent with the detection's brightness. 
    
    Future work will also account for the joint dependence of astrometry and photometry in the simulations. 
    That is, the knowledge of a planet's astrometry will be less accurate if its phase angle is higher and the planet is therefore dimmer, and vice versa.
    A detailed study of the effect of non-detections on confusion rates and the scheduling (and number) of visits per system would also directly inform yield calculations for future missions.
    Adjusting the observing cadence of a system may support reducing confusion that remains due to closely separated planet detections. 
    Deviating from uniformly spaced observations, which were previously suggested for single-planet systems \cite{horning2019minimum}, might enable better distinction in separation and phase space.
    With additional orbit ranking metrics and simulations, we can provide a more complete assessment of confusion rates and their impact on yield for a mission like HWO. 

\subsection*{Disclosures}
The authors declare that there are no financial interests, commercial affiliations, or other potential conflicts of interest that could have influenced the objectivity of this research or the writing of this paper. 

\subsection* {Code, Data, and Materials Availability}\label{sec:data_avail}
The code developed in this manuscript is available on GitHub: \url{https://github.com/MIT-STARLab/deconfuser}.
~Orbital parameters for the systems described in this work are provided in supplemental machine-readable files and can be found on GitHub (\url{https://github.com/MIT-STARLab/deconfuser/tree/main/tutorials}) and Zenodo (\url{https://doi.org/10.5281/zenodo.18301600}). 
A Jupyter Notebook tutorial is available on GitHub to generate plots of the orbit fits for all the test case systems (like Figs.~\ref{fig:all_confused_options} \&~\ref{fig:compare_top_ranked}).

\subsection* {Acknowledgments}

    S.\,N.\,H. would like to thank Eric Nielsen for helpful discussion on MCMC orbit-fitting and Vanessa Bailey for support in specifying the Roman Coronagraph Instrument parameters.
    The authors would like to thank those who reviewed the manuscript for their helpful feedback and suggestions.
    This research was carried out at the Massachusetts Institute of Technology and in part at the Jet Propulsion Laboratory, California Institute of Technology. AI tools were used to clean up language and grammar. 


\appendix    

\section{Likelihood Ranking}\label{appendix_A}

    Here we describe the likelihood ranking scheme developed for deconfusion with photometry.
    In this work, we consider each planet detection to be an independent event, as described in Sec.~\ref{sec:ranking}. 
    Each planet detection will have a unique photon count that is dependent on factors including the orbital parameters of the system, the observation time, the background contribution, and the planet's properties. 
    The number of photons detected also depends on the parameters of the observing instrument, which motivates the inclusion of the detector model described in Secs.~\ref{sec:phot_model} and~\ref{sec:ranking}.
    
    Here we describe the number of counts collected by the observing detector with a slightly simplified approach for clarity, adopting the observation time as $\Delta t$, the background counts as $d$, and the photon flux as $\phi$. 
    The background here includes contributions from both the detector and astrophysical sources. 
    The expected number of photons for a detection ($n_p$) is obtained by combining these possible contributions to the photon count, which can be described by Eq.~(\ref{A_eqn:expected_counts}). 
    
    \begin{equation}\label{A_eqn:expected_counts}
        n_p(a, e, i, \Omega, \omega, M_0) = \phi_p(a, e, i, \Omega, \omega, M_0) \cdot \Delta t + d
    \end{equation}
    
    \noindent The orbital parameter definitions are listed in Table~\ref{tab:orb_def}. 

    \begin{table}[h!]
        \caption{Orbital parameter variables and their definitions.}                
        \label{tab:orb_def}    
        \centering        
        \small
        \begin{tabular}{cl}      
        \hline\hline               
        Parameter & Description \\         
        \hline                      
            $a$ & Semi-major axis (AU) \\
            $e$ & Eccentricity \\
            $i$ & Inclination \\
            $f$ & True anomaly \\
            $\omega_p$ & Argument of periastron \\
            $\Omega$ & Longitude of ascending node\\
            $E$ & Eccentric anomaly \\
            $M_0$ & Mean anomaly \\
            $\tau$ & Epoch of periastron passage \\
        \hline                                  
        \end{tabular}
    \end{table}
    
    The number of photons measured at the detector gives the planet's measured brightness (or contrast). 
    For each detection, the probability of measuring a certain number of photons, given the orbital parameters of the system, can be described by the probability function,
    
    \begin{equation}\label{A_eqn:prob_function}
        L_\mathrm{detection} = f(y | n_p(a, e, i, \Omega, \omega, M_0)),
    \end{equation}
    
    \noindent where $f(y |n_{p})$ is the distribution returned by the noise model as described in Sec.~\ref{sec:ranking}. 
    This distribution represents the measured photons from the combination of the planet detection and detector noise.
    Each set of planet detections must be considered together as part of a system in order to determine the likelihood of the whole system's orbit matches ($L_\mathrm{system}$). 
    The likelihood of measuring all detections for a system given the parameters of the matched orbits can be described by the product of the individual detections,
    
    \begin{equation}\label{A_eqn:likelihood_func}
        L_\mathrm{system} = \prod_{j=1}^{N} f(y_j |n_{p,j} (a_j , e_j , i_j , \Omega_j , \omega_j , M_{0,j} )),
    \end{equation}
    
    \noindent where $y$ is the measured photon count and $N$ is the number of measurements (detections). 
    The likelihood of a single planet's orbit in a system ($L_\mathrm{orbit}$) is the product of only the detections belonging to the orbit of the planet of interest, assuming the detections were correctly identified by the deconfuser.

\section{The Solar System as a Confused System}\label{appendix_B}

    We also apply the photometry likelihood ranking scheme to a group of Solar System planets as if they were a directly-imaged exoplanetary system. 
    We have chosen to model the four planets that fall within the semimajor axis range defined for the rest of the simulated systems in this paper, which includes Venus, Earth, Mars, and Jupiter. 
    Orbital parameters were adopted from JPL Horizons Solar System Dynamics page (\href{https://ssd.jpl.nasa.gov/planets/approx_pos.html}{https://ssd.jpl.nasa.gov/planets/approx\_pos.html}) \cite{standish1992orbital}. 
    We follow the same simulation methods as are described in Sec.~\ref{sec:sims}, with three observations of the system spread equally over the span of one Earth year. 
    Venus, Earth, Mars, and Jupiter do not exhibit confusion at low inclinations. 
    We present a medium  and high inclination case here.
    
    We maintain the range of the true inclinations of the planets with respect to one another and increase the overall viewing inclination by randomly sampling an inclination value in the medium  and high inclination groups and incorporating it for all planets. 
    The orbital parameters are listed in Table~\ref{tab:solar_system_sim}.
    We assign albedos and radii in the same manner as presented in Sec.~\ref{subsubsec:sim_detections} above, assuming albedos of 0.3 and Earth-like radii. 
    We also tested the photometry model using the true geometric albedos and radii of the planets (0.67, 0.3, 0.15, and 0.52 for Venus, Earth, Mars, and Jupiter, respectively\cite{Cox2000PhT....53j..77C, Mallama2006Icar..182...10M, Mallama2009Icar..204...11M, Mallama2012Icar..220..211M}); the order of the photometry ranking results are the same as the case with the constant Earth-like albedos and radii.
    Changing the albedos and radii of planets in this Solar System case does not change the orbits that are fit by the deconfuser (because photometry does not influence the orbit-fitting portion of the algorithm). 
    It also does not affect the results of photometry ranking, because the majority of the confusion in the Solar System cases stems from the detections of Earth and Venus being at small separations relative to one another on-sky.
    
    \begin{table*}[h!]
        \caption[Orbital parameters of the simulated Solar System planets (Venus, Earth, Mars, Jupiter) and the number of confused orbit options]{Orbital parameters of the simulated Solar System planets (Venus, Earth, Mars, Jupiter) and the number of confused orbit options per system. Detections and confused orbits are shown in Figs.~\ref{fig:SS_medi_confused} and~\ref{fig:SS_highi_confused} for the medium and high inclination cases, respectively. }                
        \label{tab:solar_system_sim}    
        \centering
        \small
        \begin{tabular}{cccccccccc}      
        \hline\hline               
        Planet & a & e & True i & ``Med'' i & ``High'' i & No.~confused & No.~confused \\ 
         & (AU) &  & (°) & (°) & (°) & (Med i) & (High i) \\
        \hline                      
        Venus   & 0.723 & 0.007 & 3.395 & 62.39 & 79.39 & \multirow{4}{*}{2} & \multirow{4}{*}{2} \\
        Earth   & 1.000 & 0.017 & 0.000 & 60.00 & 76.00 & &  \\
        Mars    & 1.524 & 0.093 & 1.850 & 60.85 & 77.85 & & \\
        Jupiter & 5.203 & 0.048 & 1.304 & 60.30 & 77.30 & & \\
        \hline                                  
        \end{tabular}
    \end{table*}

\begin{figure*}[h!]
        \centering
        \includegraphics[width=\linewidth]{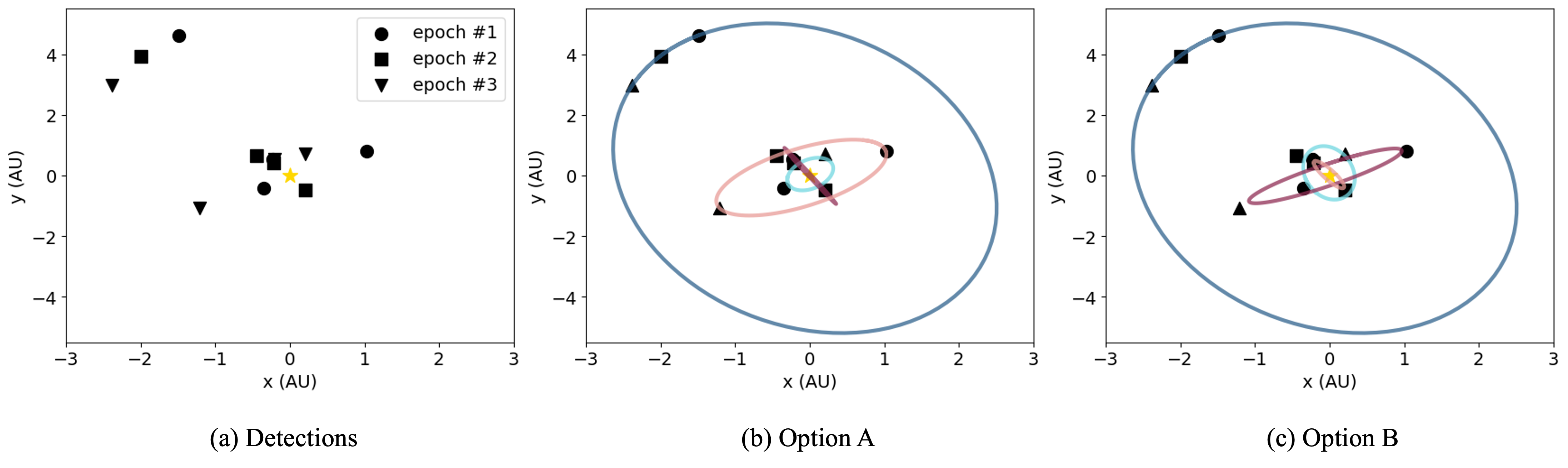}
        \caption{(a) Medium inclination simulated detections of Venus, Earth, Mars, and Jupiter. Note that some detections are overlapping. For example, detections from epochs one (circle) and two (square) of the inner two planets to the left of the star. (b) and (c) show the two confused orbit options returned by the deconfuser.} 
        \label{fig:SS_medi_confused}
    \end{figure*}

\begin{figure*}[h!]
        \centering
        \includegraphics[width=\linewidth]{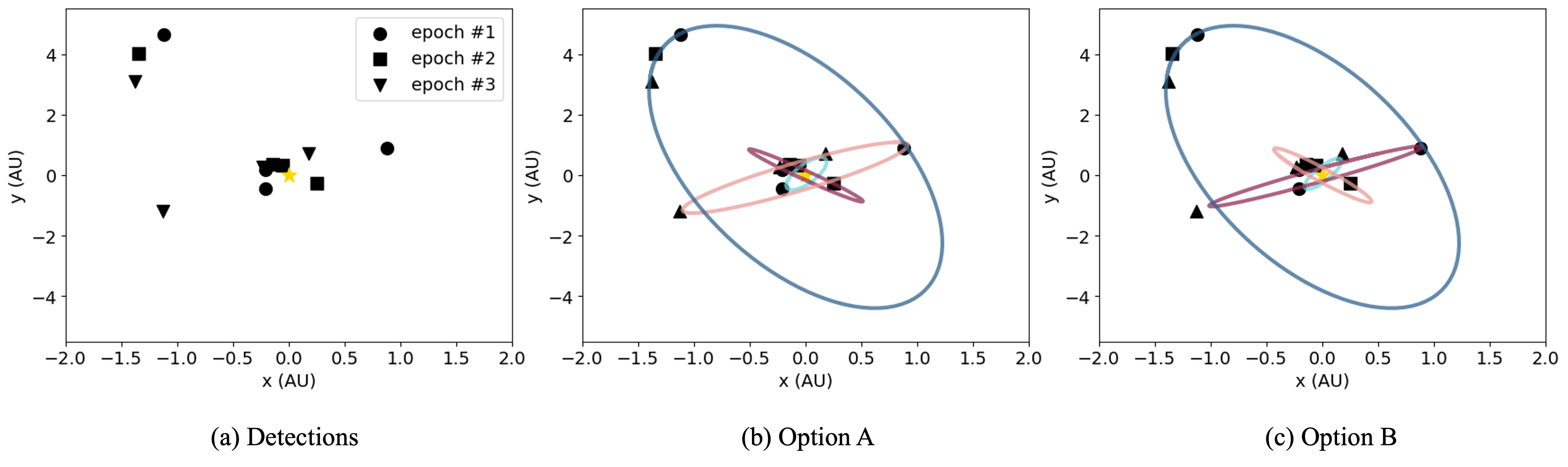}
        \caption{(a) High inclination simulated detections of Venus, Earth, Mars, and Jupiter. (b) and (c) show the two confused orbit options returned by the astrometric deconfuser.} 
        \label{fig:SS_highi_confused}
    \end{figure*}

    In both the medium  and high inclination scenarios, the astrometric deconfuser returns two confused orbit options for the four Solar System planets. 
    The medium inclination detections and confused options are shown in Fig.~\ref{fig:SS_medi_confused}. 
    The high inclination detections and confused options are shown in Fig.~\ref{fig:SS_highi_confused}. 

\begin{figure}[h!]
    \begin{center}
    \begin{tabular}{c}
        \includegraphics[width=0.8\linewidth]{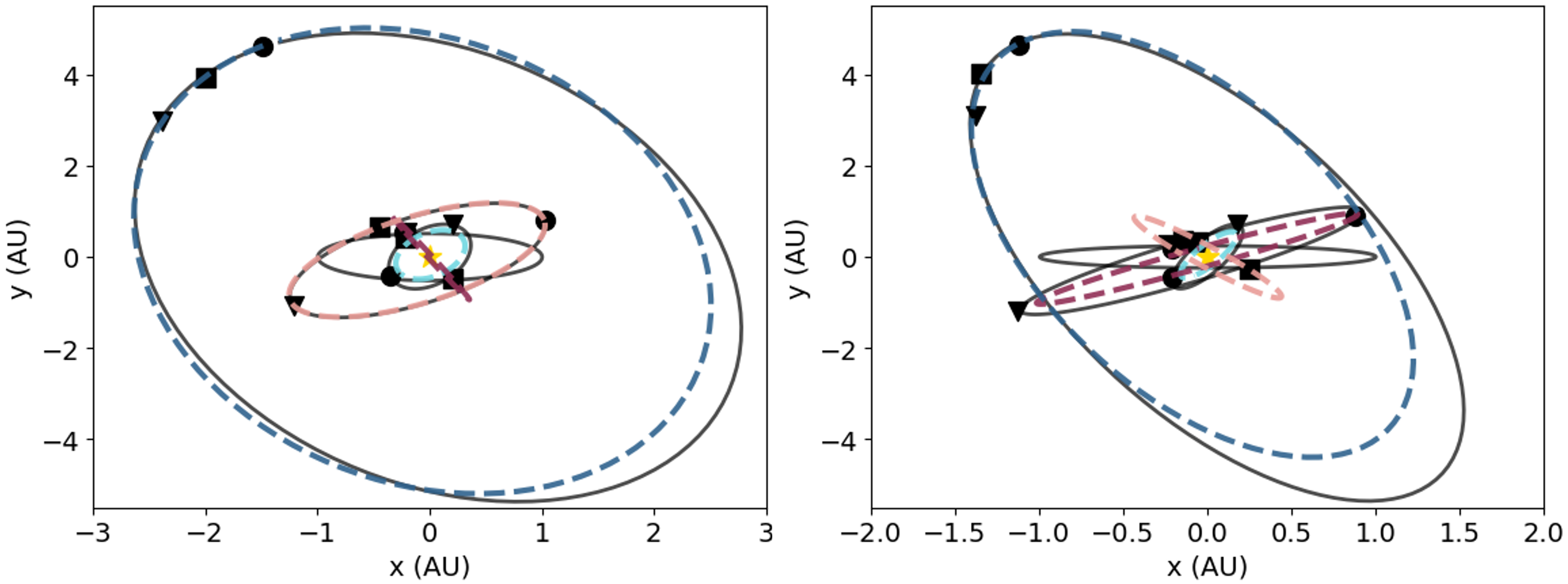}
        \\
        (a) \hspace{5.8cm} (b)
    \end{tabular}
    \end{center}
        \caption{Top ranked orbit options (dashed lines) compared to the true simulated Solar System orbits (solid black lines) at medium  (a) and high (b) inclination.} 
        \label{fig:SS_compared}
    \end{figure}

    The top-ranked orbit options are compared to the true orbits in Fig.~\ref{fig:SS_compared}.
    The confused orbit options and their rankings are given in Table~\ref{tab:solar_system_rankings}.
    In the medium inclination case, the photometry ranking scheme ranks option A (panel b in Fig.~\ref{fig:SS_medi_confused}) higher than option B (panel c in Fig.~\ref{fig:SS_medi_confused}). 
    Option A also contains the orbits which have the best match to the true system. 
    That is, the semimajor axes, eccentricities, and inclinations in option A have the smallest percent difference between the two options, compared to the true system. 

    \begin{table*}
        \caption{Solar system photometry ranking results. Orbital parameters and updated rankings for the confused orbit options returned by the combined deconfuser for the medium, and high inclination Solar system case. The confused orbit options are shown in Figs.~\ref{fig:SS_medi_confused} and~\ref{fig:SS_highi_confused}.}
        \label{tab:solar_system_rankings} 
        \centering
        \footnotesize
        \begin{tabular}{cccccccccc}
        \hline\hline             
         &  & & \multicolumn3c{``Confused'' orbit parameters} & Deconfuser &   &   & \\
        Incl. & System & Planet no. & $a$ & $e$ & $i$ & RMS fit & Orbit & System & System \\
        Group & option &  & (AU) & & (°) & error (AU) & likelihood & likelihood & ranking \\
        \hline
        \multirow{4}{*}{Med} & \multirow{4}{*}{A} & 1 & 0.553 & 0.083 & 120.70 & 0.113 & $1.221\mathrm{e}{\minus 9}$  & \multirow{4}{*}{$1.431\mathrm{e}{\minus 14}$} & \multirow{4}{*}{1} \\
         & & 2 & 0.991 & 0.078 & 88.74 & 0.045 & $1.144\mathrm{e}{\minus 2}$  &  &  \\
        &  & 3 & 1.523 & 0.097 & 60.71 & 0.007 & $7.227\mathrm{e}{\minus 3}$  & &  \\
        &  & 4 & 5.139 & 0.034 & 60.78 & 0.005 & $1.455\mathrm{e}{\minus 1}$  & &  \\
         \hline
        \multirow{4}{*}{Med} & \multirow{4}{*}{B} & 1 & 0.886 & 0.096 & 112.43 & 0.255 & $3.944\mathrm{e}{\minus 8}$  & \multirow{4}{*}{$1.076\mathrm{e}{\minus 18}$} & \multirow{4}{*}{2} \\
         & & 2 & 1.340 & 0.098 & 102.35 & 0.253 & $3.152\mathrm{e}{\minus 3}$  &  &  \\
        &  & 3 & 0.485 & 0.064 & 79.04 & 0.068 & $5.658\mathrm{e}{\minus 5}$  & &  \\
        &  & 4 & 5.139 & 0.034 & 60.78 & 0.005 & $1.530\mathrm{e}{\minus 3}$  & &  \\
        \hline
        \multirow{4}{*}{High} & \multirow{4}{*}{A} & 1 & 0.553 & 0.083 & 102.86 & 0.090 & $1.585\mathrm{e}{\minus 7}$ & \multirow{4}{*}{$3.182\mathrm{e}{\minus 15}$} & \multirow{4}{*}{2}  \\
         & & 2 & 1.006 & 0.033 & 85.28 & 0.009 & $9.966\mathrm{e}{\minus 3}$ & & \\
         & & 3 & 1.513 & 0.097 & 77.58 & 0.022 & $5.820\mathrm{e}{\minus 3}$ & & \\
         & & 4 & 4.726 & 0.071 & 76.60 & 0.024 & $3.460\mathrm{e}{\minus 4}$ & & \\
        \hline
        \multirow{4}{*}{High} & \multirow{4}{*}{B} & 1 & 0.525 & 0.090 & 79.96 & 0.195 & $1.255\mathrm{e}{\minus 6}$ & \multirow{4}{*}{$7.356\mathrm{e}{\minus 14}$} & \multirow{4}{*}{1} \\
         & & 2 & 1.376 & 0.099 & 96.50 & 0.235 & $4.948\mathrm{e}{\minus 3}$ & &  \\
         & & 3 & 0.991 & 0.011 & 96.62 & 0.012 & $6.511\mathrm{e}{\minus 2}$ & & \\
         & & 4 & 4.726 & 0.071 & 76.60 & 0.024 & $1.819\mathrm{e}{\minus 4}$ & &  \\
        \hline
        \end{tabular}
    \end{table*}

    In the high inclination case, the photometry ranking scheme ranks option B (panel c in Fig.~\ref{fig:SS_highi_confused}) as the top option, despite it being the ``worst'' fit of the two options. 
    The statistical nature of photon detections compound with the small separations between Venus and Earth in this scenario, as discussed in Sec.~\ref{sec:discussion}.
    The likelihood of the orbit fit for Jupiter also varies between options one and two, despite the parameters being the same, due to the limited number of photons received. 
        
    The results of this Solar System test agree with the results from the simulated systems presented in Secs.~\ref{sec:deconf_w_phot} and~\ref{subsec:disccussion_phot_deconf} above. 
    Photometry largely shows promise for deconfusion of otherwise closely ranked orbit options, but can be complicated by low-photon statistics and small separations between multiple planets in a system.

\section{Comparing the deconfuser and orbitize}\label{appendixC_orbitize}

    To validate the orbit fits returned by the deconfuser, we compared two of the simulated systems and their fits presented in this work to orbit fits generated by running the MCMC implementation of \texttt{orbitize!}~\cite{blunt2020orbitize, Blunt2024JOSS....9.6756B, Foreman-Mackey2013PASP..125..306F, Vousden2016MNRAS.455.1919V}. 
    \texttt{orbitize!}~is a widely-used and well-tested orbit-fitting software which infers posterior distributions of orbital parameters given their relative astrometry ($\Delta$R.A./$\Delta$decl.) on the sky. 
    We emphasize again here that the deconfuser is meant to quickly provide discrete solutions of possible orbit fits compared to other orbit-fitting tools which provide posterior distributions of orbital parameters.

    To compare the simulated system detections generated in this work, we first convert the on-sky positions from Cartesian coordinates to relative right ascension and declination. 
    We follow the conversion process employed by \texttt{orbitize!}, which was initially presented in Green 1985\cite{Green1985spas.book.....G} and is shown in Equations~\ref{eqn:RA} and~\ref{eqn:decl}.  

    \begin{equation}\label{eqn:RA}
        \Delta \mathrm{R.A.} = \pi a(1-e\cos E) \Big[ 
        \cos^2{\tfrac{i}{2}}\,\sin(f+\omega_p+\Omega) - \sin^2{\tfrac{i}{2}}\,\sin(f+\omega_p-\Omega) \Big]
    \end{equation}
    
    \begin{equation}\label{eqn:decl}
        \Delta \mathrm{decl.} = \pi a(1-e\cos E)\Big[
        \cos^2{\tfrac{i}{2}}\,\cos(f+\omega_p+\Omega) + \sin^2{\tfrac{i}{2}}\,\cos(f+\omega_p-\Omega) \Big]
    \end{equation}

    \noindent The orbital parameters are defined in Table~\ref{tab:orb_def}.
    
    MCMC and other Bayesian processes are undefined if the error is zero. Therefore, we must include some error on the relative astrometry in order to employ MCMC for orbit-fitting. We apply a small random Gaussian error to the astrometry with $\sigma =$ 1\,mas. This is comparable to current real-life astrometry error \cite{Wang2016AJ....152...97W}. 
        
    We generate the input files required by \texttt{orbitize!}~by converting the epochs of observation to Modified Julian Day (MJD) format (JD-2400000.5). 
    The chosen date in MJD is arbitrary, but the spacing of the observation epochs remains consistent with the epoch spacing simulated in this work. For example, MJD 59215.0, 59397.5, and 59580.0 represent three equally spaced observations spread over one year.
    
    We initialize the MCMC sampler with 20 temperatures, 1,000 walkers, and 50,000 steps per walker. 
    For MCMC, temperature scales the target function to adjust the efficiency of exploring the parameter space. 
    Higher temperatures allow for more efficient exploration of the parameter space without becoming stuck in regions with high probability density; 
    lower temperatures allow chains to efficiently sample the high probability regions \cite{Earl2005PCCP....7.3910E, Vousden2016MNRAS.455.1919V}.
    We visually inspect the walkers to assess convergence and reject the first 1,000 steps before computing the posterior distribution. 
    We use the default priors in \texttt{orbitize!}~(see Table 1 in Blunt et al.~2020\cite{blunt2020orbitize}), except for the prior on semimajor axis. 
    We reduce the maximum range on the semimajor axis prior from 10,000\,AU to 10\,AU to reduce computation time. 
    This scaled-down maximum would still fall well outside the 9.7\,$\lambda/D$ outer working angle (OWA) of the Roman Hybrid-Lyot Coronagraph \cite{Riggs2021SPIE11823E..1YR} for a system at 10\,pc.
    The semimajor axis prior then becomes LogUniform(1,10). 
    Of the known exoplanet systems at the time of writing, there are only 124 planets with semimajor axes greater than 10\,AU, and only 79 of those have semimajor axes less than 100\,AU \cite{Christiansen2025PSJ.....6..186C}.
    
\begin{figure*}[h!]
   \centering
    \includegraphics[width=\textwidth]{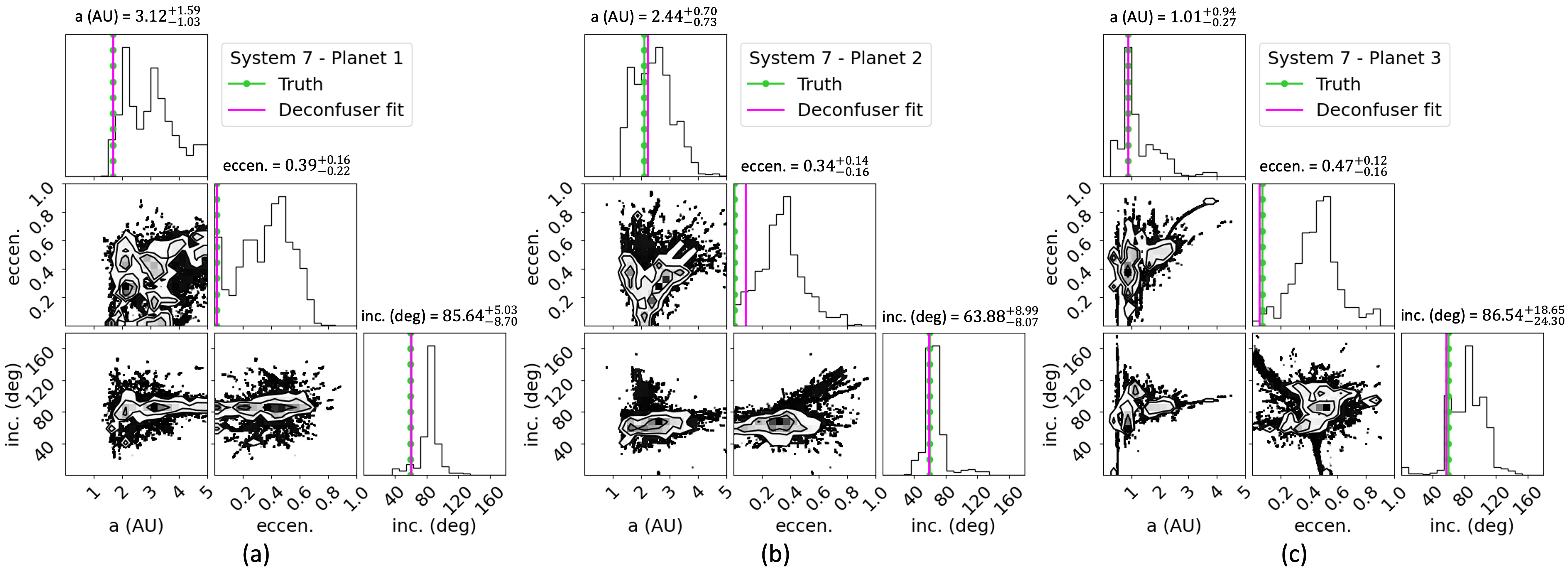}
      \caption{Posterior distributions of orbital parameters and orbit fits estimated by MCMC with \texttt{orbitize!}~for planets (a) one, (b) two, and (c) three in system seven. Posterior distributions are shown with the true parameters of the simulated system plotted in green and the top ranked deconfuser fit chosen by the photometry ranking in magenta. Mean values estimated by \texttt{orbitize!}~are also listed above each histogram and in Table~\ref{tab:sys7_orbitize_comparison}. }
     \label{fig:orbitize_all_corners}
   \end{figure*}

    Figure~\ref{fig:orbitize_all_corners} shows the posterior distributions of orbital parameters in medium inclination system seven returned by \texttt{orbitize!}~for planets one, two, and three, respectively. 
    The simulated system values (``truth'') and the deconfuser fit values are shown in green and magenta, respectively, on top of the posterior histograms.
    The comparison between the deconfuser fits and truth values emphasize that the deconfuser is fitting the shape parameters and inclination of orbits well, with the exception of the eccentricity of the second planet.
    For all three planets in medium inclination system seven, the semimajor axis is fit by the deconfuser to within 0.1-26\% of the true value, the eccentricity is fit to within 4-180\% of the true value, and the inclination is fit within 0.9-64\% of truth.
    The reduced accuracy of the fit of the second planet's eccentricity by the deconfuser could be attributed to the reduced fraction of the orbit that is covered (less than half) by the detections, compared to planets one and three.
    In comparison, the mean of the \texttt{orbitize!}~distributions are up to an 85\% difference for the planet's semimajor axes, up to 6700\% from their true eccentricities, and up to 46\% from their true inclinations. 
    A comparison of the true orbital parameters, top ranked deconfuser fit, and \texttt{orbitize!}~means are shown in Table~\ref{tab:sys7_orbitize_comparison}.

    \begin{table}[ht]
        \caption{Comparison between \texttt{orbitize!}~and deconfuser orbital parameters and their true values for medium inclination system seven. Corner plots are shown in Figure~\ref{fig:orbitize_all_corners} and orbit solutions are shown in Figure~\ref{fig:orbitize_orbits}. Semimajor axes ($a$) are given in units of AU, inclinations ($i$) are given in units of degrees. Deconfuser orbits were fit with a maximum tolerance of 0.05\,AU.}  
        \label{tab:sys7_orbitize_comparison}
        \begin{center}       
        \small
        \begin{tabular}{ccccr} 
        \hline\hline
        Planet & Orbit & Truth & Combined deconfuser & \texttt{orbitize!} \\
        no. & parameter & value & top-ranked fit & fit mean \\
        \hline\hline
        \rule[-1ex]{0pt}{3.5ex}  & $a$ & 1.681 & 1.680 & 3.12$^{+1.59}_{-1.03}$ \\
        \rule[-1ex]{0pt}{3.5ex}Planet 1 & $e$ & 0.015 & 0.014 & 0.39$^{+0.16}_{-0.22}$\\
        \rule[-1ex]{0pt}{3.5ex}         & $i$ & 59.00 & 68.16 & 85.64$^{+5.03}_{-8.70}$ \\
        \hline
        \rule[-1ex]{0pt}{3.5ex}          & $a$ & 2.090  & 2.222 & 2.44$^{+0.70}_{0.73}$\\
        \rule[-1ex]{0pt}{3.5ex} Planet 2 & $e$ &  0.005 & 0.087 & 0.34$^{+0.14}_{-0.16}$ \\
        \rule[-1ex]{0pt}{3.5ex}          & $i$ & 60.00  & 58.50 & 63.88$^{+8.99}_{-8.07}$ \\
         \hline
        \rule[-1ex]{0pt}{3.5ex}           & $a$ & 0.864 & 0.871 & 1.01$^{+0.94}_{-0.27}$ \\
        \rule[-1ex]{0pt}{3.5ex} Planet 3  & $e$ & 0.071 & 0.049 & 0.47$^{+0.12}_{-0.16}$ \\
        \rule[-1ex]{0pt}{3.5ex}           & $i$ & 60.00 & 56.76 & 86.54$^{+18.65}_{-24.30}$ \\
        \hline 
        \end{tabular}
        \end{center}
    \end{table} 

\begin{figure*}[h!]
        \centering
        \includegraphics[width=\textwidth]{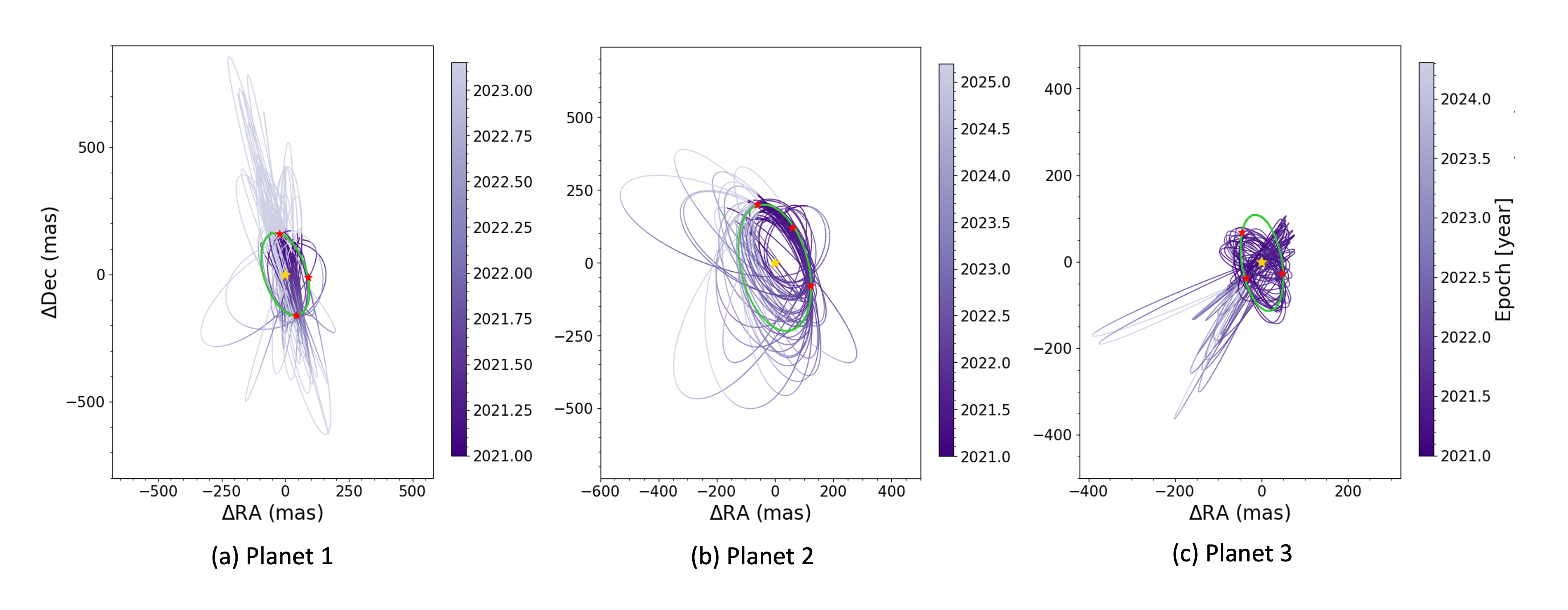}
        \caption{A subset of 50 random orbits fit by \texttt{orbitize!}~(purple) for the three planets in system seven: (a) planet one, (b) planet two, and (c) planet three. The top ranked deconfuser fit from the combined deconfuser is shown for comparison in green. Planet detections are indicated by red stars.}
        \label{fig:orbitize_orbits}
    \end{figure*}

    In most cases, the fit deconfuser value for the semimajor axis, eccentricity, and inclination falls within or near the peak of the posterior distribution fit by \texttt{orbitize!}, validating the deconfuser's fits to these parameters. 
    The cases where the deconfuser fit parameter falls outside of the peak of the posterior distribution are also cases where the true parameter is outside of the distribution, and therefore \texttt{orbitize!}~also struggled to find the true fit. 
    For all parameters (semimajor axis, eccentricity, and inclination), the deconfuser fit falls within three sigma of the \texttt{orbitize!}~mean. 
    However, the \texttt{orbitize!} distributions are relatively widespread (especially for $a$ and $e$).
    The orbit fits for each of the three planets are shown in Figure~\ref{fig:orbitize_orbits} with the top ranked combined deconfuser fit (i.e., from photometry) plotted for reference. 

\clearpage
\newpage

\section{Photometry Ranking Results for All Simulated Systems}\label{appendix_D_tables}
\singlespacing
\footnotesize
        \begin{center}
        \captionof{table}{Photometry ranking results for all simulated systems. Orbital parameters from the astrometric deconfuser and updated rankings from the combined deconfuser for the low, medium, and high inclination systems. RMS errors are also listed for each individual orbit fit by the astrometric deconfuser.}
        \label{tab:all_confused_rankings}
        \end{center}
        \tablehead{
        \hline
        \hline
         & & & \multicolumn{3}{c}{``Confused" orbit parameters} & Deconfuser & & \\
        Incl. & System & Planet & $a$ & $e$ & $i$ & RMS fit & Orbit & System & System \\
        Group & option & no. & (AU) & & (°) & error (AU) & likelihood & likelihood & ranking \\
        \hline
        \hline
        }
        
        \begin{supertabular}{cccccccccc}
        \hline
            & 1$_\mathrm{A}$ & 1 & 5.139 & 0.075 & 37.9 & 0.003 & $2.051\mathrm{e\textrm{-}}{2}$ &  & \\ 
        Low & 1$_\mathrm{A}$ & 2 & 2.752 & 0.092 & 36.25 & 0.006 & $1.500\mathrm{e\textrm{-}}{4}$ & $5.594\mathrm{e\textrm{-}}{9}$ & 1 \\ 
            & 1$_\mathrm{A}$ & 3 & 2.881 & 0.095 & 33.96 & 0.003 & $1.810\mathrm{e\textrm{-}}{3}$ & &  \\ 
        \hline
            & 1$_\mathrm{B}$ & 1 & 5.139 & 0.075 & 37.9 & 0.003 & $2.150\mathrm{e\textrm{-}}{2}$ &  & \\ 
        Low & 1$_\mathrm{B}$ & 2 & 2.733 & 0.093 & 33.1 & 0.042 & $2.600\mathrm{e\textrm{-}}{4}$ & $5.026\mathrm{e\textrm{-}}{9}$ & 2 \\ 
            & 1$_\mathrm{B}$ & 3 & 2.625 & 0.098 & 112.84 & 1.137 & $9.000\mathrm{e\textrm{-}}{4}$ & &  \\ 
        \hline
            & 1$_\mathrm{C}$ & 1 & 5.139 & 0.075 & 37.9 & 0.003 & $1.961\mathrm{e\textrm{-}}{2}$ &  & \\ 
        Low & 1$_\mathrm{C}$ & 2 & 2.538 & 0.098 & 63.7 & 1.227 & $1.600\mathrm{e\textrm{-}}{4}$ & $4.775\mathrm{e\textrm{-}}{9}$ & 3 \\ 
            & 1$_\mathrm{C}$ & 3 & 2.625 & 0.098 & 65.37 & 1.152 & $1.500\mathrm{e\textrm{-}}{3}$ & &  \\ 
        \hline
            & 1$_\mathrm{D}$ & 1 & 5.139 & 0.075 & 37.9 & 0.003 & $2.596\mathrm{e\textrm{-}}{2}$ &  & \\ 
        Low & 1$_\mathrm{D}$ & 2 & 2.661 & 0.075 & 31.7 & 0.029 & $2.500\mathrm{e\textrm{-}}{4}$ & $1.262\mathrm{e\textrm{-}}{9}$ & 4 \\ 
            & 1$_\mathrm{D}$ & 3 & 2.555 & 0.098 & 119.04 & 1.289 & $2.000\mathrm{e\textrm{-}}{4}$ & &  \\ 
        \hline
            & 2$_\mathrm{A}$ & 1 & 1.314 & 0.067 & 36.74 & 0.011 & $1.700\mathrm{e\textrm{-}}{4}$ &  & \\ 
        Low & 2$_\mathrm{A}$ & 2 & 0.937 & 0.027 & 39.82 & 0.005 & $1.000\mathrm{e\textrm{-}}{4}$ & $4.261\mathrm{e\textrm{-}}{10}$ & 1 \\ 
            & 2$_\mathrm{A}$ & 3 & 2.469 & 0.098 & 34.65 & 0.019 & $2.483\mathrm{e\textrm{-}}{2}$ & &  \\ 
        \hline
            & 2$_\mathrm{B}$ & 1 & 1.423 & 0.097 & 51.05 & 0.046 & $6.000\mathrm{e\textrm{-}}{5}$ &  & \\ 
        Low & 2$_\mathrm{B}$ & 2 & 0.814 & 0.092 & 61.44 & 0.332 & $4.000\mathrm{e\textrm{-}}{5}$ & $5.730\mathrm{e\textrm{-}}{11}$ & 2 \\ 
            & 2$_\mathrm{B}$ & 3 & 2.469 & 0.098 & 34.65 & 0.019 & $2.793\mathrm{e\textrm{-}}{2}$ & &  \\ 
        \hline
            & 3$_\mathrm{A}$ & 1 & 2.862 & 0.084 & 37.81 & 0.008 & $8.800\mathrm{e\textrm{-}}{4}$ &  & \\ 
        Low & 3$_\mathrm{A}$ & 2 & 2.190 & 0.052 & 35.04 & 0.003 & $4.900\mathrm{e\textrm{-}}{4}$ & $3.113\mathrm{e\textrm{-}}{9}$ & 2 \\ 
            & 3$_\mathrm{A}$ & 3 & 2.555 & 0.074 & 38.17 & 0.008 & $7.250\mathrm{e\textrm{-}}{3}$ & &  \\ 
        \hline
            & 3$_\mathrm{B}$ & 1 & 2.956 & 0.089 & 45.09 & 0.002 & $1.010\mathrm{e\textrm{-}}{3}$ &  & \\ 
        Low & 3$_\mathrm{B}$ & 2 & 2.825 & 0.099 & 114.44 & 1.073 & $2.900\mathrm{e\textrm{-}}{3}$ & $1.944\mathrm{e\textrm{-}}{8}$ & 1 \\ 
            & 3$_\mathrm{B}$ & 3 & 2.19 & 0.052 & 35.04 & 0.003 & $6.630\mathrm{e\textrm{-}}{3}$ & &  \\ 
        \hline
            & 4$_\mathrm{A}$ & 1 & 4.532 & 0.096 & 39.09 & 0.009 & $2.250\mathrm{e\textrm{-}}{3}$ &  & \\ 
        Low & 4$_\mathrm{A}$ & 2 & 1.779 & 0.035 & 35.30 & 0.006 & $3.360\mathrm{e\textrm{-}}{3}$ &  $6.160\mathrm{e\textrm{-}}{9}$ & 2  \\ 
            & 4$_\mathrm{A}$ & 3 & 1.233 & 0.03 & 38.39 & 0.010 & $8.100\mathrm{e\textrm{-}}{4}$ & & \\ 
        \hline
            & 4$_\mathrm{B}$ & 1 & 4.532 & 0.096 & 39.09 & 0.009 & $1.750\mathrm{e\textrm{-}}{3}$ &  & \\ 
        Low & 4$_\mathrm{B}$ & 2 & 1.272 & 0.098 & 111.74 & 0.626 & $4.730\mathrm{e\textrm{-}}{3}$ &  $9.306\mathrm{e\textrm{-}}{9}$ & 1 \\ 
            & 4$_\mathrm{B}$ & 3 & 1.349 & 0.081 & 142.94 & 0.006 & $1.130\mathrm{e\textrm{-}}{3}$ & &   \\ 
        \hline
            & 5$_\mathrm{A}$ & 1 & 0.969 & 0.075 & 163.09 & 0.020 & $2.020\mathrm{e\textrm{-}}{3}$ &  & \\ 
        Low & 5$_\mathrm{A}$ & 2 & 2.486 & 0.079 & 27.26 & 0.003 & $1.810\mathrm{e\textrm{-}}{3}$ & $2.578\mathrm{e\textrm{-}}{7}$ & 1  \\ 
            & 5$_\mathrm{A}$ & 3 & 2.844 & 0.084 & 19.74 & 0.003 & $7.062\mathrm{e\textrm{-}}{2}$ & & \\ 
        \hline
            & 5$_\mathrm{B}$ & 1 & 0.969 & 0.075 & 163.09 & 0.020 & $1.880\mathrm{e\textrm{-}}{3}$ &  & \\ 
        Low & 5$_\mathrm{B}$ & 2 & 2.900 & 0.086 & 35.11 & 0.015 & $2.080\mathrm{e\textrm{-}}{3}$ & $6.538\mathrm{e\textrm{-}}{8}$ & 2  \\ 
            & 5$_\mathrm{B}$ & 3 & 2.862 & 0.099 & 116.68 & 1.196 & $1.667\mathrm{e\textrm{-}}{2}$ & & \\ 
        \hline
            & 6$_\mathrm{A}$ & 1 & 1.367 & 0.09 & 48.93 & 0.007 & $5.640\mathrm{e\textrm{-}}{3}$ &  & \\ 
        Low & 6$_\mathrm{A}$ & 2 & 3.876 & 0.051 & 42.08 & 0.002 & $2.560\mathrm{e\textrm{-}}{3}$ &  $9.151\mathrm{e\textrm{-}}{9}$ & 1 \\ 
            & 6$_\mathrm{A}$ & 3 & 0.906 & 0.09 & 42.14 & 0.009 & $6.300\mathrm{e\textrm{-}}{4}$ & &  \\ 
        \hline
            & 6$_\mathrm{B}$ & 1 & 1.492 & 0.097 & 67.31 & 0.010 & $7.310\mathrm{e\textrm{-}}{3}$ &  & \\ 
        Low & 6$_\mathrm{B}$ & 2 & 0.861 & 0.07 & 0.0 & 0.163 & $3.500\mathrm{e\textrm{-}}{3}$ & $1.547\mathrm{e\textrm{-}}{9}$ & 2 \\ 
            & 6$_\mathrm{B}$ & 3 & 3.876 & 0.051 & 42.08 & 0.002 & $6.000\mathrm{e\textrm{-}}{5}$ & &  \\ 
        \hline
            & 7$_\mathrm{A}$ & 1 & 1.202 & 0.058 & 24.6 & 0.007 & $1.244\mathrm{e\textrm{-}}{2}$ &  & \\ 
        Low & 7$_\mathrm{A}$ & 2 & 4.001 & 0.098 & 42.53 & 0.003 & $4.410\mathrm{e\textrm{-}}{3}$ & $6.208\mathrm{e\textrm{-}}{9}$ & 1 \\ 
            & 7$_\mathrm{A}$ & 3 & 1.085 & 0.072 & 130.46 & 0.014 & $1.100\mathrm{e\textrm{-}}{4}$ & &  \\ 
        \hline
            & 7$_\mathrm{B}$ & 1 & 1.492 & 0.097 & 67.0 & 0.022 & $2.420\mathrm{e\textrm{-}}{3}$ &  & \\ 
        Low & 7$_\mathrm{B}$ & 2 & 1.180 & 0.093 & 133.53 & 0.018 & $4.340\mathrm{e\textrm{-}}{3}$ & $5.626\mathrm{e\textrm{-}}{12}$ & 2 \\ 
            & 7$_\mathrm{B}$ & 3 & 4.001 & 0.098 & 42.53 & 0.003 & $5.354\mathrm{e\textrm{-}}{7}$ & &  \\ 
        \hline
            & 8$_\mathrm{A}$ & 1 & 1.729 & 0.035 & 40.01 & 0.006 & $2.700\mathrm{e\textrm{-}}{4}$ &  & \\ 
        Low & 8$_\mathrm{A}$ & 2 & 2.098 & 0.007 & 40.27 & 0.005 & $1.140\mathrm{e\textrm{-}}{3}$ &  $7.248\mathrm{e\textrm{-}}{11}$ & 1  \\ 
            & 8$_\mathrm{A}$ & 3 & 0.861 & 0.091 & 40.68 & 0.005 & $2.300\mathrm{e\textrm{-}}{4}$ & & \\ 
        \hline
            & 8$_\mathrm{B}$ & 1 & 1.376 & 0.099 & 99.97 & 0.488 & $2.900\mathrm{e\textrm{-}}{4}$ &  & \\ 
        Low & 8$_\mathrm{B}$ & 2 & 1.297 & 0.076 & 61.53 & 0.004 & $6.310\mathrm{e\textrm{-}}{3}$ & $1.360\mathrm{e\textrm{-}}{12}$ & 2\\ 
            & 8$_\mathrm{B}$ & 3 & 2.098 & 0.007 & 40.27 & 0.005 & $7.410\mathrm{e\textrm{-}}{7}$ & &  \\ 
        \hline
            & 9$_\mathrm{A}$ & 1 & 5.358 & 0.075 & 21.5 & 0.010 & $2.239\mathrm{e\textrm{-}}{2}$ &  & \\ 
        Low & 9$_\mathrm{A}$ & 2 & 0.814 & 0.033 & 20.24 & 0.003 & $1.000\mathrm{e\textrm{-}}{4}$ & $3.964\mathrm{e\textrm{-}}{8}$ & 1 \\ 
            & 9$_\mathrm{A}$ & 3 & 1.297 & 0.02 & 21.29 & 0.007 & $1.697\mathrm{e\textrm{-}}{2}$ &  &  \\ 
        \hline
            & 9$_\mathrm{B}$ & 1 & 5.358 & 0.075 & 21.5 & 0.010 & $2.339\mathrm{e\textrm{-}}{2}$ &  & \\ 
        Low & 9$_\mathrm{B}$ & 2 & 1.376 & 0.092 & 53.61 & 0.027 & $3.757\mathrm{e\textrm{-}}{7}$ & $2.531\mathrm{e\textrm{-}}{10}$ & 2  \\ 
            & 9$_\mathrm{B}$ & 3 & 1.24 & 0.08 & 124.12 & 0.007 & $2.879\mathrm{e\textrm{-}}{2}$ & & \\ 
        \hline
            & 10$_\mathrm{A}$ & 1 & 3.938 & 0.065 & 25.91 & 0.004 & $5.684\mathrm{e\textrm{-}}{2}$ &  & \\ 
        Low & 10$_\mathrm{A}$ & 2 & 1.599 & 0.029 & 25.17 & 0.014 & $1.800\mathrm{e\textrm{-}}{4}$ & $1.485\mathrm{e\textrm{-}}{9}$ & 1 \\ 
            & 10$_\mathrm{A}$ & 3 & 0.537 & 0.025 & 27.94 & 0.005 & $1.500\mathrm{e\textrm{-}}{4}$ & &  \\ 
        \hline
            & 10$_\mathrm{B}$ & 1 & 3.938 & 0.065 & 25.91 & 0.004 & $6.708\mathrm{e\textrm{-}}{2}$ &  & \\ 
        Low & 10$_\mathrm{B}$ & 2 & 1.394 & 0.099 & 78.79 & 0.228 & $3.588\mathrm{e\textrm{-}}{4}$ & $1.907\mathrm{e\textrm{-}}{14}$ & 2  \\ 
            & 10$_\mathrm{B}$ & 3 & 6.02 & 0.097 & 91.12 & 0.285 & $7.923\mathrm{e\textrm{-}}{10}$ & & \\ 
        \hline
            & 1$_\mathrm{A}$ & 1 & 4.532 & 0.057 & 59.97 & 0.003 & $2.530\mathrm{e\textrm{-}}{2}$ &  & \\ 
        Med & 1$_\mathrm{A}$ & 2 & 2.661 & 0.07 & 61.08 & 0.001 & $9.000\mathrm{e\textrm{-}}{5}$ & $4.229\mathrm{e\textrm{-}}{8}$ & 4  \\ 
            & 1$_\mathrm{A}$ & 3 & 2.469 & 0.098 & 56.09 & 0.004 & $1.862\mathrm{e\textrm{-}}{2}$ & & \\ 
        \hline
            & 1$_\mathrm{B}$ & 1 & 4.532 & 0.057 & 59.97 & 0.003 & $4.771\mathrm{e\textrm{-}}{2}$ &  & \\ 
        Med & 1$_\mathrm{B}$ & 2 & 2.608 & 0.062 & 58.93 & 0.028 & $1.890\mathrm{e\textrm{-}}{3}$ & $1.440\mathrm{e\textrm{-}}{6}$ & 2 \\ 
            & 1$_\mathrm{B}$ & 3 & 2.555 & 0.098 & 102.94 & 0.763 & $1.603\mathrm{e\textrm{-}}{2}$ & &  \\ 
        \hline
            & 1$_\mathrm{C}$ & 1 & 4.532 & 0.057 & 59.97 & 0.003 & $2.607\mathrm{e\textrm{-}}{2}$ &  & \\ 
        Med & 1$_\mathrm{C}$ & 2 & 2.608 & 0.098 & 70.40 & 1.217 & $8.300\mathrm{e\textrm{-}}{4}$ & $2.460\mathrm{e\textrm{-}}{7}$ & 3 \\ 
            & 1$_\mathrm{C}$ & 3 & 2.52 & 0.098 & 76.17 & 0.787 & $1.132\mathrm{e\textrm{-}}{2}$ & &  \\ 
        \hline
            & 1$_\mathrm{D}$ & 1 & 4.532 & 0.057 & 59.97 & 0.003 & $7.685\mathrm{e\textrm{-}}{2}$ &  & \\ 
        Med & 1$_\mathrm{D}$ & 2 & 2.679 & 0.098 & 58.95 & 0.028 & $3.010\mathrm{e\textrm{-}}{3}$ &  $4.690\mathrm{e\textrm{-}}{6}$ & 1 \\ 
            & 1$_\mathrm{D}$ & 3 & 2.679 & 0.098 & 111.4 & 1.266 & $2.029\mathrm{e\textrm{-}}{2}$ & &  \\ 
        \hline
            & 2$_\mathrm{A}$ & 1 & 1.314 & 0.052 & 62.38 & 0.004 & $7.000\mathrm{e\textrm{-}}{5}$ &  & \\ 
        Med & 2$_\mathrm{A}$ & 2 & 0.958 & 0.025 & 51.52 & 0.005 & $4.100\mathrm{e\textrm{-}}{4}$ & $1.047\mathrm{e\textrm{-}}{9}$ & 1 \\ 
            & 2$_\mathrm{A}$ & 3 & 2.825 & 0.089 & 65.94 & 0.0003 & $3.541\mathrm{e\textrm{-}}{2}$ &  & \\ 
        \hline
            & 2$_\mathrm{B}$ & 1 & 1.349 & 0.096 & 67.78 & 0.038 & $9.000\mathrm{e\textrm{-}}{5}$ &  & \\ 
        Med & 2$_\mathrm{B}$ & 2 & 0.814 & 0.092 & 72.82 & 0.268 & $1.100\mathrm{e\textrm{-}}{4}$ & $3.040\mathrm{e\textrm{-}}{10}$ & 2 \\ 
            & 2$_\mathrm{B}$ & 3 & 2.825 & 0.089 & 65.94 & 0.0003 & $3.287\mathrm{e\textrm{-}}{2}$ & &  \\ 
        \hline
            & 3$_\mathrm{A}$ & 1 & 2.788 & 0.085 & 67.71 & 0.006 & $8.100\mathrm{e\textrm{-}}{4}$ &  & \\ 
        Med & 3$_\mathrm{A}$ & 2 & 2.334 & 0.019 & 68.57 & 0.003 & $2.300\mathrm{e\textrm{-}}{4}$ & $3.353\mathrm{e\textrm{-}}{9}$ & 3 \\ 
            & 3$_\mathrm{A}$ & 3 & 2.752 & 0.047 & 68.66 & 0.007 & $1.839\mathrm{e\textrm{-}}{2}$ & &  \\ 
        \hline
            & 3$_\mathrm{B}$ & 1 & 2.844 & 0.089 & 68.58 & 0.018 & $1.780\mathrm{e\textrm{-}}{3}$ &  & \\ 
        Med & 3$_\mathrm{B}$ & 2 & 2.679 & 0.098 & 101.49 & 0.498 & $4.700\mathrm{e\textrm{-}}{4}$ & $2.418\mathrm{e\textrm{-}}{8}$ & 1  \\ 
            & 3$_\mathrm{B}$ & 3 & 2.334 & 0.019 & 68.57 & 0.003 & $2.878\mathrm{e\textrm{-}}{2}$ & & \\ 
        \hline
            & 3$_\mathrm{C}$ & 1 & 3.11 & 0.089 & 70.86 & 0.015 & $1.950\mathrm{e\textrm{-}}{3}$ &  & \\ 
        Med & 3$_\mathrm{C}$ & 2 & 2.919 & 0.099 & 103.2 & 0.922 & $4.000\mathrm{e\textrm{-}}{4}$ & $1.682\mathrm{e\textrm{-}}{8}$ & 2  \\ 
            & 3$_\mathrm{C}$ & 3 & 2.334 & 0.019 & 68.57 & 0.003 & $2.148\mathrm{e\textrm{-}}{2}$ & & \\ 
        \hline
            & 4$_\mathrm{A}$ & 1 & 4.618 & 0.097 & 64.29 & 0.016 & $3.190\mathrm{e\textrm{-}}{3}$ &  & \\ 
        Med & 4$_\mathrm{A}$ & 2 & 1.753 & 0.0 & 62.49 & 0.003 & $5.180\mathrm{e\textrm{-}}{3}$ & $1.301\mathrm{e\textrm{-}}{7}$ & 2 \\ 
            & 4$_\mathrm{A}$ & 3 & 1.225 & 0.029 & 62.2 & 0.009 & $7.870\mathrm{e\textrm{-}}{3}$ & &  \\ 
        \hline
            & 4$_\mathrm{B}$ & 1 & 4.618 & 0.097 & 64.29 & 0.016 & $2.830\mathrm{e\textrm{-}}{3}$ &  & \\ 
        Med & 4$_\mathrm{B}$ & 2 & 1.297 & 0.098 & 101.57 & 0.513 & $8.630\mathrm{e\textrm{-}}{3}$ & $1.600\mathrm{e\textrm{-}}{7}$ & 1 \\ 
            & 4$_\mathrm{B}$ & 3 & 0.847 & 0.049 & 0.0 & 0.151 & $6.540\mathrm{e\textrm{-}}{3}$ & &  \\ 
        \hline
            & 5$_\mathrm{A}$ & 1 & 0.964 & 0.067 & 113.96 & 0.008 & $3.800\mathrm{e\textrm{-}}{4}$ &  & \\ 
        Med & 5$_\mathrm{A}$ & 2 & 2.956 & 0.097 & 64.84 & 0.005 & $1.100\mathrm{e\textrm{-}}{4}$ & $3.743\mathrm{e\textrm{-}}{11}$ & 1 \\ 
            & 5$_\mathrm{A}$ & 3 & 3.227 & 0.096 & 62.72 & 0.018 & $9.400\mathrm{e\textrm{-}}{4}$ & &  \\ 
        \hline
            & 5$_\mathrm{B}$ & 1 & 1.035 & 0.09 & 67.07 & 0.013 & $1.500\mathrm{e\textrm{-}}{4}$ &  & \\ 
        Med & 5$_\mathrm{B}$ & 2 & 2.919 & 0.087 & 64.81 & 0.003 & $1.100\mathrm{e\textrm{-}}{4}$ & $2.523\mathrm{e\textrm{-}}{11}$ & 2 \\ 
            & 5$_\mathrm{B}$ & 3 & 2.679 & 0.098 & 108.97 & 1.248 & $1.580\mathrm{e\textrm{-}}{3}$ & &  \\ 
        \hline
            & 6$_\mathrm{A}$ & 1 & 1.331 & 0.056 & 58.22 & 0.004 & $1.830\mathrm{e\textrm{-}}{3}$ &  & \\ 
        Med & 6$_\mathrm{A}$ & 2 & 3.918 & 0.09 & 56.20 & 0.007 & $8.330\mathrm{e\textrm{-}}{3}$ & $9.997\mathrm{e\textrm{-}}{8}$ & 1  \\ 
            & 6$_\mathrm{A}$ & 3 & 1.078 & 0.10 & 121.06 & 0.008 & $6.570\mathrm{e\textrm{-}}{3}$ & & \\ 
        \hline
            & 6$_\mathrm{B}$ & 1 & 1.492 & 0.097 & 71.34 & 0.039 & $5.600\mathrm{e\textrm{-}}{4}$ &  & \\ 
        Med & 6$_\mathrm{B}$ & 2 & 0.866 & 0.047 & 47.80 & 0.129 & $5.720\mathrm{e\textrm{-}}{3}$ &  $3.893\mathrm{e\textrm{-}}{10}$ & 2 \\ 
            & 6$_\mathrm{B}$ & 3 & 3.918 & 0.09 & 56.20 & 0.007 & $1.200\mathrm{e\textrm{-}}{4}$ & &  \\ 
        \hline
            & 7$_\mathrm{A}$ & 1 & 1.656 & 0.022 & 59.36 & 0.006 & $1.420\mathrm{e\textrm{-}}{3}$ &  & \\ 
        Med & 7$_\mathrm{A}$ & 2 & 2.159 & 0.058 & 58.47 & 0.004 & $1.023\mathrm{e\textrm{-}}{2}$ & $6.254\mathrm{e\textrm{-}}{8}$ & 1 \\ 
            & 7$_\mathrm{A}$ & 3 & 1.124 & 0.02 & 112.9 & 0.008 & $4.310\mathrm{e\textrm{-}}{3}$ & &  \\ 
        \hline
            & 7$_\mathrm{B}$ & 1 & 1.394 & 0.099 & 97.75 & 0.408 & $8.000\mathrm{e\textrm{-}}{5}$ &  & \\ 
        Med & 7$_\mathrm{B}$ & 2 & 0.805 & 0.082 & 133.56 & 0.071 & $1.200\mathrm{e\textrm{-}}{3}$ & $4.980\mathrm{e\textrm{-}}{13}$ & 2 \\ 
            & 7$_\mathrm{B}$ & 3 & 2.159 & 0.058 & 58.47 & 0.004 & $1.000\mathrm{e\textrm{-}}{5}$ & &  \\ 
        \hline
            & 8$_\mathrm{A}$ & 1 & 4.834 & 0.097 & 44.92 & 0.002 & $1.014\mathrm{e\textrm{-}}{1}$ &  & \\ 
        Med & 8$_\mathrm{A}$ & 2 & 0.805 & 0.045 & 48.61 & 0.008 & $8.960\mathrm{e\textrm{-}}{3}$ & $2.024\mathrm{e\textrm{-}}{6}$ & 1  \\ 
            & 8$_\mathrm{A}$ & 3 & 1.289 & 0.01 & 49.19 & 0.006 & $2.230\mathrm{e\textrm{-}}{3}$ & & \\ 
        \hline
            & 8$_\mathrm{B}$ & 1 & 4.834 & 0.097 & 44.92 & 0.002 & $8.442\mathrm{e\textrm{-}}{2}$ &  & \\ 
        Med & 8$_\mathrm{B}$ & 2 & 1.233 & 0.099 & 56.52 & 0.042 & $3.000\mathrm{e\textrm{-}}{5}$ & $4.205\mathrm{e\textrm{-}}{9}$ & 2 \\ 
            & 8$_\mathrm{B}$ & 3 & 0.801 & 0.10 & 51.85 & 0.162 & $1.970\mathrm{e\textrm{-}}{3}$ & &  \\ 
        \hline
            & 9$_\mathrm{A}$  & 1 & 1.059 & 0.098 & 91.05 & 0.326 & $7.000\mathrm{e\textrm{-}}{5}$ &  & \\ 
        Med & 9$_\mathrm{A}$  & 2 & 1.61 & 0.011 & 62.58 & 0.004 & $2.100\mathrm{e\textrm{-}}{4}$ & $1.584\mathrm{e\textrm{-}}{9}$ & 1  \\ 
            & 9$_\mathrm{A}$  & 3 & 3.425 & 0.095 & 62.39 & 0.010 & $1.115\mathrm{e\textrm{-}}{1}$ & & \\ 
        \hline
            & 9$_\mathrm{B}$ & 1 & 1.195 & 0.10 & 79.54 & 0.482 & $2.600\mathrm{e\textrm{-}}{4}$ &  & \\ 
        Med & 9$_\mathrm{B}$ & 2 & 0.876 & 0.095 & 57.43 & 0.161 & $1.000\mathrm{e\textrm{-}}{5}$ & $1.225\mathrm{e\textrm{-}}{10}$ & 2 \\ 
            & 9$_\mathrm{B}$ & 3 & 3.425 & 0.095 & 62.39 & 0.010 & $3.242\mathrm{e\textrm{-}}{2}$ & &  \\ 
        \hline
            & 10$_\mathrm{A}$ & 1 & 3.980 & 0.086 & 59.39 & 0.017 & $3.419\mathrm{e\textrm{-}}{3}$ &  & \\ 
        Med & 10$_\mathrm{A}$ & 2 & 1.600 & 0.042 & 59.33 & 0.014 & $4.973\mathrm{e\textrm{-}}{3}$ &   $1.726\mathrm{e\textrm{-}}{8}$ & 1 \\ 
            & 10$_\mathrm{A}$ & 3 & 0.533 & 0.082 & 62.29 & 0.008 & $1.015\mathrm{e\textrm{-}}{3}$ & &  \\ 
        \hline
            & 10$_\mathrm{B}$ & 1 & 3.980 & 0.086 & 59.39 & 0.017 & $2.474\mathrm{e\textrm{-}}{3}$ &  & \\ 
        Med & 10$_\mathrm{B}$ & 2 & 1.195 & 0.098 & 74.36 & 0.055 & $9.212\mathrm{e\textrm{-}}{3}$ & $7.446\mathrm{e\textrm{-}}{16}$ & 2 \\ 
            & 10$_\mathrm{B}$ & 3 & 6.02 & 0.097 & 90.64 & 0.313 & $3.267\mathrm{e\textrm{-}}{11}$ & &  \\ 
        \hline
             & 1$_\mathrm{A}$ & 1 & 4.404 & 0.089 & 80.55 & 0.004 & $2.205\mathrm{e\textrm{-}}{2}$ &  & \\ 
        High & 1$_\mathrm{A}$ & 2 & 2.418 & 0.098 & 79.59 & 0.022 & $2.640\mathrm{e\textrm{-}}{3}$ & $2.435\mathrm{e\textrm{-}}{7}$ & 2 \\ 
             & 1$_\mathrm{A}$ & 3 & 2.862 & 0.064 & 80.32 & 0.012 & $4.190\mathrm{e\textrm{-}}{3}$ & &  \\ 
        \hline
             & 1$_\mathrm{B}$ & 1 & 4.404 & 0.089 & 80.55 & 0.004 & $1.783\mathrm{e\textrm{-}}{2}$ &  & \\ 
        High & 1$_\mathrm{B}$ & 2 & 2.302 & 0.076 & 78.44 & 0.009 & $3.330\mathrm{e\textrm{-}}{3}$ & $1.837\mathrm{e\textrm{-}}{7}$ & 3 \\ 
             & 1$_\mathrm{B}$ & 3 & 2.222 & 0.097 & 94.82 & 0.391 & $3.100\mathrm{e\textrm{-}}{3}$ & &  \\ 
        \hline
             & 1$_\mathrm{C}$ & 1 & 4.404 & 0.089 & 80.55 & 0.004 & $2.214\mathrm{e\textrm{-}}{2}$ &  & \\ 
        High & 1$_\mathrm{C}$ & 2 & 2.59 & 0.098 & 82.28 & 1.214 & $1.700\mathrm{e\textrm{-}}{3}$ &  $7.024\mathrm{e\textrm{-}}{8}$ & 4 \\ 
             & 1$_\mathrm{C}$ & 3 & 2.19 & 0.097 & 85.46 & 0.436 & $1.860\mathrm{e\textrm{-}}{3}$ & &  \\ 
        \hline
             & 1$_\mathrm{D}$ & 1 & 4.404 & 0.089 & 80.55 & 0.004 & $1.434\mathrm{e\textrm{-}}{2}$ &  & \\ 
        High & 1$_\mathrm{D}$ & 2 & 2.608 & 0.055 & 79.65 & 0.013 & $4.970\mathrm{e\textrm{-}}{3}$ & $2.539\mathrm{e\textrm{-}}{7}$ & 1  \\ 
             & 1$_\mathrm{D}$ & 3 & 2.469 & 0.098 & 99.06 & 1.235 & $3.560\mathrm{e\textrm{-}}{3}$ & & \\ 
        \hline
             & 2$_\mathrm{A}$ & 1 & 1.297 & 0.055 & 84.88 & 0.004 & $1.800\mathrm{e\textrm{-}}{4}$ &  & \\ 
        High & 2$_\mathrm{A}$ & 2 & 0.937 & 0.022 & 81.71 & 0.014 & $1.145\mathrm{e\textrm{-}}{2}$ & $4.938\mathrm{e\textrm{-}}{8}$ & 1  \\ 
             & 2$_\mathrm{A}$ & 3 & 2.52 & 0.098 & 85.29 & 0.008 & $2.443\mathrm{e\textrm{-}}{2}$ & & \\ 
        \hline
             & 2$_\mathrm{B}$ & 1 & 1.297 & 0.055 & 84.88 & 0.004 & $1.900\mathrm{e\textrm{-}}{4}$ &  & \\ 
        High & 2$_\mathrm{B}$ & 2 & 0.975 & 0.089 & 148.58 & 0.023 & $1.171\mathrm{e\textrm{-}}{2}$ & $8.227\mathrm{e\textrm{-}}{9}$ & 4 \\ 
             & 2$_\mathrm{B}$ & 3 & 1.217 & 0.099 & 90.17 & 0.647 & $3.640\mathrm{e\textrm{-}}{3}$ & &  \\ 
        \hline
             & $2_\mathrm{C}$ & 1 & 1.314 & 0.078 & 85.53 & 0.006 & $2.000\mathrm{e\textrm{-}}{4}$ &  & \\ 
        High & $2_\mathrm{C}$ & 2 & 1.472 & 0.098 & 92.49 & 0.114 & $2.310\mathrm{e\textrm{-}}{3}$ & $1.354\mathrm{e\textrm{-}}{8}$ & 2 \\ 
             & $2_\mathrm{C}$ & 3 & 2.52 & 0.098 & 85.29 & 0.008 & $2.909\mathrm{e\textrm{-}}{2}$ & &  \\ 
        \hline
             & 2$_\mathrm{D}$ & 1 & 1.314 & 0.078 & 85.53 & 0.006 & $2.200\mathrm{e\textrm{-}}{4}$ &  & \\ 
        High & 2$_\mathrm{D}$ & 2 & 3.547 & 0.075 & 89.23 & 0.008 & $3.250\mathrm{e\textrm{-}}{3}$ &  $1.158\mathrm{e\textrm{-}}{8}$ & 3  \\ 
             & 2$_\mathrm{D}$ & 3 & 1.017 & 0.09 & 32.78 & 0.022 & $1.637\mathrm{e\textrm{-}}{2}$ & & \\ 
        \hline
             & 3$_\mathrm{A}$ & 1 & 2.807 & 0.047 & 75.96 & 0.001 & $1.970\mathrm{e\textrm{-}}{3}$ &  & \\ 
        High & 3$_\mathrm{A}$ & 2 & 2.19 & 0.094 & 75.52 & 0.003 & $8.400\mathrm{e\textrm{-}}{4}$ & $3.004\mathrm{e\textrm{-}}{9}$ & 1 \\ 
             & 3$_\mathrm{A}$ & 3 & 2.807 & 0.052 & 77.33 & 0.012 & $1.820\mathrm{e\textrm{-}}{3}$ & &  \\ 
        \hline
             & 3$_\mathrm{B}$ & 1 & 2.919 & 0.099 & 76.63 & 0.024 & $1.920\mathrm{e\textrm{-}}{3}$ &  & \\ 
        High & 3$_\mathrm{B}$ & 2 & 2.697 & 0.09 & 97.15 & 0.303 & $6.400\mathrm{e\textrm{-}}{4}$ &  $1.108\mathrm{e\textrm{-}}{9}$ & 2 \\ 
             & 3$_\mathrm{B}$ & 3 & 2.19 & 0.094 & 75.52 & 0.003 & $9.000\mathrm{e\textrm{-}}{4}$ & &  \\ 
        \hline
             & 3$_\mathrm{C}$ & 1 & 2.994 & 0.074 & 77.69 & 0.005 & $3.270\mathrm{e\textrm{-}}{3}$ &  & \\ 
        High & 3$_\mathrm{C}$ & 2 & 2.919 & 0.099 & 97.65 & 0.887 & $2.900\mathrm{e\textrm{-}}{4}$ & $1.072\mathrm{e\textrm{-}}{9}$ & 3 \\ 
             & 3$_\mathrm{C}$ & 3 & 2.19 & 0.094 & 75.52 & 0.003 & $1.140\mathrm{e\textrm{-}}{3}$ & &  \\ 
        \hline
             & 4$_\mathrm{A}$ & 1 & 4.468 & 0.096 & 89.03 & 0.013 & $1.000\mathrm{e\textrm{-}}{5}$ &  & \\ 
        High & 4$_\mathrm{A}$ & 2 & 2.083 & 0.099 & 89.87 & 0.013 & $3.500\mathrm{e\textrm{-}}{4}$ & $4.482\mathrm{e\textrm{-}}{11}$ & 2  \\ 
             & 4$_\mathrm{A}$ & 3 & 1.256 & 0.063 & 88.18 & 0.015 & $1.711\mathrm{e\textrm{-}}{2}$ & & \\ 
        \hline
             & 4$_\mathrm{B}$ & 1 & 4.468 & 0.096 & 89.03 & 0.013 & $1.000\mathrm{e\textrm{-}}{5}$ &  & \\ 
        High & 4$_\mathrm{B}$ & 2 & 1.323 & 0.097 & 90.26 & 0.445 & $1.521\mathrm{e\textrm{-}}{2}$ & $1.589\mathrm{e\textrm{-}}{9}$ & 1 \\ 
             & 4$_\mathrm{B}$ & 3 & 1.159 & 0.086 & 90.79 & 0.032 & $1.395\mathrm{e\textrm{-}}{2}$ & &  \\ 
        \hline
             & 4$_\mathrm{C}$ & 1 & 1.534 & 0.1 & 91.14 & 0.048 & $1.000\mathrm{e\textrm{-}}{5}$ &  & \\ 
        High & 4$_\mathrm{C}$ & 2 & -- & -- & -- & -- & -- & $2.823\mathrm{e\textrm{-}}{7}$ & 3  \\ 
             & 4$_\mathrm{C}$ & 3 & 1.256 & 0.063 & 88.18 & 0.015 & $2.823\mathrm{e\textrm{-}}{2}$ & & \\ 
        \hline
             & 5$_\mathrm{A}$ & 1 & 0.969 & 0.061 & 85.58 & 0.017 & $5.830\mathrm{e\textrm{-}}{3}$ &  & \\ 
        High & 5$_\mathrm{A}$ & 2 & 2.52 & 0.066 & 90.26 & 0.007 & $2.000\mathrm{e\textrm{-}}{5}$ & $1.004\mathrm{e\textrm{-}}{8}$ & 1 \\ 
             & 5$_\mathrm{A}$ & 3 & 3.09 & 0.076 & 90.30 & 0.014 & $8.306\mathrm{e\textrm{-}}{2}$ & &  \\ 
        \hline
             & 5$_\mathrm{B}$ & 1 & 0.969 & 0.061 & 85.58 & 0.017 & $5.090\mathrm{e\textrm{-}}{3}$ &  & \\ 
        High & 5$_\mathrm{B}$ & 2 & 3.033 & 0.086 & 90.16 & 0.002 & $4.000\mathrm{e\textrm{-}}{5}$ & $2.220\mathrm{e\textrm{-}}{9}$ & 2 \\ 
             & 5$_\mathrm{B}$ & 3 & 1.017 & 0.096 & 0.0 & 1.254 & $1.172\mathrm{e\textrm{-}}{2}$ & &  \\ 
        \hline
             & 6$_\mathrm{A}$ & 1 & 1.256 & 0.01 & 87.43 & 0.004 & $1.045\mathrm{e\textrm{-}}{2}$ &  & \\ 
        High & 6$_\mathrm{A}$ & 2 & 3.897 & 0.035 & 89.27 & 0.002 & $7.630\mathrm{e\textrm{-}}{3}$ & $2.820\mathrm{e\textrm{-}}{9}$ & 2 \\ 
             & 6$_\mathrm{A}$ & 3 & 1.078 & 0.063 & 91.43 & 0.016 & $4.000\mathrm{e\textrm{-}}{5}$ & &  \\ 
        \hline
             & 6$_\mathrm{B}$ & 1 & 4.022 & 0.05 & 90.56 & 0.026 & $3.930\mathrm{e\textrm{-}}{3}$ &  & \\ 
        High & 6$_\mathrm{B}$ & 2 & 1.297 & 0.091 & 90.8 & 0.033 & $4.300\mathrm{e\textrm{-}}{4}$ & $2.059\mathrm{e\textrm{-}}{11}$ & 3 \\ 
             & 6$_\mathrm{B}$ & 3 & 1.078 & 0.063 & 91.43 & 0.016 & $1.000\mathrm{e\textrm{-}}{5}$ & &  \\ 
        \hline
             & 6$_\mathrm{C}$ & 1 & 4.022 & 0.05 & 90.56 & 0.026 & $1.323\mathrm{e\textrm{-}}{2}$ &  & \\ 
        High & 6$_\mathrm{C}$ & 2 & 1.151 & 0.097 & 89.34 & 0.026 & $1.010\mathrm{e\textrm{-}}{3}$ & $4.316\mathrm{e\textrm{-}}{8}$ & 1 \\ 
             & 6$_\mathrm{C}$ & 3 & 0.452 & 0.094 & 95.97 & 0.158 & $3.230\mathrm{e\textrm{-}}{3}$ & &  \\ 
        \hline
             & 6$_\mathrm{D}$ & 1 & 1.091 & 0.098 & 89.86 & 0.050 & $1.133\mathrm{e\textrm{-}}{2}$ &  & \\ 
        High & 6$_\mathrm{D}$ & 2 & 1.151 & 0.097 & 89.34 & 0.026 & $1.410\mathrm{e\textrm{-}}{3}$ & $9.713\mathrm{e\textrm{-}}{12}$ & 4 \\ 
             & 6$_\mathrm{D}$ & 3 & 3.897 & 0.035 & 89.27 & 0.002 & $6.093\mathrm{e\textrm{-}}{7}$ & &  \\ 
        \hline
             & 7$_\mathrm{A}$ & 1 & 1.779 & 0.088 & 85.12 & 0.004 & $1.706\mathrm{e\textrm{-}}{2}$ &  & \\ 
        High & 7$_\mathrm{A}$ & 2 & 2.286 & 0.085 & 86.52 & 0.008 & $1.820\mathrm{e\textrm{-}}{3}$ &  $9.011\mathrm{e\textrm{-}}{9}$ & 1  \\ 
             & 7$_\mathrm{A}$ & 3 & 1.124 & 0.02 & 91.13 & 0.005 & $2.900\mathrm{e\textrm{-}}{4}$ & & \\ 
        \hline
             & 7$_\mathrm{B}$ & 1 & 1.394 & 0.099 & 90.19 & 0.344 & $4.210\mathrm{e\textrm{-}}{3}$ &  & \\ 
        High & 7$_\mathrm{B}$ & 2 & 1.151 & 0.067 & 86.04 & 0.016 & $3.830\mathrm{e\textrm{-}}{3}$ & $1.521\mathrm{e\textrm{-}}{13}$ & 2 \\ 
             & 7$_\mathrm{B}$ & 3 & 2.286 & 0.085 & 86.52 & 0.008 & $9.441\mathrm{e\textrm{-}}{9}$ &&  \\ 
        \hline
             & 8$_\mathrm{A}$ & 1 & 4.813 & 0.099 & 85.6 & 0.015 & $4.000\mathrm{e\textrm{-}}{5}$ &  & \\ 
        High & 8$_\mathrm{A}$ & 2 & 0.801 & 0.082 & 87.57 & 0.010 & $6.000\mathrm{e\textrm{-}}{4}$ & $1.546\mathrm{e\textrm{-}}{13}$ & 2  \\ 
             & 8$_\mathrm{A}$ & 3 & 1.28 & 0.032 & 87.46 & 0.003 & $1.000\mathrm{e\textrm{-}}{5}$ & &  \\ 
        \hline
             & 8$_\mathrm{B}$ & 1 & 4.813 & 0.099 & 85.6 & 0.015 & $6.000\mathrm{e\textrm{-}}{5}$ &  & \\ 
        High & 8$_\mathrm{B}$ & 2 & 1.217 & 0.082 & 87.87 & 0.006 & $1.900\mathrm{e\textrm{-}}{4}$ & $1.151\mathrm{e\textrm{-}}{11}$ & 1 \\ 
             & 8$_\mathrm{B}$ & 3 & 1.18 & 0.099 & 91.56 & 0.010 & $9.700\mathrm{e\textrm{-}}{4}$ & &  \\ 
        \hline
             & 9$_\mathrm{A}$ & 1 & 0.561 & 0.082 & 95.42 & 0.058 & $4.410\mathrm{e\textrm{-}}{3}$ &  & \\ 
        High & 9$_\mathrm{A}$ & 2 & 1.588 & 0.093 & 82.88 & 0.012 & $4.870\mathrm{e\textrm{-}}{3}$ & $1.049\mathrm{e\textrm{-}}{6}$ & 1 \\ 
             & 9$_\mathrm{A}$ & 3 & 3.547 & 0.087 & 84.0 & 0.002 & $4.881\mathrm{e\textrm{-}}{2}$ & &  \\ 
        \hline
             & 9$_\mathrm{B}$ & 1 & 1.195 & 0.10 & 88.04 & 0.476 & $1.000\mathrm{e\textrm{-}}{5}$ &  & \\ 
        High & 9$_\mathrm{B}$ & 2 & 0.574 & 0.09 & 90.06 & 0.152 & $4.120\mathrm{e\textrm{-}}{3}$ &  $1.406\mathrm{e\textrm{-}}{9}$ & 2 \\ 
             & 9$_\mathrm{B}$ & 3 & 3.547 & 0.087 & 84.0 & 0.002 & $4.621\mathrm{e\textrm{-}}{2}$ &  &  \\ 
        \hline
             & 10$_\mathrm{A}$ & 1 & 3.207 & 0.093 & 77.09 & 0.011 & $2.880\mathrm{e\textrm{-}}{2}$ &  & \\ 
        High & 10$_\mathrm{A}$ & 2 & 1.555 & 0.057 & 78.53 & 0.006 & $3.905\mathrm{e\textrm{-}}{5}$ & $5.322\mathrm{e\textrm{-}}{9}$ & 1 \\ 
             & 10$_\mathrm{A}$ & 3 & 0.543 & 0.083 & 78.17 & 0.006 & $4.733\mathrm{e\textrm{-}}{3}$ & &  \\ 
        \hline
             & 10$_\mathrm{B}$ & 1 & 3.207 & 0.093 & 77.09 & 0.011 & $2.893\mathrm{e\textrm{-}}{2}$ &  & \\ 
        High & 10$_\mathrm{B}$ & 2 & 1.124 & 0.086 & 83.29 & 0.038 & $3.454\mathrm{e\textrm{-}}{4}$ & $4.500\mathrm{e\textrm{-}}{16}$ & 2  \\ 
             & 10$_\mathrm{B}$ & 3 & 6.02 & 0.097 & 90.49 & 0.322 & $4.497\mathrm{e\textrm{-}}{11}$ &  & \\
        \hline
        \hline
        \end{supertabular}

\doublespacing
\normalsize
\bibliography{references}   

@ARTICLE{pogorelyuk2022deconfusing,
       author = {{Pogorelyuk}, Leonid and {Fitzgerald}, Riley and {Vlahakis}, Sophia and {Morgan}, Rhonda and {Cahoy}, Kerri},
        title = "{Deconfusing Detections in Directly Imaged Multiplanet Systems}",
      journal = {\apj},
         year = 2022,
        month = oct,
       volume = {937},
       number = {2},
          eid = {66},
        pages = {66},
          doi = {10.3847/1538-4357/ac8d56},
       adsurl = {https://ui.adsabs.harvard.edu/abs/2022ApJ...937...66P}
}

@BOOK{astro2020,
       author = {{National Academies of Sciences, Engineering, and Medicine}},
        title = "{Pathways to Discovery in Astronomy and Astrophysics for the 2020s}",
        publisher = {The National Academies Press},
         year = 2021,
        address = {Washington, DC},
          doi = {10.17226/26141},
       adsurl = {https://ui.adsabs.harvard.edu/abs/2021pdaa.book.....N}
}

@ARTICLE{cahoy2010exoplanet,
  title     = "{EXOPLANET ALBEDO SPECTRA AND COLORS AS A FUNCTION OF PLANET
               PHASE, SEPARATION, AND METALLICITY}",
  author    = "Cahoy, Kerri L and Marley, Mark S and Fortney, Jonathan J",
  journal   = "ApJ",
  publisher = "IOP Publishing",
  volume    =  724,
  number    =  1,
  pages     = "189",
  month     =  oct,
  year      =  2010,
  url       = "https://iopscience.iop.org/article/10.1088/0004-637X/724/1/189/meta",
  language  = "en",
  issn      = "0004-637X",
  doi       = "10.1088/0004-637X/724/1/189"
}

@ARTICLE{madhusudhan2012analytic,
       author = {{Madhusudhan}, Nikku and {Burrows}, Adam},
        title = "{Analytic Models for Albedos, Phase Curves, and Polarization of Reflected Light from Exoplanets}",
      journal = {\apj},
         year = 2012,
        month = mar,
       volume = {747},
       number = {1},
          eid = {25},
        pages = {25},
          doi = {10.1088/0004-637X/747/1/25},
archivePrefix = {arXiv},
       eprint = {1112.4476},
 primaryClass = {astro-ph.EP},
       adsurl = {https://ui.adsabs.harvard.edu/abs/2012ApJ...747...25M}
}

@INCOLLECTION{traub2010direct,
       author = {{Traub}, W.~A. and {Oppenheimer}, B.~R.},
        title = "{Direct Imaging of Exoplanets}",
    booktitle = {Exoplanets},
    publisher = {University of Arizona Press},
         year = 2010,
       editor = {{Seager}, S.},
        pages = {111-156},
       adsurl = {https://ui.adsabs.harvard.edu/abs/2010exop.book..111T}
}

@ARTICLE{smith2020utilizing,
       author = {{Smith}, Adam J.~R.~W. and {Mandell}, Avi M. and {Villanueva}, Geronimo L. and {Moore}, Michaele Dan},
        title = "{Utilizing a Database of Simulated Geometric Albedo Spectra for Photometric Characterization of Rocky Exoplanet Atmospheres}",
      journal = {\aj},
         year = 2020,
        month = nov,
       volume = {160},
       number = {5},
          eid = {204},
        pages = {204},
          doi = {10.3847/1538-3881/abb4eb},
archivePrefix = {arXiv},
       eprint = {2009.01330},
 primaryClass = {astro-ph.EP},
       adsurl = {https://ui.adsabs.harvard.edu/abs/2020AJ....160..204S}
}

@ARTICLE{Salvador2024ApJ...969L..22S,
       author = {{Salvador}, Arnaud and {Robinson}, Tyler D. and {Fortney}, Jonathan J. and {Marley}, Mark S.},
        title = "{Influence of Orbit and Mass Constraints on Reflected Light Characterization of Directly Imaged Rocky Exoplanets}",
      journal = {\apjl},
         year = 2024,
        month = jul,
       volume = {969},
       number = {1},
          eid = {L22},
        pages = {L22},
          doi = {10.3847/2041-8213/ad54c5},
archivePrefix = {arXiv},
       eprint = {2406.07749},
 primaryClass = {astro-ph.EP},
       adsurl = {https://ui.adsabs.harvard.edu/abs/2024ApJ...969L..22S}
}

@ARTICLE{batalha2018color,
       author = {{Batalha}, Natasha E. and {Smith}, Adam J.~R.~W. and {Lewis}, Nikole K. and {Marley}, Mark S. and {Fortney}, Jonathan J. and {Macintosh}, Bruce},
        title = "{Color Classification of Extrasolar Giant Planets: Prospects and Cautions}",
      journal = {\aj},
         year = 2018,
        month = oct,
       volume = {156},
       number = {4},
          eid = {158},
        pages = {158},
          doi = {10.3847/1538-3881/aad59d},
archivePrefix = {arXiv},
       eprint = {1807.08453},
 primaryClass = {astro-ph.EP},
       adsurl = {https://ui.adsabs.harvard.edu/abs/2018AJ....156..158B}
}

@INPROCEEDINGS{traub2003extrasolar,
       author = {{Traub}, Wesley A.},
        title = "{Extrasolar planet characteristics in the visible wavelength range}",
    booktitle = {Earths: DARWIN/TPF and the Search for Extrasolar Terrestrial Planets},
         year = 2003,
       editor = {{Fridlund}, Malcolm and {Henning}, Thomas and {Lacoste}, Huguette},
       series = {ESA Special Publication},
       volume = {539},
        month = oct,
        pages = {231-239},
       adsurl = {https://ui.adsabs.harvard.edu/abs/2003ESASP.539..231T}
}

@BOOK{sobolev1975light,
       author = {{Sobolev}, V.~V.},
        title = "{Light scattering in planetary atmospheres}",
         year = 1975,
        publisher = {Pergamon Press},
       adsurl = {https://ui.adsabs.harvard.edu/abs/1975lpsa.book.....S}
}

@ARTICLE{Wang2016AJ....152...97W,
       author = {{Wang}, Jason J. and {Graham}, James R. and {Pueyo}, Laurent and {Kalas}, Paul and {Millar-Blanchaer}, Maxwell A. and {Ruffio}, Jean-Baptiste and {De Rosa}, Robert J. and {Ammons}, S. Mark and {Arriaga}, Pauline and {Bailey}, Vanessa P. and {Barman}, Travis S. and {Bulger}, Joanna and {Burrows}, Adam S. and {Cardwell}, Andrew and {Chen}, Christine H. and {Chilcote}, Jeffrey K. and {Cotten}, Tara and {Fitzgerald}, Michael P. and {Follette}, Katherine B. and {Doyon}, Ren{\'e} and {Duch{\^e}ne}, Gaspard and {Greenbaum}, Alexandra Z. and {Hibon}, Pascale and {Hung}, Li-Wei and {Ingraham}, Patrick and {Konopacky}, Quinn M. and {Larkin}, James E. and {Macintosh}, Bruce and {Maire}, J{\'e}r{\^o}me and {Marchis}, Franck and {Marley}, Mark S. and {Marois}, Christian and {Metchev}, Stanimir and {Nielsen}, Eric L. and {Oppenheimer}, Rebecca and {Palmer}, David W. and {Patel}, Rahul and {Patience}, Jenny and {Perrin}, Marshall D. and {Poyneer}, Lisa A. and {Rajan}, Abhijith and {Rameau}, Julien and {Rantakyr{\"o}}, Fredrik T. and {Savransky}, Dmitry and {Sivaramakrishnan}, Anand and {Song}, Inseok and {Soummer}, Remi and {Thomas}, Sandrine and {Vasisht}, Gautam and {Vega}, David and {Wallace}, J. Kent and {Ward-Duong}, Kimberly and {Wiktorowicz}, Sloane J. and {Wolff}, Schuyler G.},
        title = "{The Orbit and Transit Prospects for {\ensuremath{\beta}} Pictoris b Constrained with One Milliarcsecond Astrometry}",
      journal = {\aj},
         year = 2016,
        month = oct,
       volume = {152},
       number = {4},
          eid = {97},
        pages = {97},
          doi = {10.3847/0004-6256/152/4/97},
archivePrefix = {arXiv},
       eprint = {1607.05272},
 primaryClass = {astro-ph.EP},
       adsurl = {https://ui.adsabs.harvard.edu/abs/2016AJ....152...97W}
}

@ARTICLE{luvoir2019arXiv,
       author = {{The LUVOIR Team}},
        title = "{The LUVOIR Mission Concept Study Final Report}",
      journal = {arXiv e-prints},
         year = 2019,
        month = dec,
          eid = {arXiv:1912.06219},
        pages = {arXiv:1912.06219},
          doi = {10.48550/arXiv.1912.06219},
archivePrefix = {arXiv},
       eprint = {1912.06219},
 primaryClass = {astro-ph.IM},
       adsurl = {https://ui.adsabs.harvard.edu/abs/2019arXiv191206219T}
}

@ARTICLE{habex2020arXiv,
       author = {{Gaudi}, B. Scott and {Seager}, Sara and {Mennesson}, Bertrand and {Kiessling}, Alina and {Warfield}, Keith and {Cahoy}, Kerri and {Clarke}, John T. and {Domagal-Goldman}, Shawn and {Feinberg}, Lee and {Guyon}, Olivier and {Kasdin}, Jeremy and {Mawet}, Dimitri and {Plavchan}, Peter and {Robinson}, Tyler and {Rogers}, Leslie and {Scowen}, Paul and {Somerville}, Rachel and {Stapelfeldt}, Karl and {Stark}, Christopher and {Stern}, Daniel and {Turnbull}, Margaret and {Amini}, Rashied and {Kuan}, Gary and {Martin}, Stefan and {Morgan}, Rhonda and {Redding}, David and {Stahl}, H. Philip and {Webb}, Ryan and {Alvarez-Salazar}, Oscar and {Arnold}, William L. and {Arya}, Manan and {Balasubramanian}, Bala and {Baysinger}, Mike and {Bell}, Ray and {Below}, Chris and {Benson}, Jonathan and {Blais}, Lindsey and {Booth}, Jeff and {Bourgeois}, Robert and {Bradford}, Case and {Brewer}, Alden and {Brooks}, Thomas and {Cady}, Eric and {Caldwell}, Mary and {Calvet}, Rob and {Carr}, Steven and {Chan}, Derek and {Cormarkovic}, Velibor and {Coste}, Keith and {Cox}, Charlie and {Danner}, Rolf and {Davis}, Jacqueline and {Dewell}, Larry and {Dorsett}, Lisa and {Dunn}, Daniel and {East}, Matthew and {Effinger}, Michael and {Eng}, Ron and {Freebury}, Greg and {Garcia}, Jay and {Gaskin}, Jonathan and {Greene}, Suzan and {Hennessy}, John and {Hilgemann}, Evan and {Hood}, Brad and {Holota}, Wolfgang and {Howe}, Scott and {Huang}, Pei and {Hull}, Tony and {Hunt}, Ron and {Hurd}, Kevin and {Johnson}, Sandra and {Kissil}, Andrew and {Knight}, Brent and {Kolenz}, Daniel and {Kraus}, Oliver and {Krist}, John and {Li}, Mary and {Lisman}, Doug and {Mandic}, Milan and {Mann}, John and {Marchen}, Luis and {Marrese-Reading}, Colleen and {McCready}, Jonathan and {McGown}, Jim and {Missun}, Jessica and {Miyaguchi}, Andrew and {Moore}, Bradley and {Nemati}, Bijan and {Nikzad}, Shouleh and {Nissen}, Joel and {Novicki}, Megan and {Perrine}, Todd and {Pineda}, Claudia and {Polanco}, Otto and {Putnam}, Dustin and {Qureshi}, Atif and {Richards}, Michael and {Eldorado Riggs}, A.~J. and {Rodgers}, Michael and {Rud}, Mike and {Saini}, Navtej and {Scalisi}, Dan and {Scharf}, Dan and {Schulz}, Kevin and {Serabyn}, Gene and {Sigrist}, Norbert and {Sikkia}, Glory and {Singleton}, Andrew and {Shaklan}, Stuart and {Smith}, Scott and {Southerd}, Bart and {Stahl}, Mark and {Steeves}, John and {Sturges}, Brian and {Sullivan}, Chris and {Tang}, Hao and {Taras}, Neil and {Tesch}, Jonathan and {Therrell}, Melissa and {Tseng}, Howard and {Valente}, Marty and {Van Buren}, David and {Villalvazo}, Juan and {Warwick}, Steve and {Webb}, David and {Westerhoff}, Thomas and {Wofford}, Rush and {Wu}, Gordon and {Woo}, Jahning and {Wood}, Milana and {Ziemer}, John and {Arney}, Giada and {Anderson}, Jay and {Ma{\'\i}z-Apell{\'a}niz}, Jes{\'u}s and {Bartlett}, James and {Belikov}, Ruslan and {Bendek}, Eduardo and {Cenko}, Brad and {Douglas}, Ewan and {Dulz}, Shannon and {Evans}, Chris and {Faramaz}, Virginie and {Feng}, Y. Katherina and {Ferguson}, Harry and {Follette}, Kate and {Ford}, Saavik and {Garc{\'\i}a}, Miriam and {Geha}, Marla and {Gelino}, Dawn and {G{\"o}tberg}, Ylva and {Hildebrandt}, Sergi and {Hu}, Renyu and {Jahnke}, Knud and {Kennedy}, Grant and {Kreidberg}, Laura and {Isella}, Andrea and {Lopez}, Eric and {Marchis}, Franck and {Macri}, Lucas and {Marley}, Mark and {Matzko}, William and {Mazoyer}, Johan and {McCandliss}, Stephan and {Meshkat}, Tiffany and {Mordasini}, Christoph and {Morris}, Patrick and {Nielsen}, Eric and {Newman}, Patrick and {Petigura}, Erik and {Postman}, Marc and {Reines}, Amy and {Roberge}, Aki and {Roederer}, Ian and {Ruane}, Garreth and {Schwieterman}, Edouard and {Sirbu}, Dan and {Spalding}, Christopher and {Teplitz}, Harry and {Tumlinson}, Jason and {Turner}, Neal and {Werk}, Jessica and {Wofford}, Aida and {Wyatt}, Mark and {Young}, Amber and {Zellem}, Rob},
        title = "{The Habitable Exoplanet Observatory (HabEx) Mission Concept Study Final Report}",
      journal = {arXiv e-prints},
         year = 2020,
        month = jan,
          eid = {arXiv:2001.06683},
        pages = {arXiv:2001.06683},
          doi = {10.48550/arXiv.2001.06683},
archivePrefix = {arXiv},
       eprint = {2001.06683},
 primaryClass = {astro-ph.IM},
       adsurl = {https://ui.adsabs.harvard.edu/abs/2020arXiv200106683G}
}

@ARTICLE{Mamajek2024arXiv240212414M,
       author = {{Mamajek}, Eric and {Stapelfeldt}, Karl},
        title = "{NASA Exoplanet Exploration Program (ExEP) Mission Star List for the Habitable Worlds Observatory (2023)}",
      journal = {arXiv e-prints},
         year = 2024,
        month = feb,
          eid = {arXiv:2402.12414},
        pages = {arXiv:2402.12414},
          doi = {10.48550/arXiv.2402.12414},
archivePrefix = {arXiv},
       eprint = {2402.12414},
 primaryClass = {astro-ph.IM},
       adsurl = {https://ui.adsabs.harvard.edu/abs/2024arXiv240212414M}
}

@ARTICLE{brown2005single,
       author = {{Brown}, Robert A.},
        title = "{Single-Visit Photometric and Obscurational Completeness}",
      journal = {\apj},
         year = 2005,
        month = may,
       volume = {624},
       number = {2},
        pages = {1010-1024},
          doi = {10.1086/429124},
archivePrefix = {arXiv},
       eprint = {astro-ph/0503077},
 primaryClass = {astro-ph},
       adsurl = {https://ui.adsabs.harvard.edu/abs/2005ApJ...624.1010B}
}

@ARTICLE{robinson2016characterizing,
  title     = "{Characterizing Rocky and Gaseous Exoplanets with 2 m Class
               Space-based Coronagraphs}",
  author    = "Robinson, Tyler D and Stapelfeldt, Karl R and Marley, Mark S",
  journal   = "PASP",
  publisher = "IOP Publishing",
  volume    =  128,
  number    =  960,
  pages     = "025003",
  month     =  jan,
  year      =  2016,
  url       = "https://iopscience.iop.org/article/10.1088/1538-3873/128/960/025003/meta",
  language  = "en",
  issn      = "1538-3873",
  doi       = "10.1088/1538-3873/128/960/025003"
}

@ARTICLE{keithly2021solar,
       author = {{Keithly}, Dean Robert and {Savransky}, Dmitry},
        title = "{The Solar System as an Exosystem: Planet Confusion}",
      journal = {\apjl},
         year = 2021,
        month = sep,
       volume = {919},
       number = {1},
          eid = {L11},
        pages = {L11},
          doi = {10.3847/2041-8213/ac20cf},
       adsurl = {https://ui.adsabs.harvard.edu/abs/2021ApJ...919L..11K}
}

@INPROCEEDINGS{horning2019minimum,
       author = {{Horning}, Andrew and {Morgan}, Rhonda and {Nielson}, Eric},
        title = "{Minimum number of observations for exoplanet orbit determination}",
    booktitle = {Society of Photo-Optical Instrumentation Engineers (SPIE) Conference Series},
         year = 2019,
       series = {Society of Photo-Optical Instrumentation Engineers (SPIE) Conference Series},
       volume = {11117},
        month = sep,
          eid = {111171C},
        pages = {111171C},
          doi = {10.1117/12.2529741},
       adsurl = {https://ui.adsabs.harvard.edu/abs/2019SPIE11117E..1CH}
}

@article{standish1992orbital,
  title={Orbital ephemerides of the Sun, Moon, and planets},
  author={Standish, E Myles and Williams, James G and others},
  journal={Explanatory supplement to the astronomical almanac},
  pages={279--323},
  year={1992},
  publisher={University Science Books Mill Valley}
}

@ARTICLE{Hirsch2013-wg,
  title     = "{A stochastic model for electron multiplication charge-coupled
               devices--from theory to practice}",
  author    = "Hirsch, Michael and Wareham, Richard J and Martin-Fernandez,
               Marisa L and Hobson, Michael P and Rolfe, Daniel J",
  journal   = "PLoS One",
  publisher = "Public Library of Science (PLoS)",
  volume    =  8,
  number    =  1,
  pages     = "e53671",
  month     =  jan,
  year      =  2013,
  url       = "https://journals.plos.org/plosone/article?id=10.1371/journal.pone.0053671",
  language  = "en",
  issn      = "1932-6203",
  pmid      = "23382848",
  doi       = "10.1371/journal.pone.0053671",
  pmc       = "PMC3561409"
}

@misc{lowfssim_bdube,
    title={LOWFSSim},
    year={2021},
    author={Brandon Dube and Brian Kern and Bijan Nemati and A.J. Riggs and Hanying Zhou and John Krist and Dan Wilson and Bala Balasubramanian and Dwight Moody},
    url={https://github.com/nasa-jpl/lowfssim/tree/a76d89e3e6c5286674da490492ccc59f5b754965},
    note={Accessed 2 January 2026},
    publisher={Jet Propulsion Laboratory, California Institute of Technology}
}

@article{kane2012habitable,
  title={The habitable zone and extreme planetary orbits},
  author={Kane, Stephen R and Gelino, Dawn M},
  journal={Astrobiology},
  volume={12},
  number={10},
  pages={940--945},
  year={2012},
  publisher={Mary Ann Liebert, Inc. 140 Huguenot Street, 3rd Floor New Rochelle, NY 10801 USA}
}

@ARTICLE{Henyey1941ApJ....93...70H,
       author = {{Henyey}, L.~G. and {Greenstein}, J.~L.},
        title = "{Diffuse radiation in the Galaxy.}",
      journal = {\apj},
         year = 1941,
        month = jan,
       volume = {93},
        pages = {70-83},
          doi = {10.1086/144246},
       adsurl = {https://ui.adsabs.harvard.edu/abs/1941ApJ....93...70H}
}

@ARTICLE{Irvine1965ApJ...142.1563I,
       author = {{Irvine}, William M.},
        title = "{Multiple Scattering by Large Particles.}",
      journal = {\apj},
         year = 1965,
        month = nov,
       volume = {142},
        pages = {1563},
          doi = {10.1086/148436},
       adsurl = {https://ui.adsabs.harvard.edu/abs/1965ApJ...142.1563I}
}

@ARTICLE{blunt2020orbitize,
       author = {{Blunt}, Sarah and {Wang}, Jason J. and {Angelo}, Isabel and {Ngo}, Henry and {Cody}, Devin and {De Rosa}, Robert J. and {Graham}, James R. and {Hirsch}, Lea and {Nagpal}, Vighnesh and {Nielsen}, Eric L. and {Pearce}, Logan and {Rice}, Malena and {Tejada}, Roberto},
        title = "{orbitize!: A Comprehensive Orbit-fitting Software Package for the High-contrast Imaging Community}",
      journal = {\aj},
         year = 2020,
        month = mar,
       volume = {159},
       number = {3},
          eid = {89},
        pages = {89},
          doi = {10.3847/1538-3881/ab6663},
archivePrefix = {arXiv},
       eprint = {1910.01756},
 primaryClass = {astro-ph.EP},
       adsurl = {https://ui.adsabs.harvard.edu/abs/2020AJ....159...89B}
}

@ARTICLE{wertz2017pyastrofit,
       author = {{Wertz}, O. and {Absil}, O. and {G{\'o}mez Gonz{\'a}lez}, C.~A. and {Milli}, J. and {Girard}, J.~H. and {Mawet}, D. and {Pueyo}, L.},
        title = "{VLT/SPHERE robust astrometry of the HR8799 planets at milliarcsecond-level accuracy. Orbital architecture analysis with PyAstrOFit}",
      journal = {\aap},
         year = 2017,
        month = feb,
       volume = {598},
          eid = {A83},
        pages = {A83},
          doi = {10.1051/0004-6361/201628730},
archivePrefix = {arXiv},
       eprint = {1610.04014},
 primaryClass = {astro-ph.EP},
       adsurl = {https://ui.adsabs.harvard.edu/abs/2017A&A...598A..83W}
}

@ARTICLE{thompson2023octofitter,
       author = {{Thompson}, William and {Lawrence}, Jensen and {Blakely}, Dori and {Marois}, Christian and {Wang}, Jason and {Giordano}, Mos{\'e} and {Brandt}, Timothy and {Johnstone}, Doug and {Ruffio}, Jean-Baptiste and {Ammons}, S. Mark and {Crotts}, Katie A. and {Do {\'O}}, Clarissa R. and {Gonzales}, Eileen C. and {Rice}, Malena},
        title = "{Octofitter: Fast, Flexible, and Accurate Orbit Modeling to Detect Exoplanets}",
      journal = {\aj},
         year = 2023,
        month = oct,
       volume = {166},
       number = {4},
          eid = {164},
        pages = {164},
          doi = {10.3847/1538-3881/acf5cc},
archivePrefix = {arXiv},
       eprint = {2402.01971},
 primaryClass = {astro-ph.EP},
       adsurl = {https://ui.adsabs.harvard.edu/abs/2023AJ....166..164T}
}

@BOOK{Green1985spas.book.....G,
       author = {{Green}, Robin M.},
        title = "{Spherical Astronomy}",
        publisher = {Cambridge University Press},
        address = {Cambridge, UK},
         year = 1985,
       adsurl = {https://ui.adsabs.harvard.edu/abs/1985spas.book.....G}
}

@INPROCEEDINGS{Riggs2021SPIE11823E..1YR,
       author = {{Riggs}, A.~J. Eldorado and {Bailey}, Vanessa and {Moody}, Dwight C. and {Sidick}, Erkin and {Balasubramanian}, Kunjithapatham and {Moore}, Douglas M. and {Wilson}, Daniel W. and {Ruane}, Garreth and {Sirbu}, Dan and {Gersh-Range}, Jessica and {Trauger}, John and {Mennesson}, Bertrand and {Siegler}, Nicholas and {Bendek}, Eduardo and {Groff}, Tyler D. and {Zimmerman}, Neil T. and {Debes}, John and {Basinger}, Scott A. and {Kasdin}, N. Jeremy},
        title = "{Flight mask designs of the Roman Space Telescope coronagraph instrument}",
    booktitle = {Techniques and Instrumentation for Detection of Exoplanets X},
         year = 2021,
       editor = {{Shaklan}, Stuart B. and {Ruane}, Garreth J.},
       series = {Society of Photo-Optical Instrumentation Engineers (SPIE) Conference Series},
       volume = {11823},
        month = sep,
          eid = {118231Y},
        pages = {118231Y},
          doi = {10.1117/12.2598599},
archivePrefix = {arXiv},
       eprint = {2108.05986},
 primaryClass = {astro-ph.IM},
       adsurl = {https://ui.adsabs.harvard.edu/abs/2021SPIE11823E..1YR}
}

@ARTICLE{Romero-Wolf2021JATIS...7b1219R,
       author = {{Romero-Wolf}, Andrew and {Bryden}, Geoffrey and {Agnes}, Greg and {Arenberg}, Jonathan W. and {Bradford}, Samuel Case and {D'Amico}, Simone and {Debes}, John and {Greenhouse}, Matt and {Hu}, Renyu and {Matousek}, Steve and {Rhodes}, Jason and {Ziemer}, John},
        title = "{Starshade Rendezvous: exoplanet orbit constraints from multi-epoch direct imaging}",
      journal = {Journal of Astronomical Telescopes, Instruments, and Systems},
         year = 2021,
        month = apr,
       volume = {7},
          eid = {021219},
        pages = {021219},
          doi = {10.1117/1.JATIS.7.2.021219},
       adsurl = {https://ui.adsabs.harvard.edu/abs/2021JATIS...7b1219R}
}

@ARTICLE{Guimond2019AJ....157..188G,
       author = {{Guimond}, Claire Marie and {Cowan}, Nicolas B.},
        title = "{Three Direct Imaging Epochs Could Constrain the Orbit of Earth 2.0 inside the Habitable Zone}",
      journal = {\aj},
         year = 2019,
        month = may,
       volume = {157},
       number = {5},
          eid = {188},
        pages = {188},
          doi = {10.3847/1538-3881/ab0f2e},
archivePrefix = {arXiv},
       eprint = {1903.06184},
 primaryClass = {astro-ph.EP},
       adsurl = {https://ui.adsabs.harvard.edu/abs/2019AJ....157..188G}
}

@ARTICLE{Blunt2024JOSS....9.6756B,
       author = {{Blunt}, Sarah and {Wang}, Jason and {Hirsch}, Lea and {Tejada}, Roberto and {Nagpal}, Vighnesh and {Surti}, Tirth and {Covarrubias}, Sofia and {McKenna}, Thea and {Ch{\'a}vez}, Rodrigo and {Llop-Sayson}, Jorge and {Arora}, Mireya and {Chavez}, Amanda and {Cody}, Devin and {Choudhary}, Saanika and {Smith}, Adam and {Balmer}, William and {Stolker}, Tomas and {Gallamore}, Hannah and {{\'O}}, Clarissa and {Nielsen}, Eric and {De Rosa}, Robert},
        title = "{orbitize! v3: Orbit fitting for the High-contrast Imaging Community}",
      journal = {The Journal of Open Source Software},
         year = 2024,
        month = sep,
       volume = {9},
       number = {101},
          eid = {6756},
        pages = {6756},
          doi = {10.21105/joss.06756},
archivePrefix = {arXiv},
       eprint = {2409.11573},
 primaryClass = {astro-ph.IM},
       adsurl = {https://ui.adsabs.harvard.edu/abs/2024JOSS....9.6756B}
}

@ARTICLE{Vousden2016MNRAS.455.1919V,
       author = {{Vousden}, W.~D. and {Farr}, W.~M. and {Mandel}, I.},
        title = "{Dynamic temperature selection for parallel tempering in Markov chain Monte Carlo simulations}",
      journal = {\mnras},
         year = 2016,
        month = jan,
       volume = {455},
       number = {2},
        pages = {1919-1937},
          doi = {10.1093/mnras/stv2422},
archivePrefix = {arXiv},
       eprint = {1501.05823},
 primaryClass = {astro-ph.IM},
       adsurl = {https://ui.adsabs.harvard.edu/abs/2016MNRAS.455.1919V}
}

@ARTICLE{Foreman-Mackey2013PASP..125..306F,
       author = {{Foreman-Mackey}, Daniel and {Hogg}, David W. and {Lang}, Dustin and {Goodman}, Jonathan},
        title = "{emcee: The MCMC Hammer}",
      journal = {\pasp},
         year = 2013,
        month = mar,
       volume = {125},
       number = {925},
        pages = {306},
          doi = {10.1086/670067},
archivePrefix = {arXiv},
       eprint = {1202.3665},
 primaryClass = {astro-ph.IM},
       adsurl = {https://ui.adsabs.harvard.edu/abs/2013PASP..125..306F}
}

@ARTICLE{Hu2019ApJ...887..166H,
       author = {{Hu}, Renyu},
        title = "{Information in the Reflected-light Spectra of Widely Separated Giant Exoplanets}",
      journal = {\apj},
         year = 2019,
        month = dec,
       volume = {887},
       number = {2},
          eid = {166},
        pages = {166},
          doi = {10.3847/1538-4357/ab58c7},
archivePrefix = {arXiv},
       eprint = {1911.06274},
 primaryClass = {astro-ph.EP},
       adsurl = {https://ui.adsabs.harvard.edu/abs/2019ApJ...887..166H}
}

@ARTICLE{Gao2017AJ....153..139G,
       author = {{Gao}, Peter and {Marley}, Mark S. and {Zahnle}, Kevin and {Robinson}, Tyler D. and {Lewis}, Nikole K.},
        title = "{Sulfur Hazes in Giant Exoplanet Atmospheres: Impacts on Reflected Light Spectra}",
      journal = {\aj},
         year = 2017,
        month = mar,
       volume = {153},
       number = {3},
          eid = {139},
        pages = {139},
          doi = {10.3847/1538-3881/aa5fab},
archivePrefix = {arXiv},
       eprint = {1701.00318},
 primaryClass = {astro-ph.EP},
       adsurl = {https://ui.adsabs.harvard.edu/abs/2017AJ....153..139G}
}

@ARTICLE{Madden2018AsBio..18.1559M,
       author = {{Madden}, J.~H. and {Kaltenegger}, Lisa},
        title = "{A Catalog of Spectra, Albedos, and Colors of Solar System Bodies for Exoplanet Comparison}",
      journal = {Astrobiology},
         year = 2018,
        month = dec,
       volume = {18},
       number = {12},
        pages = {1559-1573},
          doi = {10.1089/ast.2017.1763},
archivePrefix = {arXiv},
       eprint = {1807.11442},
 primaryClass = {astro-ph.EP},
       adsurl = {https://ui.adsabs.harvard.edu/abs/2018AsBio..18.1559M}
}

@ARTICLE{Dulz2020ApJ...893..122D,
       author = {{Dulz}, Shannon D. and {Plavchan}, Peter and {Crepp}, Justin R. and {Stark}, Christopher and {Morgan}, Rhonda and {Kane}, Stephen R. and {Newman}, Patrick and {Matzko}, William and {Mulders}, Gijs D.},
        title = "{Joint Radial Velocity and Direct Imaging Planet Yield Calculations. I. Self-consistent Planet Populations}",
      journal = {\apj},
         year = 2020,
        month = apr,
       volume = {893},
       number = {2},
          eid = {122},
        pages = {122},
          doi = {10.3847/1538-4357/ab7b73},
archivePrefix = {arXiv},
       eprint = {2003.01739},
 primaryClass = {astro-ph.EP},
       adsurl = {https://ui.adsabs.harvard.edu/abs/2020ApJ...893..122D}
}

@ARTICLE{Gladman1993Icar..106..247G,
       author = {{Gladman}, Brett},
        title = "{Dynamics of Systems of Two Close Planets}",
      journal = {\icarus},
         year = 1993,
        month = nov,
       volume = {106},
       number = {1},
        pages = {247-263},
          doi = {10.1006/icar.1993.1169},
       adsurl = {https://ui.adsabs.harvard.edu/abs/1993Icar..106..247G}
}

@ARTICLE{Smith2009Icar..201..381S,
       author = {{Smith}, Andrew W. and {Lissauer}, Jack J.},
        title = "{Orbital stability of systems of closely-spaced planets}",
      journal = {\icarus},
         year = 2009,
        month = may,
       volume = {201},
       number = {1},
        pages = {381-394},
          doi = {10.1016/j.icarus.2008.12.027},
       adsurl = {https://ui.adsabs.harvard.edu/abs/2009Icar..201..381S}
}

@ARTICLE{carrion-gonzalez2020A&A...640A.136C,
       author = {{Carri{\'o}n-Gonz{\'a}lez}, {\'O}. and {Garc{\'\i}a Mu{\~n}oz}, A. and {Cabrera}, J. and {Csizmadia}, Sz. and {Santos}, N.~C. and {Rauer}, H.},
        title = "{Directly imaged exoplanets in reflected starlight: the importance of knowing the planet radius}",
      journal = {\aap},
         year = 2020,
        month = aug,
       volume = {640},
          eid = {A136},
        pages = {A136},
          doi = {10.1051/0004-6361/202038101},
archivePrefix = {arXiv},
       eprint = {2006.08784},
 primaryClass = {astro-ph.EP},
       adsurl = {https://ui.adsabs.harvard.edu/abs/2020A&A...640A.136C}
}

@ARTICLE{Dyudina2005ApJ...618..973D,
       author = {{Dyudina}, Ulyana A. and {Sackett}, Penny D. and {Bayliss}, Daniel D.~R. and {Seager}, S. and {Porco}, Carolyn C. and {Throop}, Henry B. and {Dones}, Luke},
        title = "{Phase Light Curves for Extrasolar Jupiters and Saturns}",
      journal = {\apj},
         year = 2005,
        month = jan,
       volume = {618},
       number = {2},
        pages = {973-986},
          doi = {10.1086/426050},
archivePrefix = {arXiv},
       eprint = {astro-ph/0406390},
 primaryClass = {astro-ph},
       adsurl = {https://ui.adsabs.harvard.edu/abs/2005ApJ...618..973D}
}

@ARTICLE{Arnold2004A&A...420.1153A,
       author = {{Arnold}, L. and {Schneider}, J.},
        title = "{The detectability of extrasolar planet surroundings. I. Reflected-light photometry of unresolved rings}",
      journal = {\aap},
         year = 2004,
        month = jun,
       volume = {420},
        pages = {1153-1162},
          doi = {10.1051/0004-6361:20035720},
archivePrefix = {arXiv},
       eprint = {astro-ph/0403330},
 primaryClass = {astro-ph},
       adsurl = {https://ui.adsabs.harvard.edu/abs/2004A&A...420.1153A}
}

@ARTICLE{Limbach2024AJ....168...57L,
       author = {{Limbach}, Mary Anne and {Lustig-Yaeger}, Jacob and {Vanderburg}, Andrew and {Vos}, Johanna M. and {Heller}, Ren{\'e} and {Robinson}, Tyler D.},
        title = "{Exomoons and Exorings with the Habitable Worlds Observatory. I. On the Detection of Earth{\textendash}Moon Analog Shadows and Eclipses}",
      journal = {\aj},
         year = 2024,
        month = aug,
       volume = {168},
       number = {2},
          eid = {57},
        pages = {57},
          doi = {10.3847/1538-3881/ad4a75},
archivePrefix = {arXiv},
       eprint = {2405.02408},
 primaryClass = {astro-ph.EP},
       adsurl = {https://ui.adsabs.harvard.edu/abs/2024AJ....168...57L}
}

@ARTICLE{Coulter2022ApJS..263...15C,
       author = {{Coulter}, Daniel J. and {Barnes}, Jason W. and {Fortney}, Jonathan J.},
        title = "{Jupiter and Saturn as Spectral Analogs for Extrasolar Gas Giants and Brown Dwarfs}",
      journal = {\apjs},
         year = 2022,
        month = nov,
       volume = {263},
       number = {1},
          eid = {15},
        pages = {15},
          doi = {10.3847/1538-4365/ac886a},
archivePrefix = {arXiv},
       eprint = {2208.05541},
 primaryClass = {astro-ph.EP},
       adsurl = {https://ui.adsabs.harvard.edu/abs/2022ApJS..263...15C}
}

@INPROCEEDINGS{Bailey2023SPIE12680E..0TB,
       author = {{Bailey}, Vanessa P. and {Bendek}, Eduardo and {Monacelli}, Brian and {Baker}, Caleb and {Bedrosian}, Gasia and {Cady}, Eric and {Douglas}, Ewan S. and {Groff}, Tyler and {Hildebrandt}, Sergi R. and {Kasdin}, N. Jeremy and {Krist}, John and {Macintosh}, Bruce and {Mennesson}, Bertrand and {Morrissey}, Patrick and {Poberezhskiy}, Ilya and {Subedi}, Hari B. and {Rhodes}, Jason and {Roberge}, Aki and {Ygouf}, Marie and {Zellem}, Robert T. and {Zhao}, Feng and {Zimmerman}, Neil T.},
        title = "{Nancy Grace Roman Space Telescope coronagraph instrument overview and status}",
    booktitle = {Society of Photo-Optical Instrumentation Engineers (SPIE) Conference Series},
         year = 2023,
       series = {Society of Photo-Optical Instrumentation Engineers (SPIE) Conference Series},
       volume = {12680},
        month = oct,
          eid = {126800T},
        pages = {126800T},
          doi = {10.1117/12.2679036},
archivePrefix = {arXiv},
       eprint = {2309.08672},
 primaryClass = {astro-ph.IM},
       adsurl = {https://ui.adsabs.harvard.edu/abs/2023SPIE12680E..0TB}
}

@ARTICLE{Mennesson2020arXiv200805624M,
       author = {{Mennesson}, B. and {Juanola-Parramon}, R. and {Nemati}, B. and {Ruane}, G. and {Bailey}, V.~P. and {Bolcar}, M. and {Martin}, S. and {Zimmerman}, N. and {Stark}, C. and {Pueyo}, L. and {Benford}, D. and {Cady}, E. and {Crill}, B. and {Douglas}, E. and {Gaudi}, B.~S. and {Kasdin}, J. and {Kern}, B. and {Krist}, J. and {Kruk}, J. and {Luchik}, T. and {Macintosh}, B. and {Mandell}, A. and {Mawet}, D. and {McEnery}, J. and {Meshkat}, T. and {Poberezhskiy}, I. and {Rhodes}, J. and {Riggs}, A.~J. and {Turnbull}, M. and {Roberge}, A. and {Shi}, F. and {Siegler}, N. and {Stapelfeldt}, K. and {Ygouf}, M. and {Zellem}, R. and {Zhao}, F.},
        title = "{Paving the Way to Future Missions: the Roman Space Telescope Coronagraph Technology Demonstration}",
      journal = {arXiv e-prints},
         year = 2020,
        month = aug,
          eid = {arXiv:2008.05624},
        pages = {arXiv:2008.05624},
          doi = {10.48550/arXiv.2008.05624},
archivePrefix = {arXiv},
       eprint = {2008.05624},
 primaryClass = {astro-ph.IM},
       adsurl = {https://ui.adsabs.harvard.edu/abs/2020arXiv200805624M}
}

@ARTICLE{He2020AJ....160..276H,
       author = {{He}, Matthias Y. and {Ford}, Eric B. and {Ragozzine}, Darin and {Carrera}, Daniel},
        title = "{Architectures of Exoplanetary Systems. III. Eccentricity and Mutual Inclination Distributions of AMD-stable Planetary Systems}",
      journal = {\aj},
         year = 2020,
        month = dec,
       volume = {160},
       number = {6},
          eid = {276},
        pages = {276},
          doi = {10.3847/1538-3881/abba18},
archivePrefix = {arXiv},
       eprint = {2007.14473},
 primaryClass = {astro-ph.EP},
       adsurl = {https://ui.adsabs.harvard.edu/abs/2020AJ....160..276H}
}

@INPROCEEDINGS{Feinberg2024SPIE13092E..1NF,
       author = {{Feinberg}, Lee and {Ziemer}, John and {Ansdell}, Megan and {Crooke}, Julie and {Dressing}, Courtney and {Mennesson}, Bertrand and {O'Meara}, John and {Pepper}, Joshua and {Roberge}, Aki},
        title = "{The Habitable Worlds Observatory engineering view: status, plans, and opportunities}",
    booktitle = {Space Telescopes and Instrumentation 2024: Optical, Infrared, and Millimeter Wave},
         year = 2024,
       editor = {{Coyle}, Laura E. and {Matsuura}, Shuji and {Perrin}, Marshall D.},
       series = {Society of Photo-Optical Instrumentation Engineers (SPIE) Conference Series},
       volume = {13092},
        month = aug,
          eid = {130921N},
        pages = {130921N},
          doi = {10.1117/12.3018328},
       adsurl = {https://ui.adsabs.harvard.edu/abs/2024SPIE13092E..1NF}
}

@misc{roman_ipac_website,
    author={{Roman Science Support Center at IPAC}},
    title={{Spacecraft and Instrument Parameters (previous releases)}},
    url="https://roman.ipac.caltech.edu/page/param-db-2?csvfile=RomanParameters_Phase_C_01-20-2023.csv&csvid=18",
    note={{\href{https://roman.ipac.caltech.edu/page/param-db-2?csvfile=RomanParameters_Phase_C_01-20-2023.csv&csvid=18}{https://roman.ipac.caltech.edu/page/param-db-2?csvfile=RomanParameters\_Phase\_C\_01-20-2023.csv\&csvid=18} (accessed 06.01.2025)}}
}

@ARTICLE{Mallama2017Icar..282...19M,
       author = {{Mallama}, Anthony and {Krobusek}, Bruce and {Pavlov}, Hristo},
        title = "{Comprehensive wide-band magnitudes and albedos for the planets, with applications to exo-planets and Planet Nine}",
      journal = {\icarus},
         year = 2017,
        month = jan,
       volume = {282},
        pages = {19-33},
          doi = {10.1016/j.icarus.2016.09.023},
archivePrefix = {arXiv},
       eprint = {1609.05048},
 primaryClass = {astro-ph.EP},
       adsurl = {https://ui.adsabs.harvard.edu/abs/2017Icar..282...19M}
}

@INPROCEEDINGS{Defrere2012SPIE.8442E..0MD,
       author = {{Defr{\`e}re}, D. and {Stark}, C. and {Cahoy}, K. and {Beerer}, I.},
        title = "{Direct imaging of exoEarths embedded in clumpy debris disks}",
    booktitle = {Space Telescopes and Instrumentation 2012: Optical, Infrared, and Millimeter Wave},
         year = 2012,
       editor = {{Clampin}, Mark C. and {Fazio}, Giovanni G. and {MacEwen}, Howard A. and {Oschmann}, Jr., Jacobus M.},
       series = {Society of Photo-Optical Instrumentation Engineers (SPIE) Conference Series},
       volume = {8442},
        month = sep,
          eid = {84420M},
        pages = {84420M},
          doi = {10.1117/12.926324},
       adsurl = {https://ui.adsabs.harvard.edu/abs/2012SPIE.8442E..0MD}
}

@ARTICLE{Currie2025arXiv250319932C,
       author = {{Currie}, Miles H. and {Debes}, John and {Hasegawa}, Yasuhiro and {Rebollido}, Isabel and {Faramaz}, Virginie and {Ertel}, Steve and {Danchi}, William and {Mennesson}, Bertrand and {Wyatt}, Mark and {SAG23 Members}, NASA},
        title = "{Exozodiacal dust as a limitation to exoplanet imaging and spectroscopy}",
      journal = {arXiv e-prints},
         year = 2025,
        month = mar,
          eid = {arXiv:2503.19932},
        pages = {arXiv:2503.19932},
          doi = {10.48550/arXiv.2503.19932},
archivePrefix = {arXiv},
       eprint = {2503.19932},
 primaryClass = {astro-ph.IM},
       adsurl = {https://ui.adsabs.harvard.edu/abs/2025arXiv250319932C}
}

@ARTICLE{Roberge2012PASP..124..799R,
       author = {{Roberge}, Aki and {Chen}, Christine H. and {Millan-Gabet}, Rafael and {Weinberger}, Alycia J. and {Hinz}, Philip M. and {Stapelfeldt}, Karl R. and {Absil}, Olivier and {Kuchner}, Marc J. and {Bryden}, Geoffrey},
        title = "{The Exozodiacal Dust Problem for Direct Observations of Exo-Earths}",
      journal = {\pasp},
         year = 2012,
        month = aug,
       volume = {124},
       number = {918},
        pages = {799},
          doi = {10.1086/667218},
archivePrefix = {arXiv},
       eprint = {1204.0025},
 primaryClass = {astro-ph.EP},
       adsurl = {https://ui.adsabs.harvard.edu/abs/2012PASP..124..799R}
}

@ARTICLE{Levasseur-Regourd1980A&A....84..277L,
       author = {{Levasseur-Regourd}, A.~C. and {Dumont}, R.},
        title = "{Absolute photometry of zodiacal light.}",
      journal = {\aap},
         year = 1980,
        month = apr,
       volume = {84},
        pages = {277-279},
       adsurl = {https://ui.adsabs.harvard.edu/abs/1980A&A....84..277L}
}

@ARTICLE{Cox2000PhT....53j..77C,
       author = {{Cox}, Arthur N. and {Pilachowski}, Catherine A.},
        title = "{Allen's Astrophysical Quantities}",
      journal = {Physics Today},
         year = 2000,
        month = oct,
       volume = {53},
       number = {10},
        pages = {77},
          doi = {10.1063/1.1325201},
       adsurl = {https://ui.adsabs.harvard.edu/abs/2000PhT....53j..77C}
}

@ARTICLE{Mesa2023A&A...672A..93M,
       author = {{Mesa}, D. and {Gratton}, R. and {Kervella}, P. and {Bonavita}, M. and {Desidera}, S. and {D'Orazi}, V. and {Marino}, S. and {Zurlo}, A. and {Rigliaco}, E.},
        title = "{AF Lep b: The lowest-mass planet detected by coupling astrometric and direct imaging data}",
      journal = {\aap},
         year = 2023,
        month = apr,
       volume = {672},
          eid = {A93},
        pages = {A93},
          doi = {10.1051/0004-6361/202345865},
archivePrefix = {arXiv},
       eprint = {2302.06213},
 primaryClass = {astro-ph.EP},
       adsurl = {https://ui.adsabs.harvard.edu/abs/2023A&A...672A..93M}
}

@ARTICLE{DoO2025ApJ...995..190D,
       author = {{Do {\'O}}, Clarissa R. and {Bae}, Jaehan and {Konopacky}, Quinn M. and {Nguyen}, Jayke S. and {Diamond}, Patrick and {Go{\'z}dziewski}, Krzysztof and {Jankowski}, Dawid},
        title = "{On the Orbital Evolution of Multiple Wide Super-Jupiters: How Disk Migration and Dispersal Shape the Stability of the PDS 70 System}",
      journal = {\apj},
         year = 2025,
        month = dec,
       volume = {995},
       number = {2},
          eid = {190},
        pages = {190},
          doi = {10.3847/1538-4357/ae12ec},
archivePrefix = {arXiv},
       eprint = {2510.11021},
 primaryClass = {astro-ph.EP},
       adsurl = {https://ui.adsabs.harvard.edu/abs/2025ApJ...995..190D}
}

@ARTICLE{Blunt2023AJ....166..257B,
       author = {{Blunt}, Sarah and {Balmer}, W.~O. and {Wang}, J.~J. and {Lacour}, S. and {Petrus}, S. and {Bourdarot}, G. and {Kammerer}, J. and {Pourr{\'e}}, N. and {Rickman}, E. and {Shangguan}, J. and {Winterhalder}, T. and {Abuter}, R. and {Amorim}, A. and {Asensio-Torres}, R. and {Benisty}, M. and {Berger}, J.-P. and {Beust}, H. and {Boccaletti}, A. and {Bohn}, A. and {Bonnefoy}, M. and {Bonnet}, H. and {Brandner}, W. and {Cantalloube}, F. and {Caselli}, P. and {Charnay}, B. and {Chauvin}, G. and {Chavez}, A. and {Choquet}, E. and {Christiaens}, V. and {Cl{\'e}net}, Y. and {Du Foresto}, V. Coud{\'e} and {Cridland}, A. and {Dembet}, R. and {Drescher}, A. and {Duvert}, G. and {Eckart}, A. and {Eisenhauer}, F. and {Feuchtgruber}, H. and {Garcia}, P. and {Garcia Lopez}, R. and {Gendron}, E. and {Genzel}, R. and {Gillessen}, S. and {Girard}, J.~H. and {Haubois}, X. and {Hei{\ss}el}, G. and {Henning}, Th. and {Hinkley}, S. and {Hippler}, S. and {Horrobin}, M. and {Houll{\'e}}, M. and {Hubert}, Z. and {Jocou}, L. and {Keppler}, M. and {Kervella}, P. and {Kreidberg}, L. and {Lagrange}, A.-M. and {Lapeyr{\`e}re}, V. and {Le Bouquin}, J.-B. and {L{\'e}na}, P. and {Lutz}, D. and {Maire}, A.-L. and {Mang}, F. and {Marleau}, G.-D. and {M{\'e}rand}, A. and {Molli{\`e}re}, P. and {Monnier}, J.~D. and {Mordasini}, C. and {Mouillet}, D. and {Nasedkin}, E. and {Nowak}, M. and {Ott}, T. and {Otten}, G.~P.~P.~L. and {Paladini}, C. and {Paumard}, T. and {Perraut}, K. and {Perrin}, G. and {Pfuhl}, O. and {Pueyo}, L. and {Rameau}, J. and {Rodet}, L. and {Rustamkulov}, Z. and {Shimizu}, T. and {Sing}, D. and {Stolker}, T. and {Straubmeier}, C. and {Sturm}, E. and {Tacconi}, L.~J. and {van Dishoeck}, E.~F. and {Vigan}, A. and {Vincent}, F. and {Ward-Duong}, K. and {Widmann}, F. and {Wieprecht}, E. and {Wiezorrek}, E. and {Woillez}, J. and {Yazici}, S. and {Young}, A. and {Exogravity Collaboration}},
        title = "{First VLTI/GRAVITY Observations of HIP 65426 b: Evidence for a Low or Moderate Orbital Eccentricity}",
      journal = {\aj},
         year = 2023,
        month = dec,
       volume = {166},
       number = {6},
          eid = {257},
        pages = {257},
          doi = {10.3847/1538-3881/ad06b7},
archivePrefix = {arXiv},
       eprint = {2310.00148},
 primaryClass = {astro-ph.EP},
       adsurl = {https://ui.adsabs.harvard.edu/abs/2023AJ....166..257B}
}

@ARTICLE{Brandt2021AJ....162..186B,
       author = {{Brandt}, Timothy D. and {Dupuy}, Trent J. and {Li}, Yiting and {Brandt}, G. Mirek and {Zeng}, Yunlin and {Michalik}, Daniel and {Bardalez Gagliuffi}, Daniella C. and {Raposo-Pulido}, Virginia},
        title = "{orvara: An Efficient Code to Fit Orbits Using Radial Velocity, Absolute, and/or Relative Astrometry}",
      journal = {\aj},
         year = 2021,
        month = nov,
       volume = {162},
       number = {5},
          eid = {186},
        pages = {186},
          doi = {10.3847/1538-3881/ac042e},
archivePrefix = {arXiv},
       eprint = {2105.11671},
 primaryClass = {astro-ph.IM},
       adsurl = {https://ui.adsabs.harvard.edu/abs/2021AJ....162..186B}
}

@ARTICLE{Earl2005PCCP....7.3910E,
       author = {{Earl}, David J. and {Deem}, Michael W.},
        title = "{Parallel tempering: Theory, applications, and new perspectives}",
      journal = {Physical Chemistry Chemical Physics (Incorporating Faraday Transactions)},
         year = 2005,
        month = jan,
       volume = {7},
       number = {23},
        pages = {3910},
          doi = {10.1039/B509983H},
archivePrefix = {arXiv},
       eprint = {physics/0508111},
 primaryClass = {physics.comp-ph},
       adsurl = {https://ui.adsabs.harvard.edu/abs/2005PCCP....7.3910E}
}

@INCOLLECTION{Marley2013cctp.book..367M,
       author = {{Marley}, M.~S. and {Ackerman}, A.~S. and {Cuzzi}, J.~N. and {Kitzmann}, D.},
        title = "{Clouds and Hazes in Exoplanet Atmospheres}",
    booktitle = {Comparative Climatology of Terrestrial Planets},
         year = 2013,
       editor = {{Mackwell}, Stephen J. and {Simon-Miller}, Amy A. and {Harder}, Jerald W. and {Bullock}, Mark A.},
        pages = {367-392},
          doi = {10.2458/azu_uapress_9780816530595-ch015},
       adsurl = {https://ui.adsabs.harvard.edu/abs/2013cctp.book..367M}
}

@ARTICLE{Venkatesan2025AsBio..25...42V,
       author = {{Venkatesan}, Vidya and {Shields}, Aomawa L. and {Deitrick}, Russell and {Wolf}, Eric T. and {Rushby}, Andrew},
        title = "{A One-Dimensional Energy Balance Model Parameterization for the Formation of CO$_{2}$ Ice on the Surfaces of Eccentric Extrasolar Planets}",
      journal = {Astrobiology},
         year = 2025,
        month = jan,
       volume = {25},
       number = {1},
        pages = {42-59},
          doi = {10.1089/ast.2023.0103},
archivePrefix = {arXiv},
       eprint = {2501.11667},
 primaryClass = {astro-ph.EP},
       adsurl = {https://ui.adsabs.harvard.edu/abs/2025AsBio..25...42V}
}

@ARTICLE{Zugger2010ApJ...723.1168Z,
       author = {{Zugger}, M.~E. and {Kasting}, J.~F. and {Williams}, D.~M. and {Kane}, T.~J. and {Philbrick}, C.~R.},
        title = "{Light Scattering from Exoplanet Oceans and Atmospheres}",
      journal = {\apj},
         year = 2010,
        month = nov,
       volume = {723},
       number = {2},
        pages = {1168-1179},
          doi = {10.1088/0004-637X/723/2/1168},
archivePrefix = {arXiv},
       eprint = {1006.3525},
 primaryClass = {astro-ph.EP},
       adsurl = {https://ui.adsabs.harvard.edu/abs/2010ApJ...723.1168Z}
}

@ARTICLE{Fujii2012ApJ...755..101F,
       author = {{Fujii}, Yuka and {Kawahara}, Hajime},
        title = "{Mapping Earth Analogs from Photometric Variability: Spin-Orbit Tomography for Planets in Inclined Orbits}",
      journal = {\apj},
         year = 2012,
        month = aug,
       volume = {755},
       number = {2},
          eid = {101},
        pages = {101},
          doi = {10.1088/0004-637X/755/2/101},
archivePrefix = {arXiv},
       eprint = {1204.3504},
 primaryClass = {astro-ph.EP},
       adsurl = {https://ui.adsabs.harvard.edu/abs/2012ApJ...755..101F}
}

@ARTICLE{Cowan2013MNRAS.434.2465C,
       author = {{Cowan}, Nicolas B. and {Fuentes}, Pablo A. and {Haggard}, Hal M.},
        title = "{Light curves of stars and exoplanets: estimating inclination, obliquity and albedo}",
      journal = {\mnras},
         year = 2013,
        month = sep,
       volume = {434},
       number = {3},
        pages = {2465-2479},
          doi = {10.1093/mnras/stt1191},
archivePrefix = {arXiv},
       eprint = {1304.6398},
 primaryClass = {astro-ph.EP},
       adsurl = {https://ui.adsabs.harvard.edu/abs/2013MNRAS.434.2465C}
}

@ARTICLE{DeCock2022AJ....163....5D,
       author = {{De Cock}, Roderick and {Livengood}, Timothy A. and {Stam}, Daphne M. and {Lisse}, Carey M. and {Hewagama}, Tilak and {Deming}, L. Drake},
        title = "{Terrestrial Planet Optical Phase Curves. I. Direct Measurements of the Earth}",
      journal = {\aj},
         year = 2022,
        month = jan,
       volume = {163},
       number = {1},
          eid = {5},
        pages = {5},
          doi = {10.3847/1538-3881/ac3234},
       adsurl = {https://ui.adsabs.harvard.edu/abs/2022AJ....163....5D}
}

@MISC{Morgan2019sdet.rept.....M,
       author = {{Morgan}, Rhonda and {Savransky}, Dmitry and {Stark}, Chris and {Nielsen}, Eric},
        title = "{The Standard Definitions and Evaluation Team Final Report: A common comparison of exoplanet yield}",
 howpublished = {The Standard Definitions and Evaluation Team Final Report: A common comparison of exoplanet yield, 2019.},
         year = 2019,
        month = oct,
       adsurl = {https://ui.adsabs.harvard.edu/abs/2019sdet.rept.....M}
}

@ARTICLE{Greco2015ApJ...808..172G,
       author = {{Greco}, Johnny P. and {Burrows}, Adam},
        title = "{The Direct Detectability of Giant Exoplanets in the Optical}",
      journal = {\apj},
         year = 2015,
        month = aug,
       volume = {808},
       number = {2},
          eid = {172},
        pages = {172},
          doi = {10.1088/0004-637X/808/2/172},
archivePrefix = {arXiv},
       eprint = {1505.07832},
 primaryClass = {astro-ph.EP},
       adsurl = {https://ui.adsabs.harvard.edu/abs/2015ApJ...808..172G}
}

@ARTICLE{Mayorga2016AJ....152..209M,
       author = {{Mayorga}, L.~C. and {Jackiewicz}, J. and {Rages}, K. and {West}, R.~A. and {Knowles}, B. and {Lewis}, N. and {Marley}, M.~S.},
        title = "{Jupiter{\textquoteright}s Phase Variations from Cassini: A Testbed for Future Direct-imaging Missions}",
      journal = {\aj},
         year = 2016,
        month = dec,
       volume = {152},
       number = {6},
          eid = {209},
        pages = {209},
          doi = {10.3847/0004-6256/152/6/209},
archivePrefix = {arXiv},
       eprint = {1610.07679},
 primaryClass = {astro-ph.EP},
       adsurl = {https://ui.adsabs.harvard.edu/abs/2016AJ....152..209M}
}

@ARTICLE{Mallama2018A&C....25...10M,
       author = {{Mallama}, A. and {Hilton}, J.~L.},
        title = "{Computing apparent planetary magnitudes for The Astronomical Almanac}",
      journal = {Astronomy and Computing},
         year = 2018,
        month = oct,
       volume = {25},
        pages = {10-24},
          doi = {10.1016/j.ascom.2018.08.002},
archivePrefix = {arXiv},
       eprint = {1808.01973},
 primaryClass = {astro-ph.EP},
       adsurl = {https://ui.adsabs.harvard.edu/abs/2018A&C....25...10M}
}

@ARTICLE{Mallama2009Icar..204...11M,
       author = {{Mallama}, Anthony},
        title = "{Characterization of terrestrial exoplanets based on the phase curves and albedos of Mercury, Venus and Mars}",
      journal = {\icarus},
         year = 2009,
        month = nov,
       volume = {204},
       number = {1},
        pages = {11-14},
          doi = {10.1016/j.icarus.2009.07.010},
       adsurl = {https://ui.adsabs.harvard.edu/abs/2009Icar..204...11M}
}

@INPROCEEDINGS{Perkins2024SPIE13092E..0RP,
       author = {{Perkins}, Jeremy S. and {Wollack}, Edward J. and {Content}, David A. and {Abel}, Joshua C. and {Baker}, Joanne L. and {Bartusek}, Lisa M. and {Bolcar}, Matthew R. and {Han}, Lawrence L. and {Harper}, Alexia M. and {Kruk}, Jeffrey W. and {Lui}, Kuo-Chia (Alice) and {Poberezhskiy}, Ilya Y. and {Rizzo}, Maxime J. and {Smith}, Jeffrey Scott and {Schlieder}, Joshua E. and {Vess}, Melissa F. and {Zimmerman}, Neil T.},
        title = "{The Roman Space Telescope observatory build, test, and verification status}",
    booktitle = {Space Telescopes and Instrumentation 2024: Optical, Infrared, and Millimeter Wave},
         year = 2024,
       editor = {{Coyle}, Laura E. and {Matsuura}, Shuji and {Perrin}, Marshall D.},
       series = {Society of Photo-Optical Instrumentation Engineers (SPIE) Conference Series},
       volume = {13092},
        month = aug,
          eid = {130920R},
        pages = {130920R},
          doi = {10.1117/12.3022616},
       adsurl = {https://ui.adsabs.harvard.edu/abs/2024SPIE13092E..0RP}
}

@ARTICLE{Christiansen2025PSJ.....6..186C,
       author = {{Christiansen}, Jessie L. and {McElroy}, Douglas L. and {Harbut}, Marcy and {Ciardi}, David R. and {Crane}, Megan and {Good}, John and {Hardegree-Ullman}, Kevin K. and {Kesseli}, Aurora Y. and {Lund}, Michael B. and {Lynn}, Meca and {Muthiar}, Ananda and {Nilsson}, Ricky and {Oluyide}, Toba and {Papin}, Michael and {Rivera}, Amalia and {Swain}, Melanie and {Susemiehl}, Nicholas D. and {Tam}, Raymond and {van Eyken}, Julian and {Beichman}, Charles},
        title = "{The NASA Exoplanet Archive and Exoplanet Follow-up Observing Program: Data, Tools, and Usage}",
      journal = {\psj},
         year = 2025,
        month = aug,
       volume = {6},
       number = {8},
          eid = {186},
        pages = {186},
          doi = {10.3847/PSJ/ade3c2},
archivePrefix = {arXiv},
       eprint = {2506.03299},
 primaryClass = {astro-ph.EP},
       adsurl = {https://ui.adsabs.harvard.edu/abs/2025PSJ.....6..186C}
}

@ARTICLE{Zellem2014ApJ...790...53Z,
       author = {{Zellem}, Robert T. and {Lewis}, Nikole K. and {Knutson}, Heather A. and {Griffith}, Caitlin A. and {Showman}, Adam P. and {Fortney}, Jonathan J. and {Cowan}, Nicolas B. and {Agol}, Eric and {Burrows}, Adam and {Charbonneau}, David and {Deming}, Drake and {Laughlin}, Gregory and {Langton}, Jonathan},
        title = "{The 4.5 {\ensuremath{\mu}}m Full-orbit Phase Curve of the Hot Jupiter HD 209458b}",
      journal = {\apj},
         year = 2014,
        month = jul,
       volume = {790},
       number = {1},
          eid = {53},
        pages = {53},
          doi = {10.1088/0004-637X/790/1/53},
archivePrefix = {arXiv},
       eprint = {1405.5923},
 primaryClass = {astro-ph.EP},
       adsurl = {https://ui.adsabs.harvard.edu/abs/2014ApJ...790...53Z}
}

@ARTICLE{Parmentier2021MNRAS.501...78P,
       author = {{Parmentier}, Vivien and {Showman}, Adam P. and {Fortney}, Jonathan J.},
        title = "{The cloudy shape of hot Jupiter thermal phase curves}",
      journal = {\mnras},
         year = 2021,
        month = jan,
       volume = {501},
       number = {1},
        pages = {78-108},
          doi = {10.1093/mnras/staa3418},
archivePrefix = {arXiv},
       eprint = {2010.06934},
 primaryClass = {astro-ph.EP},
       adsurl = {https://ui.adsabs.harvard.edu/abs/2021MNRAS.501...78P}
}

@ARTICLE{Mayorga2020AJ....160..238M,
       author = {{Mayorga}, L.~C. and {Charbonneau}, David and {Thorngren}, D.~P.},
        title = "{Reflected Light Observations of the Galilean Satellites from Cassini: A Test Bed for Cold Terrestrial Exoplanets}",
      journal = {\aj},
         year = 2020,
        month = nov,
       volume = {160},
       number = {5},
          eid = {238},
        pages = {238},
          doi = {10.3847/1538-3881/abb8df},
archivePrefix = {arXiv},
       eprint = {2009.05467},
 primaryClass = {astro-ph.EP},
       adsurl = {https://ui.adsabs.harvard.edu/abs/2020AJ....160..238M}
}

@INPROCEEDINGS{Morgan2022SPIE12180E..20M,
       author = {{Morgan}, Rhonda and {Savransky}, Dmitry and {Turmon}, Michael and {Genszler}, Grace and {Mamajek}, Eric E. and {Robinson}, Tyler D. and {Stapelfeldt}, Karl},
        title = "{An exploration of expected number of exoplanets for a 6m class direct imaging observatory}",
    booktitle = {Space Telescopes and Instrumentation 2022: Optical, Infrared, and Millimeter Wave},
         year = 2022,
       editor = {{Coyle}, Laura E. and {Matsuura}, Shuji and {Perrin}, Marshall D.},
       series = {Society of Photo-Optical Instrumentation Engineers (SPIE) Conference Series},
       volume = {12180},
        month = aug,
          eid = {1218020},
        pages = {1218020},
          doi = {10.1117/12.2630609},
       adsurl = {https://ui.adsabs.harvard.edu/abs/2022SPIE12180E..20M}
}

@ARTICLE{Stark2016JATIS...2d1204S,
       author = {{Stark}, Christopher C. and {Shaklan}, Stuart and {Lisman}, Doug and {Cady}, Eric and {Savransky}, Dmitry and {Roberge}, Aki and {Mandell}, Avi M.},
        title = "{Maximized exoEarth candidate yields for starshades}",
      journal = {Journal of Astronomical Telescopes, Instruments, and Systems},
         year = 2016,
        month = oct,
       volume = {2},
          eid = {041204},
        pages = {041204},
          doi = {10.1117/1.JATIS.2.4.041204},
archivePrefix = {arXiv},
       eprint = {1605.04915},
 primaryClass = {astro-ph.IM},
       adsurl = {https://ui.adsabs.harvard.edu/abs/2016JATIS...2d1204S}
}

@ARTICLE{Irwin2025MNRAS.540.1719I,
       author = {{Irwin}, Patrick G.~J. and {Wenkert}, Daniel D. and {Simon}, Amy A. and {Dahl}, Emma and {Hammel}, Heidi B.},
        title = "{The bolometric Bond albedo and energy balance of Uranus}",
      journal = {\mnras},
         year = 2025,
        month = jun,
       volume = {540},
       number = {2},
        pages = {1719-1729},
          doi = {10.1093/mnras/staf800},
archivePrefix = {arXiv},
       eprint = {2502.18971},
 primaryClass = {astro-ph.EP},
       adsurl = {https://ui.adsabs.harvard.edu/abs/2025MNRAS.540.1719I}
}

@ARTICLE{Palle2016GeoRL..43.4531P,
       author = {{Palle}, E. and {Goode}, P.~R. and {Monta{\~n}{\'e}s-Rodr{\'\i}guez}, P. and {Shumko}, A. and {Gonzalez-Merino}, B. and {Martinez-Lombilla}, C. and {Jimenez-Ibarra}, F. and {Shumko}, S. and {Sanroma}, E. and {Hulist}, A. and {Miles-Paez}, P. and {Murgas}, F. and {Nowak}, G. and {Koonin}, S.~E.},
        title = "{Earth's albedo variations 1998-2014 as measured from ground-based earthshine observations}",
      journal = {\grl},
         year = 2016,
        month = may,
       volume = {43},
       number = {9},
        pages = {4531-4538},
          doi = {10.1002/2016GL068025},
archivePrefix = {arXiv},
       eprint = {1604.05880},
 primaryClass = {astro-ph.EP},
       adsurl = {https://ui.adsabs.harvard.edu/abs/2016GeoRL..43.4531P}
}

@ARTICLE{Lockwood2006Icar..180..442L,
       author = {{Lockwood}, G.~W. and {Jerzykiewicz}, Miko{\l}aj},
        title = "{Photometric variability of Uranus and Neptune, 1950 2004}",
      journal = {\icarus},
         year = 2006,
        month = feb,
       volume = {180},
       number = {2},
        pages = {442-452},
          doi = {10.1016/j.icarus.2005.09.009},
       adsurl = {https://ui.adsabs.harvard.edu/abs/2006Icar..180..442L}
}

@ARTICLE{Mallama2007Icar..192..404M,
       author = {{Mallama}, Anthony},
        title = "{The magnitude and albedo of Mars}",
      journal = {\icarus},
         year = 2007,
        month = dec,
       volume = {192},
       number = {2},
        pages = {404-416},
          doi = {10.1016/j.icarus.2007.07.011},
       adsurl = {https://ui.adsabs.harvard.edu/abs/2007Icar..192..404M}
}

@ARTICLE{Mallama2012Icar..220..211M,
       author = {{Mallama}, Anthony and {Schmude}, Richard W.},
        title = "{Cloud band variations and the integrated luminosity of Jupiter}",
      journal = {\icarus},
         year = 2012,
        month = jul,
       volume = {220},
       number = {1},
        pages = {211-215},
          doi = {10.1016/j.icarus.2012.04.031},
       adsurl = {https://ui.adsabs.harvard.edu/abs/2012Icar..220..211M}
}

@book{lambert1760photometria,
  title={Lamberts Photometrie: (Photometria, sive De mensura et gradibus luminis, colorum et umbrae)},
  author={Lambert, Johann Heinrich},
  year={1760}
}

@ARTICLE{Tinetti2006AsBio...6..881T,
       author = {{Tinetti}, Giovanna and {Meadows}, Victoria S. and {Crisp}, David and {Kiang}, Nancy Y. and {Kahn}, Brian H. and {Fishbein}, Evan and {Velusamy}, Thangasamy and {Turnbull}, Margaret},
        title = "{Detectability of Planetary Characteristics in Disk-Averaged Spectra II: Synthetic Spectra and Light-Curves of Earth}",
      journal = {Astrobiology},
         year = 2006,
        month = dec,
       volume = {6},
       number = {6},
        pages = {881-900},
          doi = {10.1089/ast.2006.6.881},
       adsurl = {https://ui.adsabs.harvard.edu/abs/2006AsBio...6..881T}
}

@ARTICLE{Stephens2015RvGeo..53..141S,
       author = {{Stephens}, Graeme L. and {O'Brien}, Denis and {Webster}, Peter J. and {Pilewski}, Peter and {Kato}, Seiji and {Li}, Jui-lin},
        title = "{The albedo of Earth}",
      journal = {Reviews of Geophysics},
         year = 2015,
        month = mar,
       volume = {53},
       number = {1},
        pages = {141-163},
          doi = {10.1002/2014RG000449},
       adsurl = {https://ui.adsabs.harvard.edu/abs/2015RvGeo..53..141S}
}

@ARTICLE{Mallama2006Icar..182...10M,
       author = {{Mallama}, Anthony and {Wang}, Dennis and {Howard}, Russell A.},
        title = "{Venus phase function and forward scattering from H $_{2}$SO $_{4}$}",
      journal = {\icarus},
         year = 2006,
        month = may,
       volume = {182},
       number = {1},
        pages = {10-22},
          doi = {10.1016/j.icarus.2005.12.014},
       adsurl = {https://ui.adsabs.harvard.edu/abs/2006Icar..182...10M}
}
\bibliographystyle{spiejour}   


\vspace{2ex}\noindent\textbf{Samantha N.~Hasler} is a Postdoctoral Associate in the Department of Earth, Atmospheric, and Planetary Sciences at the Massachusetts Institute of Technology (MIT). Her research focuses on using phase curves and color to detect and characterize exoplanets in reflected light direct imaging. She also supports the development of analysis tools for exoplanet and Solar System observations. She holds a PhD in Planetary Science from MIT and BS in Physics from Southeast Missouri State University. 

\vspace{2ex}\noindent\textbf{Riley Fitzgerald} is an Assistant Professor of Aerospace and Ocean Engineering at Virginia Tech, where he is the Ryan and Krista Frederic Junior Faculty Fellow. His research and teaching focus on astrodynamics, spacecraft GNC, and space system design. He holds PhD and MS degrees in Aeronautics and Astronautics from the Massachusetts Institute of Technology, and a BSE in Mechanical and Aerospace Engineering from Princeton University. 

\vspace{2ex}\noindent\textbf{Kerri L.~Cahoy} is the Sheila Evans Widnall 1960 Professor in the Department of Aeronautics and Astronautics at MIT. She received her BS degree (2000) in electrical engineering from Cornell University, and her MS (2002) and PhD (2008) degrees in electrical engineering from Stanford University. She joined the MIT faculty in 2011 and leads the space telecommunications, astronomy, and radiation laboratory (STAR Lab), focusing on nanosatellite and space telescope technology demonstrations as well as satellite remote sensing and autonomy.

\vspace{1ex}
\noindent Biographies of the other authors are not available.

\listoffigures

\end{spacing}
\end{document}